\documentclass[a4paper,11pt]{article}

\pdfoutput=1
\usepackage{jheppub}

\usepackage{graphicx,color,amsmath,amssymb,mathtools}
\usepackage{siunitx}

\usepackage{slashed}
\usepackage{tcolorbox}

\usepackage[bb=dsserif]{mathalpha}
\usepackage{bm}

\usepackage[export]{adjustbox}
\usepackage{tikz}
\usepackage{tkz-euclide}
\usetikzlibrary{decorations.pathmorphing}	
\tikzset{
    v/.style={decorate, decoration={snake, segment length=3mm, amplitude=0.75mm}, draw},
    f/.style={draw=black, postaction={decorate},
        decoration={markings,mark=at position .6 with {\arrow[very thick]{latex}}}},
    fb/.style={draw=black, postaction={decorate},
        decoration={markings,mark=at position .4 with {\arrowreversed[very thick]{latex}}}},
    fnar/.style={draw=black},
    g/.style={decorate, draw=black,
        decoration={coil,amplitude=3pt, segment length=3.5pt}},
    s/.style={dashed,draw=black, postaction={decorate},
        decoration={markings,mark=at position .55 with {\arrow[very thick]{latex}}}},
    sb/.style={dashed,draw=black, postaction={decorate},
        decoration={markings,mark=at position .55 with {\arrowreversed[draw=black,very thick]{latex}}}},
    snar/.style={dashed,draw=black,line width =1.25pt},
}

\definecolor{deepblue}{rgb}{0.2,0.2,0.8}

\definecolor{deepred}{rgb}{0.8,0.2,0.2}

\newcommand{\di}{\text{d}}
\newcommand{\pare}[1]{\left(#1\right)}
\newcommand{\parea}[1]{\left[#1\right]}

\newcommand{\km}{{\, {\rm km}}}

\newcommand{\cm}{{\, {\rm cm}}}

\newcommand{\keV}{{\, {\rm keV}}}
\newcommand{\MeV}{{\, {\rm MeV}}}
\newcommand{\GeV}{{\, {\rm GeV}}}
\newcommand{\TeV}{{\, {\rm TeV}}}
\newcommand{\kelvin}{{\, {\rm K}}}

\newcommand{\kms}{{\, {\rm km \, s^{-1}}}}
\newcommand{\OO}{{\mathcal{O}}}
\newcommand{\LL}{{\mathcal{L}}}

\DeclareMathOperator{\tr}{tr}
\DeclareMathOperator{\Imag}{Im}
\DeclareMathOperator{\Real}{Re}
\DeclareMathOperator{\sgn}{sgn}

\DeclareMathOperator{\erf}{erf}
\DeclareMathOperator{\erfc}{erfc}

\definecolor{mypurple}{RGB}{164,64,214}

\begin{document}

\title{Dark matter scattering in astrophysical media: collective effects}

\author[a]{William DeRocco,}
\emailAdd{wderocco@ucsc.edu}
\affiliation[a]{Santa Cruz Institute for Particle Physics and Department of Physics, University of California, Santa
Cruz, 1156 High Street, Santa Cruz, CA 95064}

\author[b]{Marios Galanis,}
\emailAdd{mgalanis@stanford.edu}
\affiliation[b]{Stanford Institute for Theoretical Physics, Stanford University, Stanford, CA 94305, USA}

\author[b]{Robert Lasenby}
\emailAdd{rlasenby@stanford.edu}

\date{\today}

\abstract{It is well-known that stars have the potential
	to be excellent dark matter detectors.
	Infalling dark matter that scatters within stars
	could lead to a range of observational
	signatures, including stellar heating, black hole formation,
	and modified heat transport.
	To make robust predictions for such 
	phenomena, it is necessary to calculate the 
	scattering rate for dark matter inside the star.
	As we show in this paper, for small enough momentum
	transfers, this requires taking into account
	\emph{collective effects} within the dense stellar medium.
	These effects have been neglected in many previous treatments;
	we demonstrate how to incorporate them systematically, and show
	that they can parametrically enhance or suppress
	dark matter scattering rates depending on how dark matter
	couples to the Standard Model.
	We show that, as a result, collective effects can significantly
	modify the potential discovery or exclusion reach
	for observations of compact objects such as white dwarfs and neutron stars.
	While the effects are more pronounced for dark matter coupling 
	through a light mediator, 
	we show that even for dark matter coupling via a heavy mediator,
	scattering rates can differ by orders of magnitude from
	their naive values for dark matter masses $\lesssim 100 \MeV$.
	We also illustrate how collective effects can be important
	for dark matter scattering in more dilute media,
	such as the Solar core.
	Our results demonstrate the need to systematically incorporate
	collective effects in a wide range of astroparticle
	contexts; to facilitate this, we provide expressions
	for in-medium self-energies for a variety of
	different media, which are applicable to many other
	processes of interest (such as particle production).
}

\maketitle


\section{Introduction}

Discovering the microphysical nature of dark matter (DM)
is one of the major goals of fundamental physics.
All existing evidence for DM comes from its gravitational interactions,
but many different types of DM models
result in very similar mass distributions
over the galactic scales on which gravitational
effects are significant, hence cannot be distinguished by their gravitational interactions alone.
Consequently, to distinguish between these models, we need
to look for signatures of non-gravitational interactions
between DM and the Standard Model (SM).

The most obvious approach is to look for interactions
of DM with laboratory systems.
Many different types of experiments,
aiming to detect various kinds of DM
candidates, have been proposed and implemented,
but so far no unambiguous evidence of DM has
been seen. In many models of DM, this is 
not unexpected.
DM that is some combination of too weakly-interacting, too
dilute
(due to individual particles being too heavy, or
being clustered in clumps),
too low-mass (so that individual interaction
events are not energetic enough to be detected),
or too strongly-interacting (so that DM is slowed
down by the Earth's atmosphere and the environment surrounding the experiment,
resulting in low-energy events)
could evade detection in such experiments.

An alternative approach, which may help to detect
some of these models, is to look for the effects of DM 
on astrophysical objects, such as stars.
These are much larger and denser than terrestrial
targets, and can consequently have larger interaction rates with DM.
The main
difficulty is how to observe the effects of such interactions.
For DM whose leading interaction with SM matter is via scattering,
scattering events inside a star can result in the DM
losing energy and being captured onto a gravitationally bound orbit.
Such DM particles then have the potential for further interactions in and around the star, leading to a variety of possible observational signatures.
These including heating old, cold stellar remnants
(either from the kinetic energy of the DM~\cite{1005.5699,1704.01577,1707.09442,1807.02840,1901.05474,1904.09803,1906.10145,1911.06334,1911.13293},
or from its annihilation~\cite{0708.2362,0709.1485,1004.0586}), formation of a black hole
inside the star~\cite{10.1103/PhysRevD.40.3221,1012.2039,1104.0382,1103.5472,1201.2400,1301.6811,1310.3509,1309.1721,1812.08773,2009.10728,2111.02414}, modification of heat transport~\cite{2009.00663,10.1086/168568,10.1086/168569,10.1088/1475-7516/2014/04/019,10.1088/1475-7516/2017/03/029,1212.2985,1505.01362,1701.03928}, or others.
Since these signatures usually rely on the accumulation
of many dark matter particles, rather than detecting a single
event, their sensitivity does not fall off at large and small
dark matter masses in the same way as laboratory experiments,
and they can potentially probe parameter space which
would be very difficult to explore on Earth.

To make predictions for these kinds of stellar observables,
one needs to calculate the scattering rate of DM inside
stars. For scatterings involving large enough momentum transfers,
this is a simple particle-by-particle process, as is familiar
from WIMP direct detection; the total scattering rate
is simply given by $\Gamma \simeq n_{\rm SM} \sigma_{\chi {\rm SM}}
v$, where $n_{\rm SM}$ is the number density of SM target
particles, $\sigma_{\chi{\rm SM}}$ is the DM-SM scattering
cross section, and $v$ is the relative velocity.
However, for smaller momentum transfers, it can be important
to take into account collective effects which arise due to
coherent interactions with multiple particles in the medium.
A very familiar example
of such an effect is the coherent scattering of particles off of
all of the nucleons within a nucleus when the transferred momentum is
$\lesssim 100 \MeV$. However, for momentum transfers smaller than
the inverse spacing between separate nuclei (or separate electrons),
more complicated effects can occur, such as screening
phenomena, scattering off of collective excitations
(e.g. phonons or plasmons), and others.
These types of effects have been studied recently in the context
of laboratory experiments searching for the scattering
of sub-GeV DM~\cite{2006.13909,2101.08263,10.1103/physrevd.104.015031,ani}, but have mostly been ignored in existing treatments
of DM scattering in stars. In this paper,
we consider such effects systematically, and show
that they can make parametrically large differences
to DM capture rates. 
Depending on the DM model,
these can be either enhancements or suppressions
of the capture rate compared to naive calculations.

Calculating these capture rates requires a model
of the stellar interior. In some circumstances, such
as the dilute, weakly-coupled plasma inside the Sun,
this task is relatively simple. In others, such as the
strongly-coupled ion lattice in cold white dwarfs, or the highly uncertain
state of matter inside neutron stars, it is much more complicated.
Given these complications, we do not attempt to provide detailed
or comprehensive analyses of stellar models; instead,
we work with simplified models of stellar media, and use
these to illustrate the physics of collective effects for
DM scattering rates. The systematic approach that we present
should be valuable for future, in-depth investigations.
Even in circumstances where collective effects
are not important, our in-medium formalism can
simplify calculations involving thermal distributions,
as we illustrate in Section~\ref{sec_solar_dpl}.

In addition to working with simplified models,
we also investigate the bounds that can be placed on scattering
and capture rates from general properties such as causality.
Specifically, for a dark photon mediator,
we show how the Kramers-Kronig relations
enable us to place bounds on scattering and capture
rates (extending the non-relativistic derivation
in~\cite{ani}), and how these bounds let us draw
phenomenologically useful conclusions.

Another contribution of this paper is to systematically 
consider the in-medium behaviour of vector
and scalar mediators. 
In particular, we present formulae for evaluating one-loop
vector and scalar self-energies in both degenerate Fermi gases and dilute
non-relativistic plasmas; as far as we are aware, these expressions
are novel.
We have made code implementing some of these formulae available online.\footnote{\href{https://github.com/wderocco/DarkScatter}{github.com/wderocco/DarkScatter}}
These and similar calculations should be useful in circumstances
beyond DM scattering in stars.
For example, in-medium self-energies are important in properly computing the emission rates
of light hidden-sector particles
from hot media such as supernovae~\cite{ed} (e.g.\
in Appendix~\ref{sec_res_scalar}, we show how 
previous calculations of electron-coupled scalar emission from supernovae
were parametrically incorrect).
Our formalism can also be applied
to DM scattering in laboratory experiments, as explored in~\cite{ani}.

The outline of the paper is as follows: in
Section~\ref{sec_srate}, we outline the
computation of scattering rates using in-medium
propagators and give specific examples of such
rates. In Section~\ref{sec_capture_rates}, we
discuss the kinematics associated with stellar
capture, after which we apply our computed capture
rates to both white dwarfs (Sec.~\ref{sec_wd})
and neutron stars (Sec.~\ref{sec_ns}). We then
discuss scattering in dilute stellar plasmas,
and ramifications for various potential
signatures in the Sun (Sec.~\ref{sec_dilute})
before concluding and discussing future directions
in Section~\ref{sec_conc}. Various formulae
for the processes discussed are collected in
Appendices~\ref{app_srates}-\ref{app_vel_dist},
while Appendix~\ref{sec_res_scalar} discusses
particle emission from supernovae.


\section{Scattering rates and in-medium propagators}
\label{sec_srate}

A systematic way to calculate the scattering rate for a particle
travelling through a medium is to use the effective
in-medium propagator for the particle.
If we consider a DM particle $\chi$ coupled to the SM
via a mediator particle $X$, then this can be related
to the effective in-medium propagator for the mediator $X$.
Diagramatically, at leading order in the 
(assumed weak) $X-{\rm SM}$ and $\chi-X$ couplings, the scattering rate can be related
to the imaginary part of the following self-energy diagram,
\begin{center}
	\begin{tikzpicture}[line width=1]
          \begin{scope}[shift={(2.4,0)}]
            \draw[f] (2*0.33,0)  node[above] {$ \chi $}
            -- (2*1.2,0);
            \node[] at (1.7,0.8) {$X$};
            \node[] at (4.3,0.8) {$X$};
            \node[] at (1.4,-0.3) {$P$};
            \node[] at (3.0,-0.3) {$P-Q$};
            \node[] at (4.5,-0.3) {$P$};
			\draw[f] (2*1,0) -- (2*2,0);
			  \draw[f] (2*2,0) -- (2*2.66,0) node[above] {$\chi$};
            \draw[v] (2*2,0) arc (0:180:2*0.5);
            \draw[pattern=crosshatch,preaction={fill=white}] (2*1.5,2*0.52) circle (2*0.13);
          \end{scope}
	\end{tikzpicture}
\end{center}
where the filled circle represents the medium (this 
usually consists of SM particles, but if there is a large density
of DM particles, e.g.\ if DM self-capture in a star is important,
then it could include DM as well).\footnote{While
we have drawn the mediator $X$ as a vector, it may be a vector
or a scalar.}
Via the optical theorem, the imaginary part of the
self-energy corresponds to the scattering rate
via mediator exchange with the medium,
\begin{gather}
	\adjustbox{valign=m}{
	\begin{tikzpicture}[line width=1,scale=1.3]
          \begin{scope}[shift={(2.4,0)}]
            \draw[f] (2*0.33,0)  
            -- (2*1.2,0);
			\draw[f] (2*1,0) -- (2*2,0);
			  \draw[f] (2*2,0) -- (2*2.66,0);
            \draw[v] (2*2,0) arc (0:180:2*0.5);
            \draw[pattern=crosshatch,preaction={fill=white}] (2*1.5,2*0.52) circle (2*0.13);
			  \draw[snar] (2.5,-0.5) -- (3.2,1.7);
          \end{scope}
	\end{tikzpicture}}
\qquad
\rightarrow
\qquad
	\adjustbox{valign=m}{
	\begin{tikzpicture}[line width=1]
          \begin{scope}[shift={(2.4,0)}]
			  \draw[f] (-1,1) node[above] {$P$}  -- (0,0.5);
			  \draw[f] (0,0.5) -- (1,1)  node[above] {$P-Q$};
			  \draw[v] (0,-0.5) -- (0,0.5);
			  \draw[-stealth] (-0.35,0.4) -- node[left] {$Q$} (-0.35,-0.4);
            \draw[pattern=crosshatch,preaction={fill=white}] (0,-0.7) circle (2*0.13);
          \end{scope}
	\end{tikzpicture}}
\label{fig_cut0}
\end{gather}
For the explicit example of Dirac fermion DM, scattering
via a vector mediator of mass $m_X$, the total
scattering rate, integrating
over momentum transfers $Q$, is given by (Appendix~\ref{app_srates})
\begin{align}
	\Gamma &=
	\frac{2 g_\chi^2}{E} \int \frac{d^3 q}{(2\pi)^3} \frac{1}{2E'}
	(1 + f(q_0)) \frac{1}{(Q^2 - m_X^2)^2} \times 
	\nonumber \\
	&\quad\left(
	2\left(E^2 - p^2 \cos^2\theta\right) \Imag \Pi_L^X(Q)
	- \left(-Q^2 + 2 p^2 \sin^2\theta\right) \Imag \Pi_T^X(Q)
	\right)
	\label{eq_gl1}
\end{align}
where $g_\chi$ is the DM-mediator coupling,
$P = (E,p)$ is the initial 4-momentum
of the DM particle, $Q = (q_0,q)$ is the scattering
4-momentum, $E' = E - q_0$ is the 
post-scattering DM energy,
$f(q_0) \equiv (e^{q_0/T}-1)^{-1}$
is the bosonic occupation number corresponding
to the temperature $T$ of the medium, $\theta$ is the angle beteween $p$ and $q$,
$\Pi_L^X(Q)$ is the in-medium self-energy for 
the longitudinal mode of $X$,
and $\Pi_T^X(Q)$ is for the transverse mode (Eq.~\eqref{eq_gl1}
assumes a uniform, isotropic medium).\footnote{Eq.~\eqref{eq_gl1}
assumes that the SM current that $X$ couples to is conserved ---
more specifically, that current-nonconserving processes
are unimportant in the medium.
If $X$ does not have flavour-changing couplings, then 
this is a good approximation at the very sub-weak-scale
temperatures in astrophysical media.}
Eq.~\eqref{eq_gl1} computes the total scattering rate 
for a DM particle passing through a spatially uniform
medium; we can compute other quantities
of interest, such as the scattering rate into particular
parts of phase space
or the momentum transfer rate, via similar expressions, as we
discuss below. There are also
similar expressions for other kinds of DM particles (e.g.\ spin-0
DM) and other types of mediators; for more details,
see Appendix~\ref{app_srates}.

We have seen how, to compute the DM scattering rate, we want to compute the
mediator self-energy in an SM medium. If we have some perturbative
description of the SM medium\footnote{This does not necessarily have
to be in terms of `bare' SM particles, but can be in terms
of weakly-interacting quasi-particles, e.g.\ Fermi
liquid theory~\cite{girvin}.}, then to leading order
 in the (assumed weak) SM-mediator coupling, the self-energy
 is given diagramatically by 
\begin{gather}
	\adjustbox{valign=m}{
	\begin{tikzpicture}[line width=1]
			  \draw[v] (-1.5,0) node[above] {$X$} -- (-0.5,0);
			  \draw[v] (0.5,0) -- (1.5,0) node[above] {$X$};
            \draw[pattern=crosshatch,preaction={fill=white}] (0,0) circle (0.5);
	\end{tikzpicture}}
	 = 
	\adjustbox{valign=m}{
	\begin{tikzpicture}[line width=1]
			  \draw[v] (-1.5,0) node[above] {$X$} -- (-0.5,0);
			  \draw[v] (0.5,0) -- (1.5,0) node[above] {$X$};
            \draw[pattern=north east lines,preaction={fill=white}] (0,0) circle (0.5);
	\end{tikzpicture}}
 + 
	\adjustbox{valign=m}{
	\begin{tikzpicture}[line width=1]
			  \draw[v] (-1.5,0) node[above] {$X$} -- (-0.5,0);
            \draw[pattern=north east lines,preaction={fill=white}] (0,0) circle (0.5);
			  \draw[line width=2,v] (0.5,0) -- (1.5,0);
            \draw[pattern=north east lines,preaction={fill=white}] (2,0) circle (0.5);
			  \draw[v] (2.5,0) -- (3.5,0) node[above] {$X$};
	\end{tikzpicture}}
\end{gather}
where the single-hatched circles on the RHS correspond to one-particle irreducible
self-energies, while the bold line corresponds to the in-medium
SM photon propagator.\footnote{In principle, we should sum over all SM
intermediate states allowed by symmetries. In
this paper, we will consider spin-0 and spin-1 mediators, so
we are interested in bosonic SM states with no conserved
SM quantum numbers. At low energies, this picks out the photon;
if we are working in an EFT including the pion,
then we can have mixing between a pseudoscalar mediator and 
the pion, but we will not consider such cases. 
If we were working in an effective description
of the medium that included excitations such as phonons,
then these should also be included in the intermediate states.}
Algebraically,
\begin{equation}
	\Pi^{XX}_{\rm tot} = \Pi^{XX} + \frac{(\Pi^{XA})^2}{Q^2 - \Pi^{AA}}
	\label{eq_pxx}
\end{equation}
where $\Pi^{XX}$ and $\Pi^{AA}$ correspond to the 1PI
self-energies, and $\Pi^{XA}$ to the 1PI mixing self-energy
(all self-energies are functions of $Q$). 
Eq.~\eqref{eq_pxx} is schematic, in that 
the SM photon propagator has vector indices,
$D_{\mu\nu}(Q)$, but for an isotropic
medium, it will split into longitudinal and transverse parts,
each of which can be put into the form of Eq.~\eqref{eq_pxx}
(see Appendix~\ref{app_srates}).

Taking the imaginary part of Eq.~\eqref{eq_pxx}, we obtain
\begin{equation}
	\Imag \Pi^{XX}_{\rm tot}
	= \Pi^{XX}_i + \frac{\Pi^{AA}_i((\Pi^{XA}_r)^2
	- (\Pi^{XA}_i)^2) + 2 \Pi^{XA}_r \Pi^{XA}_i (Q^2 - \Pi^{AA}_r)}
	{(Q^2 - \Pi^{AA}_r)^2 + (\Pi^{AA}_i)^2}
	\label{eq_pxxitot}
\end{equation}
where $\Pi_r, \Pi_i$ indicate the real and imaginary
parts of each self-energy.
A naive calculation, ignoring collective effects,
usually corresponds to the leading-order
parts of the $\Pi_i^{XX}$ term.
Diagramatically, cutting the simplest two-loop self-energy
diagram corresponds to the leading-order 
particle-by-particle scattering rate,
\begin{gather}
	\adjustbox{valign=m}{
	\begin{tikzpicture}[line width=1,scale=1.3]
          \begin{scope}[shift={(2.4,0)}]
            \draw[f] (2*0.33,0)  node[above] {$ \chi $}
            -- (2*1.2,0);
			\draw[f] (2*1,0) -- (2*2,0);
			  \draw[f] (2*2,0) -- (2*2.66,0) node[above] {$\chi$};
            \draw[v] (2*2,0) arc (0:180:2*0.5);
            \draw[preaction={fill=white}] (2*1.5,2*0.52) circle (2*0.13);
			  \draw[snar] (2.5,-0.5) -- (3.2,1.7);
            \node[] at (2.69,1.40) {$f$};
          \end{scope}
	\end{tikzpicture}}
\qquad
\rightarrow
\qquad
	\adjustbox{valign=m}{
	\begin{tikzpicture}[line width=1]
          \begin{scope}[shift={(2.4,0)}]
			  \draw[f] (-1,1) node[above] {$\chi$} -- (0,0.5);
			  \draw[f] (0,0.5) -- (1,1) node[above] {$\chi$};
			  \draw[f] (-1,-1) node[below] {$f$} -- (0,-0.5);
			  \draw[f] (0,-0.5) -- (1,-1) node[below] {$f$};
			  \draw[v] (0,-0.5) -- (0,0.5);
          \end{scope}
	\end{tikzpicture}}
\label{fig_cut1}
\end{gather}
with in-medium initial-state and final-state
occupation number factors for $f$.
The other term in Eq.~\eqref{eq_pxxitot} corresponds
to the `mixing' contributions to the scattering
rate. These can sometimes cancel against the 
$\Pi_i^{XX}$ term, resulting in a suppressed scattering rate,
or can enhance it, e.g.\ if the denominator becomes small
(giving `resonant' effects). This is similar
to the situation in particle emission calculations~\cite{ed}.

A simple model of phenomenological interest
is a `dark photon' mediator~\cite{10.1016/0370-2693(86)91377-8}, where $X$ is 
a massive vector which couples to the SM
electromagnetic current, $\LL \supset \kappa X_\mu J^{\mu}_{\rm EM}$,
with coupling suppressed by the `kinetic mixing' parameter $\kappa$.
In this case, we simply have that $\Pi^{AX} = \kappa \Pi^{AA}$
and
$\Pi^{XX} = \kappa^2 \Pi^{AA}$, so writing $\Pi \equiv \Pi^{AA}$, 
we have
\begin{equation}
	\Imag \Pi^{XX}_{\rm tot} 
	= \frac{\kappa^2 Q^4 \Imag \Pi}
	{(\Real \Pi - Q^2)^2 + (\Imag \Pi)^2}
	= \frac{\kappa^2 Q^4 \Imag \Pi}
	{|Q^2 - \Pi|^2}
	\label{eq_pidp1}
\end{equation}
We can see that, for large momentum transfers with $|Q^2| 
\gg |\Pi|$, we have
$\Imag \Pi^{XX}_{\rm tot} \simeq \kappa^2 \Imag \Pi$;
at leading order, this corresponds to the naive particle-by-particle
scattering rate from Eq.~\eqref{fig_cut1}. Collective effects are important when the $\Pi$ terms
in the denominator of Eq.~\eqref{eq_pidp1} are significant.
These can either lead to suppression,
if $|\Pi| \gg |Q^2|$ (generally referred to 
as `screening'), or
enhancement, if $Q^2 - \Real \Pi \simeq 0$ contributes significantly
to the scattering (`resonant' effects). 

Analogous analyses apply to other types of mediator, such as scalars.
One important general point is that whether medium effects are important
depends on the comparison of $|Q^2|$ to
$|\Pi(Q)|$, 
\emph{not} on the comparison between $|Q^2|$ and $m_X^2$,
as is sometimes assumed. In particular, we will see how collective
effects can sometimes be very important for scattering
via a heavy mediator (corresponding to contact
interactions in the low-energy theory).


\subsection{EM sum rules}
\label{sec_emsum}

For a dark photon mediator, the expression in
Eq.~\eqref{eq_pidp1} corresponds to
$\Imag (\Pi^{XX}_{\rm tot})_{\mu\nu} =  \kappa^2 Q^2 \Imag 
\left( -i 
D_{\mu\nu}\right)$, where
$D_{\mu\nu}$ is the in-medium propagator
for the SM photon, in Lorenz gauge (more precisely,
it is the analytic continuation of the in-medium
imaginary-time propagator --- see
Appendix~\ref{app_sumrules}).
We can also see this directly~\cite{1305.2920} by changing variables
to put the mediator-SM interaction
into kinetic mixing form, $\LL \supset
- \frac{\kappa}{2} F_{\mu\nu}F'^{\mu\nu}$, where
$F'$ is the field strength corresponding to $A'$.
Then,
\begin{gather}
	\adjustbox{valign=m}{
	\begin{tikzpicture}[line width=1]
			  \draw[v] (-1.5,0) node[above] {$X$} -- (-0.5,0);
			  \draw[v] (0.5,0) -- (1.5,0) node[above] {$X$};
            \draw[pattern=crosshatch,preaction={fill=white}] (0,0) circle (0.5);
	\end{tikzpicture}}
	 = 
	\adjustbox{valign=m}{
	\begin{tikzpicture}[line width=1]
		\draw[v] (-1.5,0) node[above] {$X$} -- (-0.5,0);
			  \draw[line width=2,v] (-0.5,0) -- node[above] {$A$} (1.0,0);
		\draw[v] (1.0,0) -- (2.0,0) node[above] {$X$};
		\node[draw,cross,rotate=45] at (-0.5,0) {};
		\node[draw,cross,rotate=45] at (1.0,0) {};
	\end{tikzpicture}}
\end{gather}
where the crosses correspond to the kinetic mixing interaction. 
Here, all of the interactions with the medium
have been subsumed into the in-medium
propagator for the SM photon.
Since the properties of the in-medium propagator
$D_{\mu\nu}$ are constrained by causality, we can
place limits on the scattering rate via a dark photon 
mediator, as discussed (for non-relativistic DM) in~\cite{ani}.
We go through the derivation of such limits in
Appendix~\ref{app_sumrules}.
For example, to derive limits on scattering rates via longitudinal
vector exchange, we can use the Kramers-Kronig relations~\cite{bellac}
to derive the sum rule
\begin{equation}
	\int_0^\infty \frac{dq_0}{q_0} \frac{Q^2}{q^2} \Imag D_L(q_0,q)
	= \frac{\pi}{2} \frac{k_S^2}{q^2(q^2 + k_S^2)}
	\label{eq_srule1}
\end{equation}
where $k_S^2 \equiv \Pi_L(0,q)$ is the 
(static) longitudinal screening scale.
In the circumstances that will be of interest to us,
we will usually have $k_S^2 \ge 0$, corresponding
to screening rather than anti-screening (see Appendix~\ref{app_sumrules}),
so we can upper-bound Eq.~\ref{eq_srule1}
by $\pi/(2q^2)$. 

Since, for a dark photon mediator, $\Imag (\Pi^{XX}_{\rm tot})_{\mu\nu} =  \kappa^2 Q^2 \Imag
\left(-i D_{\mu\nu}\right)$, and the DM scattering rate 
can be written in terms of an integral over $\Imag \Pi^X(Q)$ for different $Q$ (Eq.~\eqref{eq_gl1}),
we can use Eq.~\eqref{eq_srule1} (and
an analogous sum rule for the transverse propagator) to bound the DM scattering rate.
Specifically, we can write the scattering
rate as a nested integral of the form
$\int \dots \int \frac{dq_0}{q_0} \frac{Q^2}{q^2} \dots \Imag D_L(q_0,q)$,
and then upper-bound the $\int \frac{dq_0}{q_0}$ integral
by assuming that $D_L(q_0,q)$ (for given $q$) is a delta-function
at the value maximizing the rest of the integrand, with
the delta-function's weight set by Eq.~\eqref{eq_srule1}
(see Appendix~\ref{app_sumrules} for details).
For a non-relativistic DM particle $\chi$
with velocity $v_\chi \ll 1$, and taking
the temperature to be negligible,
this implies that the scattering rate via longitudinal mediator
exchange is upper-bounded by
\begin{equation}
	\Gamma_L \lesssim \begin{dcases}
		\frac{g_\chi^2 \kappa^2}{4\pi} m_\chi v_\chi & m_X \ll m_\chi v_\chi \\
		\frac{16}{15} \frac{g_\chi^2 \kappa^2}{4\pi} m_\chi v_\chi 
		\left(\frac{m_\chi v_\chi}{m_X}\right)^4
		& 
		m_X \gg m_\chi v_\chi
	\end{dcases}
\end{equation}
where we give the forms for very light and very heavy 
mediators (compared to the DM momentum scale).
The scattering rate limit via transverse mediator exchange
is further suppressed by $v_\chi^2$ compared to the
longitudinal rates:
\begin{equation}
	\Gamma_T \lesssim \begin{dcases}
	0.53 \times \frac{g_\chi^2 \kappa^2}{4\pi}
		m_\chi v_\chi^3 & m_X \ll m_\chi v_\chi \\
1.16 \times \frac{g_\chi^2 \kappa^2}{4\pi}
		\frac{m_\chi^5 v_\chi^7}{m_X^4} & m_X \gg m_\chi v_\chi
	\end{dcases}
\end{equation}
as follows from the form of $\Gamma_T$ (Appendix~\ref{app_srates}).
For relativistic $\chi$, the rate limits will be larger than these
formulae; see Appendix~\ref{app_sumrules}.
We can also use similar arguments to place sum rule
limits on other quantities, such as total capture
rates, as we discuss in the next Section.
As we will see in Sections~\ref{sec_wd} and
\ref{sec_ns}, these limits allow us to set
reliable bounds on the scattering rate
through a dark photon mediator, even 
without using an explicit model of the stellar medium.


\section{Stellar capture rates}
\label{sec_capture_rates}

We can use the formalism for calculating scattering rates
outlined in Section~\ref{sec_srate} to 
calculate quantities of physical interest, such
as the capture
rate for halo DM into gravitationally bound orbits in and
around a star.
For situations in which the star is optically thin, 
and capture is dominated by individual scattering events,
the capture rate can be derived from 
the DM velocity distribution at each point inside
the star, that arises from the infall under gravity of halo DM. 

At a given point inside the star, an infalling halo DM particle with
energy $E$ (as measured by a stationary observer at that point)
had energy $E_\infty = \sqrt{B} E$ far from the star,
where $\sqrt{B}$ is the gravitational redshift at
that point ($B = g_{00}$ in Schwarzschild coordinates).
In most circumstances of interest, an isotropic halo DM
velocity distribution leads to an isotropic distribution of infalling
DM at points inside the star.\footnote{The halo DM velocity distribution
at a point inside the star is isotropic
as long as there are not bound orbits inside
the star with $E_\infty > m_\chi$ (i.e.\ with $P_0 > m_\chi$, where
$P$ is the 4-momentum in Schwarzschild coordinates),
which we will call `positive orbits'.
In Newtonian gravity, positive orbits are not possible.
At a given point inside the star, all test particle trajectories with a given
energy either escape to infinity, if the total
energy (kinetic plus potential) is non-negative, or
enter bound orbits, if the total energy is negative.
In particular, the direction of the trajectory
does not affect whether it is escaping or bound.
This is not necessarily true in GR, for sufficiently compact
objects. If some of the trajectories passing through a point
inside the star form positive orbits,
then infalling halo DM cannot travel along these trajectories,
even though the total energy might be $\ge m_\chi$;
it can only travel along escaping trajectories.
As we discuss at the end
of this Section, heavy neutron stars may be just 
compact enough for positive orbits to exist.
However, even for e.g.\ the heaviest
neutron star model from~\cite{bell_ns},
positive orbits can only exist in a small volume of phase space
near the edge of the star. Consequently, their existence 
cannot have a large effect on our capture rate calculations,
and we ignore them here.}
Consequently, if we assume for simplicity that the halo
DM velocity distribution is isotropic, with phase
space density $f_\infty (E_\infty)$, then since
(from Liouville's theorem) the phase space
density is preserved during gravitational infall,
the phase space density at a point inside
the star is set by $f(E) = f_\infty(\sqrt{B} E)$.

We can relate $f_\infty$ to the velocity distribution
of halo DM by noting that we must have
\begin{equation}
	n_\chi = g_s \int \frac{d^3 p}{(2\pi)^3} f_\infty = \frac{g_s m_\chi^3}{(2\pi)^3} \int d^3 v f_\infty
	\quad \Rightarrow \quad f_\infty(p) = (2\pi)^3 \frac{n_\chi}{g_s m_\chi^3}
	p_v(p/m_\chi)
	\label{eq_finfty}
\end{equation}
where we have taken the halo DM to be non-relativistic,
$g_s$ is the spin
degeneracy of the DM,
and $p_v(v)$ is the halo DM velocity distribution.
Going forwards, we will define $\hat f_\infty \equiv g_s f_\infty/(2\pi)^3 = \frac{n_\chi}{m_\chi^3}
	p_v(p/m_\chi)$
for notational convenience
(and more generally, $\hat A \equiv \frac{g_s}{(2\pi)^3} A$
for other quantities $A$).

In terms of $f(E)$, the scattering rate per volume at
a point within the star, considering only scatterings
from unbound to bound orbits, is
(as measured by a stationary local observer)
\begin{equation}
	\frac{\Gamma}{\rm vol} = g_s \int \frac{d^3 p}{(2\pi)^3} \,
	f(p) 
\Gamma_{E \rightarrow E' < E_{\rm esc}}
	= 4\pi \int dE\, E p \, \hat f(E) \Gamma_{E \rightarrow E' < E_{\rm esc}}
\end{equation}
where $\Gamma_{E \rightarrow E' < E_{\rm esc}}$
indicates the scattering rate into bound orbits
(with energy $< E_{\rm esc} = m_\chi/\sqrt{B}$),
and we have assumed that the DM velocity distribution is 
isotropic, so that $f$ is only a function 
of $E$ (if the DM distribution is non-isotropic
but the medium's response is isotropic, then we can equivalently
consider averaging over different orientations to obtain
an isotropized velocity distribution --- see Appendix~\ref{app_vel_dist}).
Writing this in terms of $E_\infty$, we have
\begin{align}
	\frac{\Gamma}{\rm vol} &= 
	4\pi B^{-3/2} \int_{m_\chi}^\infty dE_\infty \, 
	E_\infty^2 
	\sqrt{1 - \frac{m_\chi^2 B}{E_\infty^2}}
	\hat f_\infty(E_\infty) \Gamma_{E \rightarrow E' < E_{\rm esc}}
	\nonumber \\
	&\simeq 
	4\pi \frac{\sqrt{1-B}}{B^{3/2}} m_\chi^2 \int_{m_\chi}^\infty dE_\infty \, 
	\hat f_\infty(E_\infty) \Gamma_{E \rightarrow E' < E_{\rm esc}}
\end{align}
where in the second line we used that the halo DM is non-relativistic,
so $E_\infty \simeq m_\chi$. This 
is the \emph{local} rate, as measured by a stationary observer
inside the star; due to time dilation, the total capture rate, as measured by a stationary 
observer at infinity, is
\begin{equation}
	C = 4 \pi m_\chi^2 \int dV \frac{v_{\rm esc}}{B}
	\int_{m_\chi}^\infty dE_\infty \hat f_\infty(E_\infty)
	\Gamma_{E \rightarrow E' <  E_{\rm esc}}
	\label{eq_ceq1}
\end{equation}
If the escape velocity $v_{\rm esc}= \sqrt{1-B}$ inside the star is much greater than the velocity
dispersion of the incoming DM, then the kinetic energy of DM in the star will be dominated by the energy acquired during its infall and almost all of the halo DM
particles at a point inside the star will have energy 
just above $E_{\rm esc}$. In some circumstances,
this will mean that $\Gamma_{E \rightarrow E' < E_{\rm esc}}$
can be treated as approximately constant over the small range of relevant $E$,
and we can take it out of the $\int dE_\infty$ integral.
This is not always possible; if soft scatterings are strongly
enhanced, as they can be for light mediators, then
how close a particle's energy is to the $E_{\rm esc}$ threshold can make a
significant difference to the capture rate. However,
if we can take
$\Gamma_{E \rightarrow E' < E_{\rm esc}}$
outside the integral (e.g.\ if hard scatterings dominate
the capture rate), then
\begin{align}
	C &\simeq 4\pi m_\chi^2 \int dV \frac{v_{\rm esc}}{B}
	\Gamma_{\rm down-scatter} \int_0^\infty v_\infty dv_\infty \hat f_\infty(v_\infty) \nonumber \\
	&= n_\chi \left\langle \frac{1}{v_\infty} \right\rangle
	\int dV 
 \frac{v_{\rm esc}}{B} 
\Gamma_{\rm down-scatter} 
\label{eq_cbell}
\end{align}
where $\Gamma_{\rm down-scatter}$ is the rate
of scatterings from higher to lower energies for
a DM particle with velocity $v_{\rm esc}$,
$v_\infty$ is the DM velocity at infinity,
and the angle brackets denote the average over the DM velocity 
distribution at infinity.
Eq.~\ref{eq_cbell} matches the capture rate expression
from~\cite{bell_ns}.

If we cannot take
$\Gamma_{E \rightarrow E' < E_{\rm esc}}$
outside the $\int dE_\infty$ integral, it is still possible to simplify
Eq.~\eqref{eq_ceq1}.
For clarity, we will consider the explicit example of spin-1/2 DM
scattering through the longitudinal mode
of a vector mediator, which gives a contribution to the capture rate of
\begin{align}
	\frac{dC}{dV} &= \frac{2 g_\chi^2}{\pi}
	\frac{v_{\rm esc}}{B} m_\chi^2
	\int_{m_\chi}^{\infty} dE_\infty
	\hat  f_\infty(E_\infty)
	\frac{1}{E p}
	\int dq \, q \nonumber \\
	&\int_{E - E_{\rm esc}}^{q_{0,\rm max}} dq_0
	(1 + f(q_0))\frac{1}{(Q^2 - m_X^2)^2}
	(E^2 - p^2 \cos^2\theta) \Imag \Pi^X_L(q_0,q)
	\label{eq_dcdv1}
\end{align}
where $q_{0,\rm max}$ is the maximum value of $q_0$ possible
for given $q$ and $E$, derived by taking the outgoing $\chi$ to be on-shell.
Here, we are taking the mediator self-energy
to be isotropic, i.e. to be independent of the direction of $q$
--- for non-isotropic media, this can be viewed
as the average over different directions for $q$.
If the escape velocity inside the star is much larger 
than the typical halo DM velocity, then $q_{0,\rm max}$ 
and $E^2 - p^2 \cos^2\theta$ will depend only weakly
on $E_\infty$. Consequently, the $E_\infty$
dependence of the $\int dq_0$ integral arises mostly from
the $E_\infty$ dependence of its lower limit,
$E - E_{\rm esc} = (E_\infty - m_\chi)/\sqrt{B}$ (if the energy transfer is below this limit, the DM will still have enough energy left over to escape the star after the collision).
So, denoting the kinetic energy of the DM far from the star as $E_K \equiv E_\infty - m_\chi$, we have
\begin{align}
	\int_{m_\chi}^\infty dE_\infty \hat f_\infty(E_\infty)
	\int_{(E_\infty - m_\chi)/\sqrt{B}}^{q_{0,\rm max}} dq_0 
	&\simeq \int_0^{q_{0,\rm max}}
	dq_0 \left(\int_0^{\sqrt{B} q_0} dE_K \hat f_{\infty}(m_\chi + E_K) \right) 
	\nonumber \\
	&\equiv \int_0^{q_{0,\rm max}} dq_0 \hat F_\infty(\sqrt{B}q_0) 
\end{align}
Consequently,
\begin{equation}
	\frac{dC}{dV} \simeq \frac{2 g_\chi^2}{\pi}
	\int dq \, q \int_{0}^{q_{0,\rm max}} dq_0 \hat F_\infty(\sqrt{B}q_0)
	(1 + f(q_0))\frac{1}{(Q^2 - m_X^2)^2}
	(E^2 - p^2 \cos^2\theta) \Imag \Pi^X_L
	\label{eq_cv1}
\end{equation}
Appendix~\ref{app_vel_dist} gives explicit expressions
for $\hat F_\infty$ for an (offset) Maxwell velocity distribution.
For $q_0$ much larger than 
typical halo DM kinetic energies,
we have
\begin{equation}
	\hat F_\infty(q_0) \simeq \hat F_\infty(\infty) =
	\frac{n_\chi}{4\pi m_\chi^2} \left\langle \frac{1}{v_\infty}\right\rangle
\end{equation}
corresponding to Eq.~\eqref{eq_cbell}.
For $q_0$ much smaller than such values, we have
$\hat F_\infty(q_0) \simeq \frac{n_\chi}{m_\chi^3} p_v(0) q_0$,
so the small-energy-transfer capture rate is linearly suppressed in $q_0$.
This corresponds to there being fewer DM particles
with energy close enough to $E_{\rm esc}$ to be
trapped by losing energy $q_0$.

For a dark photon mediator, we can use these formulae to place sum
rule bounds on the total capture rate, similarly 
to the bounds on the scattering rate per volume
from Section~\ref{sec_emsum}.
These are applied in Sections~\ref{sec_wd_dpl} and
\ref{sec_ns}.

Eq.~\eqref{eq_cv1} contains a factor
of $1 + f(q_0) = (1 - e^{-q_0/T})^{-1}$, so for energy transfers
$\lesssim$ the temperature $T$ of the stellar medium,
the down-scattering rate from unbound to bound
orbits is enhanced by $\sim \frac{T}{q_0}$.
However, this also means that up-scatterings from
bound to unbound orbits can be important.
When $T$ is large enough that the $q_0 \lesssim T$ phase
space becomes important, we are not necessarily just
interested in the rate of unbound-to-bound scatterings.
Often, we are interested in the properties of the dynamic equilibrium
which arises once enough dark matter has been captured
that up-scatterings from bound orbits become significant.
To understand this behaviour properly, one would generally
need a full model of the capture and evolution of DM within
the star. 
Here, we will not attempt to do that, but will note when
energy transfers $q_0 \lesssim T$ are important
for capture rates (e.g.\ Solar capture via a light
mediator, Sec.~\ref{sec_solar_dpl}).

\subsection{Geometric capture rate}

The $C = \int dV \frac{dC}{dV}$ formula 
for the total capture rate derived above holds when the star is optically
thin to DM. If, on the other hand, the star is optically thick, then the maximum
possible value for the capture rate is given by the geometric
rate, i.e. the rate at which halo DM particles intersect with the star.
To compute this correctly, one needs to take into account gravitational
focusing. For $R > 4 G M$, where $R$ is the radius
of the star and $M$ is its mass, a trajectory
grazing the star's surface can come from and escape
to infinity. The angular momentum (per unit mass) of such a trajectory
is $L = \sqrt\frac{2 R_g R}{1 - 2 R_g/R}$, where $R_g \equiv G M$.
For a DM particle with velocity at infinity $v_\infty$,
the corresponds to impact parameter $b_{\rm max} = L/v_\infty$,
so the geometric capture rate is~\cite{10.1103/PhysRevD.40.3221}
\begin{equation}
	C =  \left \langle \pi b_{\rm max}^2 n_\chi v_\infty
	\right \rangle
	= \frac{2\pi R_g R}{1 - 2 R_g/R}
	n_\chi
	\left\langle
	\frac{1}{v_\infty}\right\rangle
	\label{eq_cgeom}
\end{equation}
where the angle brackets denote averaging over
the DM velocity distribution at infinity.
If the star is more compact, so
$R \le 4 G M$, then trajectories grazing the star's surface
are bound; all incoming-from-infinity DM trajectories
have inwards radial velocity, and the geometric capture
rate is the same as an object with radius $R = 4 G M$,
giving
\begin{equation}
	C = 16 \pi R_g^2 n_\chi
	\left\langle
	\frac{1}{v_\infty}\right\rangle
\end{equation}
i.e. the same rate as a black hole.
Only the heaviest neutron stars have
radius close to the $4 G M $ threshold,
so in general, the expression from Eq.~\eqref{eq_cgeom}
will be fine for our purposes.
For dense stellar remnants such as white
dwarfs and neutron stars, capture rates not too far
from the geometric limit are often needed
to obtain observable signatures, as we discuss in subsequent sections.


\section{White dwarfs}
\label{sec_wd}

White dwarfs are the densest astrophysical objects
whose physics we understand fairly well (as we discuss
in Section~\ref{sec_ns}, the physics of neutron
star cores is very poorly understood). Consequently,
they represent a promising target for signatures
of DM scattering, with the potential for large and reliably-computable scattering rates.
In this Section, we will discuss how to perform such computations
for some representative DM models,
and how such scatterings might lead to observational signatures.
We begin by briefly summarizing some
of the signatures of DM capture in white dwarfs
proposed in the existing literature,
before outlining our calculations
of the DM capture rate in white dwarfs. 
At low DM masses, we find that this rate
is dominated by collective scattering with
longitudinal phonons, which we calculate analytically.
We discuss a number of different DM mediator models;
heavy vector mediators, light dark photon mediators,
and heavy scalar mediators.

\subsection{Observational signatures}

DM scattering in a white dwarf (WD) can lead to a variety of
possible observational consequences. The simplest
such signature is heating of the star.
Old WDs are expected to cool down to temperatures
$\OO(10^5 \kelvin)$~\cite{10.1086/317235} in their
cores; their thin outer layers are significantly
cooler, at temperatures $\sim {\rm few} 
\times 10^3 \kelvin$~\cite{10.1088/0004-637X/789/2/119}.
If enough of the energy carried by DM can be deposited
into a WD, then in regions of sufficiently high
DM density, the WD's temperature could be appreciably raised.

Since the escape velocity in a WD is non-relativistic, the 
kinetic energy carried by infalling DM is only
a small fraction (at most $\sim 10^{-2}$) of its
total energy. Consequently, 
very large ambient DM densities would
be required for purely kinetic heating of WDs to be significant,
even at the geometric capture rate.
One way to deposit more energy is for captured
DM to annihilate inside the WD~\cite{0709.1485,1512.00456,2005.11563,bell_wd}.
After the initial scattering event that
captures a DM particle into a bound
orbit, further scatterings will cause the particle
to lose more energy, thermalizing it down into a smaller volume
within the WD, where it may annihilate
with other captured DM particles.
In most models, the cross sections required to capture enough
DM for detectable heating mean that these subsequent stages of thermalization
and annihilation always happen fast enough
to set up an equilibrium between DM capture
and annihilation. In particular, to capture enough DM,
capture rates not too far from the geometric limit
are usually required~\cite{bell_wd}.
Consequently, to calculate the heating rate,
we just need to calculate the DM capture rate,
as discussed in Section~\ref{sec_srate}.

Other possible signatures include, for example,
 the formation of black hole inside the WD,
destroying it from the inside~\cite{1012.2039}, which is possible for sufficiently heavy, bosonic, asymmetric DM.
To understand this process, one needs
to understand the initial capture of the DM,
and its subsequent scattering inside the WD,
both with the SM medium and with other DM particles.
Alternatively, the decay of very heavy DM particles inside
WDs~\cite{1805.07381} (or other kinds of 
energy injection processes~\cite{1911.08883,1505.04444})
could ignite supernovae.
We leave analysis of such possibilities to future work.

\subsection{Scattering calculations}
\label{sec_wd_scalc}

Depending on its composition and temperature,
the matter in a WD core can exist in different states~\cite{10.1146/annurev.astro.41.081401.155117,0806.3692}.
In this paper, our focus will be 
 on illustrating
the physics of collective effects, rather
than performing detailed or comprehensive phenomenological
investigations. Accordingly, we will focus on a particular
nominal WD model, and leave investigations of
broader parameter space to future work.

As discussed in~\cite{bell_wd}, the best prospects for
detecting WD heating signatures appear to come from 
the heaviest WDs. Consequently, we will take
as our nominal model the heaviest WD model considered
in~\cite{bell_wd}.
This has basic parameters
\begin{equation}
	M_* \simeq 1.38 M_\odot
	\quad 
	R_* \simeq 1250 {\rm \, km}
	\quad 
	v_{\rm esc, core} \simeq 0.1
	\quad 
	\rho_{\rm core} \simeq 10^{10} {\rm \, g \, cm^{-3}}
	\quad 
	\mu_{e,\rm core} \simeq 8 \MeV
\end{equation}
where $M_*$ is the WD mass, $R_*$ is its radius,
and $v_{\rm esc,core}$, $\rho_{\rm core}$, and
$\mu_{e,\rm core}$ are the escape velocity,
density, and electron chemical potential in the WD core.
We will assume that the WD is old,
with core temperature $T_{\rm core} \simeq 10^5 \kelvin$.
\cite{bell_wd} takes the core to be composed
entirely of carbon ions and electrons; at such
low temperatures; the core consists of a 
lattice of carbon nuclei, embedded in a degenerate electron gas.
This is very analogous to a typical metal;
the major differences are that the ions
in the WD are bare nuclei, and the electrons
are at much higher (relativistic) velocities.

The most comprehensive previous study of DM scattering 
in WDs appears to be~\cite{bell_wd}.
This takes into account Pauli blocking for
scattering off electrons, but does not consider
other collective effects.
As discussed in Section~\ref{sec_srate},
this corresponds to calculating the scattering
rate using the leading-order part of the 1PI
self-energy $\Imag \Pi_L^{XX}$ (Eq.~\eqref{fig_cut1}), and ignoring the `mixing' contributions.
We will see that, even for the heavy-mediator case
that~\cite{bell_wd} considered, collective effects
can make significant differences to the scattering
rates for DM lighter than $\sim 100 \MeV$. For 
DM scattering through a light mediator, collective effects
are important up to much higher masses.

\begin{figure}[t]
	\begin{center}
		DM capture rate in WD (heavy vector mediator)\par\medskip
\includegraphics[width=0.7\textwidth]{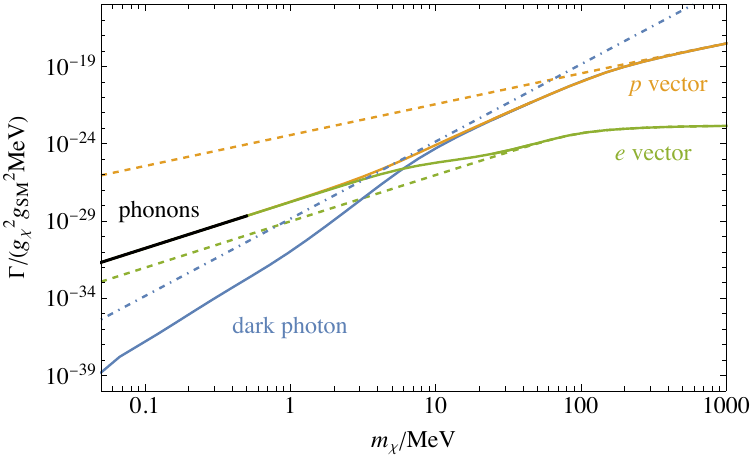}
		\caption{Rates for down-scattering to a gravitationally-bound
		orbit in the core of a heavy WD, for a DM particle with halo velocity $v_\infty = 10^{-3}$, 
		coupling through a heavy ($m_X = {\rm TeV}$) vector mediator.
		The blue curve corresponds to a dark photon mediator,
		the green curve to an electron-coupled vector, and the orange curve to a proton-coupled
		vector.
		The dashed green and orange curves correspond to calculations
		ignoring collective effects, while the solid
		curves take collective effects into account
		--- as can be seen from the plot, the solid
		curves match up with the dashed curves at high
		enough DM masses, where scattering is dominated by
		momentum transfers above screening scales.
		The dot-dashed blue curve corresponds to the sum
		rule limit (Section~\ref{sec_emsum}) on the scattering
		rate for a dark photon mediator; as expected,
		the dark photon curve lies below this limit.
		The dark photon rate is also well below
		those for other mediators at small DM masses,
		due to charge screening in the WD.
		The black curve labelled `phonons' corresponds to the rate
		for resonantly scattering into longitudinal phonons,
		given by Eq.~\eqref{eq_phrate}.
		As derived in Section~\ref{sec_wd_phonon},
		the scattering rate for 
		proton- and electron-coupled vector mediators
		is dominated by scattering into longitudinal phonons
		at small DM masses,
		accounting for the green and orange curves
		matching onto the black curve at low DM masses.}
		\label{fig_wdhv}
	\end{center}
\end{figure}

To compute scattering rates properly, the first
quantity we need to calculate is the photon self-energy
in the WD medium.
The leading-order electron contribution is fairly simple,
being very well-approximated by that of 
a fully degenerate ($T \ll \mu_e$), free electron gas
(see Appendix~\ref{app_degen_gas}).
To check that the free approximation is good, we can compare
the Coulomb interaction energy between electrons to
their kinetic energy; for a relativistic electron
gas, this gives
\begin{equation}
	\frac{e^2/r_s}{E_F}
	\sim e^2 \left(\frac{2}{9\pi}\right)^{1/3} v_F
	\simeq 4 \times 10^{-2} \ll 1
\end{equation}
where $E_F$ is the electron Fermi energy, $v_F$ is the 
Fermi velocity, and
$r_s$ is the `typical distance' between electrons,
defined via $n_e^{-1} = \frac{4}{3}\pi r_s^3$.
This ratio being small corresponds to
interactions acting as a small perturbation~\cite{raffelt_stars}.\footnote{This 
in contrast to the situation in e.g.\ metals, where
this ratio is typically $\sim 2-3$~\cite{10.1515/9781400837021}.
The electron gas in a metal 
can still be analyzed in terms of weakly-interacting
quasi-particles with renormalized parameters,
using Fermi liquid theory~\cite{girvin}.}
The leading-order formulae for the contribution to
$\Pi^{AA}(Q)$ from a degenerate fermion gas
are given in Appendix~\ref{app_degen_gas}
--- as far as we are aware, the exact forms
for the leading-order self-energy that
we present have not been derived previously.

The contribution to $\Pi^{AA}(Q)$ from the ions 
is somewhat more complicated, since Coulomb interactions
are strong enough to make the ions form a lattice~\cite{raffelt_stars};
approximating them as a free, non-relativistic gas will
not always be viable. In particular, for $q_0$ smaller
than the lattice band gap (which for
our nominal parameters will be $\OO(50 \keV)$~\cite{shapiro}), we expect the ion contribution
to the imaginary part to vanish as incoming DM cannot transfer sufficient energy to excite ions above the lattice band gap.
Additionally, for $q$ comparable to lattice
scales, the self-energy will be non-isotropic, since the lattice
picks out preferred directions.

One could treat the ion lattice properly via e.g.\
a density functional theory calculation, of 
the type performed in~\cite{10.1103/PhysRevD.101.055004,2102.09567}. Here, we will work with simpler 
approximations. For large momentum transfers, we will treat the
ions as a free, dilute, non-relativistic gas (see 
Appendix~\ref{app_dnr}).
For small momentum transfers,
 we will use the zero-velocity approximation
 $\Pi_L(Q) \simeq \omega_i^2 Q^2 / q_0^2$, 
 where $\omega_i^2 \equiv e^2 Z_i^2 n_i / m_i$
 is the ion plasma frequency,
 with $Z_i$ the ion charge, $n_i$ the ion
 number density, and $m_i$ the ion mass
 (this will be valid for $q$ much smaller than
 inverse lattice scales, and $q_0$ much smaller than the
 lattice band gap).
 This is how the ion lattice in a metal is treated
 in basic condensed-matter calculations
 for longitudinal excitations in metals~\cite{ashcroft}.
Figure~\ref{fig_wdhv} shows the resulting
scattering rates for different kinds of heavy vector mediators\footnote{These
rates are computed by numerical integration of 
Eq.~\eqref{eq_glqq0}, with $\Imag \Pi^X_L(Q)$ computed using
the formulae given in Appendix~\ref{app_vse}.
The $\Pi^{XX}$ and $\Pi^{XA}$ contributions
are computed analogously to $\Pi^{AA}$.},
where for DM masses $m_\chi \ge 25 \MeV$ we use the free
ion gas approximation,
while for $m_\chi \le 45 \MeV$ we use the zero-velocity approximation.
In the overlap regimes, the different curves match up well
enough (to within $\sim 1\%$) that the difference
is invisible on the plot,
suggesting that a proper treatment of the ion
lattice would also show similar behaviour.

The very low WD core temperatures we are considering,
along with the large WD escape velocity, mean that
DM evaporation will only be important for rather small
DM masses~\cite{bell_wd,2104.12757}. Quantitatively, if DM thermalizes in 
the WD core, its thermal velocity dispersion is
$\sigma_\chi \simeq 10^{-2} \sqrt{T/10^5\kelvin} \sqrt{100 \keV / m_\chi}$,
compared to the escape velocity $v_{\rm esc} \simeq 0.1$.
\cite{bell_wd} finds that, for this nominal WD model, evaporation becomes
significant for DM masses $m_\chi \lesssim 50 \keV$,
in agreement with these parametrics.

\begin{figure}[t]
	\begin{center}
		Imaginary part of longitudinal photon propagator in WD core\par\medskip
\includegraphics[width=0.49\textwidth]{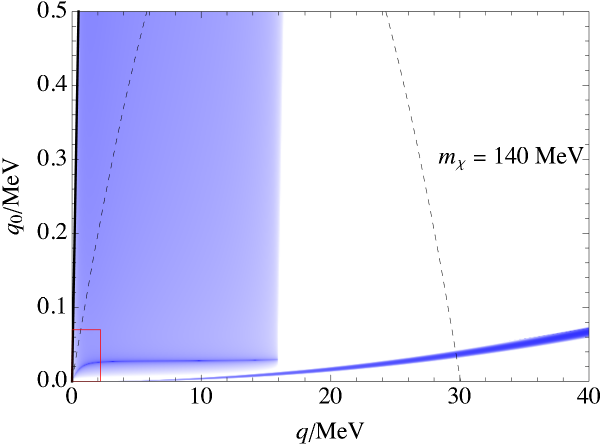}
\includegraphics[width=0.49\textwidth]{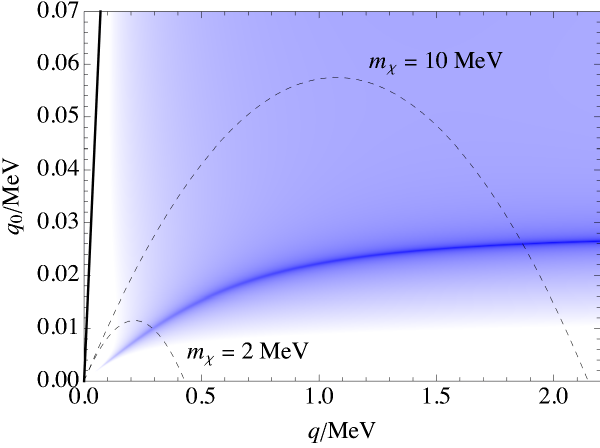}
		\caption{Density plot of the imaginary part of the longitudinal
		photon propagator $\Imag D_L(q_0,q)$ for a nominal white dwarf core
		model (see Section~\ref{sec_wd_scalc}).
		Darker shades correspond to larger
		values of the imaginary part, on a logarithmic scale.
		In the left-hand panel, the band on the left corresponds
		to scattering with the relativistic electrons,
		while the parabolic band corresponds to
		scattering with (non-relativistic) ions.
		The electron band is cut off at $q \simeq 2 E_F$
		by kinematics, where $E_F \simeq 8 \MeV$
		is the electron Fermi energy.
		The solid black line corresponds
		to the light cone, while
		the dashed black line corresponds to
		the maximum possible value
		$q_{0,\rm max}$ of $q_0$ for given $q$, for a
		DM particle of mass $m_\chi = 140 \MeV$
		(travelling at the escape velocity $v_{\rm esc,core} \simeq 0.1 c$).
		The right panel shows the imaginary part at smaller
		momentum transfers, corresponding to the red rectangle
		on the left-hand plot, along with $q_{0,\text{max}}$ curves for $m_\chi = 10$ MeV and 2 MeV.
		The main feature at these small momenta is the longitudinal
		phonon dispersion relation. }
		\label{fig_phonon}
	\end{center}
\end{figure}

While numerical calculations like those
in Figure~\ref{fig_wdhv} are, in some sense, all that we need
to make predictions, it is important to understand the physics behind
these behaviours. To that end, Figure~\ref{fig_phonon}
plots the imaginary part of the longitudinal photon propagator.
For large momentum transfers
(for which we approximate the ions as a free gas), illustrated in the 
left-hand panel, the imaginary part is dominated
by an electron scattering component at small $q$, and an ion scattering component
with $q_0 \simeq \frac{q^2}{2 m_i}$ (the latter dominates
the rate for $m_\chi \gg E_F$).\footnote{Properly, at momentum transfers
$q \gtrsim 100 \MeV$, the finite size of the carbon
ions will start to become important. We neglect these effects 
for simplicity; as we will see below, they are generally
not important for parameter
space in which collective effects are significant.
For example, in Figure~\ref{fig_wdhv}, we 
see that the no-mixing calculations
match up well with full calculations
for $m_\chi \gtrsim {\rm few} \times 100 \MeV$.
Since the escape velocity in the WD core is $\simeq 0.1$,
we need DM masses $m_\chi \gtrsim \GeV$ for 
ion form factors to be important.}
The electron Fermi momentum sets the width 
of the electron-scattering band;
for $q \gg m_e$, we must have $q - q_0 \lesssim 2 E_F$
in order for up-scattering from inside the Fermi
sea to be kinematically possible (see Appendix~\ref{app_degen_gas}).
For smaller momentum transfers, illustrated
in the right-hand panel, the dominant feature
is the longitudinal phonon pole 
(this plot looks almost the same for both free gas 
and zero-velocity ion models). 
As we derive below, resonant scattering into these
longitudinal phonons dominates the scattering
rate at small DM masses, and we can obtain analytic 
expressions for this rate.

For a vector mediator, there will also be contributions to
the scattering rate from transverse modes.
However, since
transverse modes couple to the DM particle
current rather than the particle density, these rates are suppressed
by $v_{\rm esc}^2 \sim 10^{-2}$ relative to the longitudinal
contributions, and only compete when 
the longitudinal modes are heavily screened 
(which can occur for light DM scattering through a dark
photon mediator, as we discuss in Section~\ref{sec_wd_dpl}).

\subsection{Longitudinal phonons}
\label{sec_wd_phonon}

 As noted above, our nominal WD core
 consists
of an ionic lattice embedded in a degenerate electron gas,
and is very analogous to a metal. Its collective excitations
are correspondingly similar to those of a metal; in particular,
the low-energy collective excitations are phonons (compression
and shear waves).
Since the ions are expected to form a BCC lattice~\cite{raffelt_stars},
only the acoustic modes are expected to be present,
rather than additional optical phonon modes.
As we discuss in this Section,
scattering from longitudinal phonons (compression waves)
can dominate the DM scattering rate at small DM masses.

Firstly, we review the basic theory of longitudinal 
phonons in a WD core~\cite{shapiro}.
For small momentum transfers, and phase velocities
($v_{\rm ph} = q_0/q$)
significantly larger than the ion velocities, 
then as discussed above, we can use
the zero-velocity approximation $\Pi_L(Q) \simeq \omega_i^2 Q^2/q_0^2$
for the ion contribution to the photon self-energy. 
The electron contribution is well-approximated
by that for a free, degenerate Fermi gas;
at small momentum transfers, this gives
$\Pi_L(Q) \simeq - \frac{4\alpha}{\pi} E_F p_F \frac{Q^2}{q^2}
\equiv \frac{-Q^2}{q^2} k_{\rm TF}^2$.
So, the inverse longitudinal photon propagator is given by
\begin{equation}
	Q^2 - \Pi_L(Q) \simeq Q^2 \left(1 + \frac{k_{\rm TF}^2}{q^2}
	- \frac{\omega_i^2}{q_0^2}\right)
	\label{eq_disp1}
\end{equation}
The longitudinal photon propagator has a pole, $Q^2 - \Pi_L \simeq 0$, 
at 
\begin{equation}
	q_0^2 = \omega(q)^2 \equiv \frac{\omega_i^2 q^2}{q^2 + k_{\rm TF}^2}
	\simeq \begin{cases}
		\frac{\omega_i^2}{k_{\rm TF}^2} q^2 & q^2 \ll k_{\rm TF}^2 \\
		\omega_i^2 & q^2 \gg k_{\rm TF}^2
	\end{cases}
	\label{eq_phc1}
\end{equation}
This gives the dispersion relation for longitudinal phonons.\footnote{As
is the case for metals, the acoustic dispersion relation for longitudinal phonons
arises because of screening from the electron gas;
an ion lattice in a \emph{fixed} negative charge background would
have gapped longitudinal oscillations~\cite{ashcroft}.}
The approximation to the electron contribution to $\Pi_L$ will
be valid for $q \lesssim E_F$, 
while the ion approximation should be valid at
$q$ small compared to lattice scales,
i.e. $q \lesssim {\rm few} \times \MeV$.
If there is only a single ion species,
then its number density is $n_i = n_e / Z_i
 = \frac{p_F^3}{3 \pi^2 Z_i}$. This gives a sound speed
 of
\begin{equation}
	c_s = \frac{\omega_i}{k_{\rm TF}} = \sqrt{\frac{Z_i}{3} \frac{E_F}{m_i}} v_F
\end{equation}
Taking our nominal WD core parameters, for which
$E_F \simeq 8 \MeV$,
$k_{\rm TF} \simeq 0.8 \MeV$ and $\omega_i \simeq 30 \keV$,
we obtain $c_s \simeq 0.036$.
This is illustrated in the right-hand panel of
Figure~\ref{fig_phonon}, which shows how $\Imag D_L(Q)$ is
maximised on the phonon dispersion relation,
for low momentum transfers.
The narrow dark-shaded region (the shading
is logarithmic in $\Imag D_L(Q)$) corresponds to
a weakly-damped pole.\footnote{At small momentum transfers, the imaginary part of the ion contribution
to $\Pi_L$ will be very small, while
the contribution from the electrons is
\begin{equation}
	\Pi_L(Q) \simeq k_{\rm TF}^2 \left(1 + i
	\frac{\pi}{2} \frac{q_0/q}{v_F}\right)
\end{equation}
so the imaginary part is suppressed
relative to the real part by at least
$c_s/v_F \sim \sqrt{E_F/m_i}$.}

Since $c_s$ is smaller than the escape velocity in the WD
core, $v_{\rm esc} \simeq 0.1$, it is kinematically possible
to scatter into phonons 
for all DM masses.
For weakly-damped phonons, the phonon
scattering rate is dominated by momenta 
close to the dispersion relation, and can be calculated analytically.
The escape velocity is non-relativistic,
so the leading-order scattering rate is
\begin{equation}
	\Gamma_L \simeq \frac{1}{2\pi^2} \frac{g_\chi^2}{v_\chi}
	\int dq dq_0 \, q  
	(1 + f(q_0)) \frac{1}{(q^2 + m_X^2)^2} \Imag \Pi^X_L(Q)
\end{equation}
(where we have assumed a vector mediator for simplicity ---
we discuss scalar mediators in Section~\ref{sec_scalar_wd}).
From Section~\ref{sec_srate}, $\Imag \Pi^X_L(Q)$ is given by
\begin{equation}
	\Imag \Pi^{X}_L(Q)
	= \Pi^{XX}_i + \frac{\Pi^{AA}_i((\Pi^{XA}_r)^2
	- (\Pi^{XA}_i)^2) + 2 \Pi^{XA}_r \Pi^{XA}_i (Q^2 - \Pi^{AA}_r)}
	{(Q^2 - \Pi^{AA}_r)^2 + (\Pi^{AA}_i)^2}
\end{equation}
So, if we fix $q$ and integrate over $q_0$, the resonant
contribution is dominated by the 
$\frac{\Pi^{AA}_i((\Pi^{XA}_r)^2 - (\Pi^{XA}_i)^2)}
	{(\Pi^{AA}_r - Q^2)^2 + (\Pi^{AA}_i)^2}$
	term for $q_0$ close to the dispersion relation
	value $\omega(q)$.
	If we expand $Q^2 - \Pi_r^{AA}$ to linear
	order around the dispersion relation $q_0 = \omega$,
	writing $Q^2 - \Pi_r^{AA} \simeq C (q_0 - \omega) \equiv
	C y$, then
\begin{align}
	\int dq_0 \Imag \Pi_L^X(q_0,q) (\dots)
	&\simeq
	\int dy 
\frac{\Pi^{AA}_i((\Pi^{XA}_r)^2 - (\Pi^{XA}_i)^2)}
	{C^2 y^2 + (\Pi^{AA}_i)^2} (\dots)
	\nonumber \\
	&\simeq \frac{\pi}{|C|} \left\{(\Pi^{XA}_r)^2 - (\Pi^{XA}_i)^2 (\dots)\right\}_{(\omega,q)}
\end{align}
From Eq.~\eqref{eq_disp1}, we have that 
\begin{align}
	\frac{d}{dq_0}(Q^2 - \Pi_L)
	&\simeq \frac{2}{q_0}\left(
	(1 + k_{\rm TF}^2)q_0^2 - \frac{q^2 \omega_i^2}{q_0^2}\right)
	\nonumber \\
	&= \frac{2(Q^2 - \Pi_L)}{Q^2} + \frac{2 \omega_i^2 Q^2}{q_0^3}
\end{align}
Evaluating this on the dispersion relation, we have
$C \simeq 2 \omega_i^2 Q^2 / \omega^3$.
Thus, the resonant scattering rate is 
\begin{equation}
	\Gamma_{\rm res} \simeq \frac{1}{4\pi} 
\frac{g_\chi^2}{v_\chi}
	\int \frac{dq}{q}(1 + f(\omega))
	\frac{\omega^3}{\omega_i^2}
	\frac{(\Pi^{AX}_r)^2 - (\Pi^{AX}_i)^2}{(q^2 + m_X^2)^2}
\end{equation}
The simplest situation in which to evaluate this integral
is for $m_\chi$ small
enough that $q_{\rm max} \simeq 2 m_\chi v_{\rm esc}
\lesssim k_{\rm TF} \sim \MeV$. In that case, we are scattering
off the acoustic (rather than flat) part of the dispersion
relation,
so $\omega \simeq c_s q \simeq \omega_i q / k_{\rm TF}$.
The value of $(\Pi_r^{AX})^2 - (\Pi_i^{AX})^2$ on the dispersion
relation depends on the mediator's couplings.
The contribution to the mixing self-energy from 
electrons is $\Pi_r^{AX} / Q^2 \simeq - \frac{g_{Xe}}{e} k_{\rm TF}^2 / q^2
\simeq - \frac{g_{Xe}}{e} \omega_i^2 / \omega^2$ (where $g_{Xe}$ is 
the mediator coupling to electrons), while the contribution
from ions is $\Pi_r^{AX}/Q^2 \simeq \frac{g_{Xp}}{e} \omega_i^2 / \omega^2$ (where
we have assumed a vector coupling to protons with coupling
$g_{Xp}$). This shows that mediator-electron and mediator-proton 
couplings give rise to the same resonant scattering rate ---
this is as we'd expect, since the ions and electrons move together
in low-frequency phonons --- while the rate for dark photons,
for which these contributions cancel, is suppressed. 
Taking the example of a heavy vector coupling purely to electrons
(or to protons), the resonant scattering rate is
\begin{equation}
	\Gamma_{\rm res} \simeq \frac{1}{4\pi} \frac{g_\chi^2 g_{eX}^2}{e^2 m_X^4 v_\chi}
	\omega_i k_{\rm TF}
	\int dq \, q^2
	\simeq 
	\frac{2}{3 \pi} \frac{g_\chi^2 g_{eX}^2}{e^2 v_\chi}
	\omega_i k_{\rm TF} \frac{m_\chi^3 (v_\chi - c_s)^3}{m_X^4}
	\label{eq_phrate}
\end{equation}
(ignoring the temperature of the medium),
since the maximum $q$ for which we can scatter into an acoustic
phonon is $q_{\rm max} = 2 m_\chi (v_\chi - c_s)$.
This analytic rate is drawn as the black line in 
Figure~\ref{fig_wdhv}, showing that it matches up with
the scattering rates for electron- and proton-coupled
vectors at small $m_\chi$.

For a dark photon mediator, we
have $\Imag \Pi^X_L(Q) = \frac{\kappa^2 Q^4 \Pi_i^{AA}}
	{(\Pi^{AA}_r - Q^2)^2 + (\Pi^{AA}_i)^2}$,
	so the resonant scattering rate is given by
\begin{equation}
	\Gamma_{\rm res} \simeq 
	\frac{g_\chi^2 \kappa^2}{4 \pi} \frac{1}{m_X^4 v_\chi}
	\int \frac{dq}{q} \frac{\omega^3}{\omega_i^2} q^4
	= \frac{32}{7\pi} \frac{g_\chi^2 \kappa^2}{v_\chi}
	\omega_i k_{\rm TF} \frac{m_\chi^4}{k_{\rm TF}^4}
	\frac{m_\chi^3 (v_\chi - c_s)^7}{m_X^4}
\end{equation}
As expected, this is suppressed by higher
powers of $m_\chi$ and $v_\chi$ compared
to Eq.~\eqref{eq_phrate} --- at small enough $m_\chi$, this
suppression is strong enough that transverse scattering can
dominate instead, as we discuss in Section~\ref{sec_wd_dpl}.

From Figure~\ref{fig_wdhv}, we can see that the resonant
scattering rate from Eq.~\eqref{eq_phrate} is larger
than the naive scattering rate for an electron-coupled vector,
but smaller than that for a proton-coupled vector. We can 
understand this behaviour parametrically.
For an electron-coupled vector, the contribution to the imaginary
part of the self-energy from the degenerate electron
gas is $\Imag \Pi_L \simeq \frac{g_{eX}^2}{2\pi} E_F^2 \frac{q_0}{q} \mathbb{1}_{q_0 < q v_F}$, where $\mathbb{1}$ denotes the indicator function.
So, assuming that $v_\chi \ll v_F$, the `no-mixing' scattering
rate (i.e.\ using
only the first term in Eq.~\eqref{eq_pxxitot}) is
\begin{equation}
	\Gamma_{\rm no-mix} \simeq \frac{1}{30 \pi^3} \frac{g_\chi^2 g_{eX}^2}{v_\chi}
	\frac{E_F^2 m_\chi^3 v_\chi^5}{m_X^4}
	\label{eq_gnme}
\end{equation}
Compared to the resonant scattering rate,
\begin{equation}
	\frac{\Gamma_{\rm res}}{\Gamma_{\rm no-mix}}
	\simeq \frac{20 \pi^2 k_{\rm TF} \omega_i}{e^2 E_F^2 v_\chi^2}
	\left(1 - \frac{c_s}{v_\chi}\right)^3
	= 20 \frac{v_F c_s}{v_\chi^2}
	\left(1 - \frac{c_s}{v_\chi}\right)^3
\end{equation}
If $c_s$ is not too much smaller than
$v_\chi$, then this ratio can be reasonably large
for high $v_F$ --- for our nominal
WD core, $\Gamma_{\rm res} / \Gamma_{\rm no-mix}
\simeq 20$.

For an ion-coupled vector, the most naive scattering rate
we can compute is to use the Yukawa cross-section for scattering
with stationary ions;
\begin{equation}
	\Gamma_{\rm no-mix} = n_i \sigma_{\chi i} v_\chi
	\quad , \quad \sigma_{\chi i} \simeq \frac{g_\chi^2 g_{Xp}^2 Z_i^2}{\pi}
	\frac{\mu_{\chi i}^2}{m_X^4}
	\label{eq_gnm}
\end{equation}
where we have assumed a heavy mediator ($m_X \gg m_\chi v_\chi$),
and $\mu_{\chi i} \equiv m_\chi m_i / (m_\chi + m_i)$
is the DM-ion reduced mass. This
corresponds to the dotted orange line
in Figure~\ref{fig_wdhv}, which matches
the full scattering rate at large $m_\chi$.
Compared to the resonant scattering rate at small $m_\chi$,
\begin{equation}
	\frac{\Gamma_{\rm res}}{\Gamma_{\rm no-mix}}
	\simeq \frac{2}{3 Z_i^2 e^2}
	\frac{k_{\rm TF} \omega_i m_\chi^3 (v_\chi - c_s)^3}{n_i \mu_{\chi i}^2 v_\chi^2}
	\simeq 
	\frac{2}{Z_i} \frac{m_\chi}{E_F}
	\frac{v_\chi c_s}{v_F^2} \left(1 - \frac{c_s}{v_\chi}\right)^3
	\label{eq_gratio}
\end{equation}
where we have used $\mu_{\chi i} \simeq m_\chi$, since we are interested
in the small-$m_\chi$ regime.
Since $c_s < v_\chi \ll v_F$, the resonant scattering rate is much
smaller than the naive rate unless $m_\chi \gg E_F$, 
as illustrated in Figure~\ref{fig_wdhv}.

For low-velocity ions, the scattering rate in
Eq.~\eqref{eq_gnm} is a good approximation to
the no-mixing scattering rate obtained from modelling
the ions as a free gas (since, as per Eq.~\eqref{eq_gnme},
the no-mixing scattering rate from electrons is much smaller).
Furthermore, since the imaginary part of the ion
contribution to $\Pi_L$ is unimportant for the
resonant scattering rate into phonons, and the real
part of the ion contribution is very similar for a free
gas or a zero-velocity model (at the momentum transfers of interest),
the resonant rate $\Gamma_{\rm res}$ computed above
is very close to the full scattering rate
for the free ion gas model.
Consequently, the small $\Gamma_{\rm res} / \Gamma_{\rm no-mix}$
ratio
in Eq.~\eqref{eq_gratio} corresponds to a strong cancellation
between the $\Pi_i^{XX}$ and mixing terms in Eq.~\eqref{eq_pxxitot}.

This cancellation arises because of the kinematics
of scattering with low-velocity ions.
To scatter off a stationary ion, we need $q_0 = q^2/(2m_i)$,
so $q_0/q = q/(2m_i) \le m_\chi v_\chi / m_i$. For small
$m_\chi$, this phase velocity is $\ll c_s$. Consequently,
the ion contribution to $\Pi_L$, which is $\propto 1/q_0^2$,
is much larger than the electron contribution.
Since the $\Pi_i^{XX}$ term is also dominated
by the ion contribution (for a free ion gas model),
we obtain a cancellation between this term
and the mixing term in Eq.~\eqref{eq_pxxitot}, similarly to a dark photon mediator.
Both this example, and the electron-coupled vector example above,
illustrate that even for mediators other than dark photons,
collective effects can give rise to parametrically large
differences between the full and naive calculations
(and that these can be either enhancements or suppressions).

\subsection{Transverse mediator modes}

The above calculations all applied to scattering via
the vector mediator's longitudinal mode.
For mediators other than dark photons, the scattering
rate via transverse modes is orders of magnitude smaller,
over the entire $m_\chi$ range in Figure~\ref{fig_wdhv}.
For an electron-coupled vector, the transverse
rate is suppressed by $v_\chi^2$ from the form of $\Gamma_T$
(Eq.~\eqref{eq_gt1}), and in addition, is not enhanced
by resonant scattering with phonons (the transverse
phonons --- shear waves --- arise
from the ion lattice, rather than through the lattice's interactions
with the electron gas, and couple more weakly to an electron-coupled
mediator, since they do not involve long-wavelength charge
density perturbations).
For an ion-coupled vector,
in addition to the $v_\chi^2$ suppression, the imaginary parts of $\Imag \Pi_T^{XX}$
and $\Pi_T^{AX}$ are suppressed by the small ion velocities;
e.g.\ from Appendix~\ref{app_dnr}, if we approximate the ions
by a free, non-relativistic gas, then
the ion contribution to vector self-energies is
\begin{equation}
	\Pi_T =
	\omega_p^2 - \left(\sigma_i^2
	- \frac{Q^2}{4 m_i^2}\right) \Pi_L
\end{equation}
where $\sigma_i = \sqrt{T/m_i} \sim 3 \times 10^{-5}$ is the typical ion velocity.
The only situation in which scattering via transverse
modes contributes significantly to the rate is
with a dark photon mediator, for which the longitudinal
modes are screened most strongly.
In contrast, there is no static screening for transverse
modes (i.e.\ $\Pi_T(q_0,q) \rightarrow 0$ as $q_0 \rightarrow 0$
--- see Appendix~\ref{app_sumrules}), and $\Pi_T$ is mostly imaginary
in relevant parts of the small-$Q$ phase space
(Eq.~\eqref{eq_pitr} and~\eqref{eq_piti}), so
the transverse rate can compete with the longitudinal one
at small enough momentum transfer. At
the smallest $m_\chi$ shown in Figure~\ref{fig_wdhv},
the dark photon rate begins to be dominated
by the transverse rate; we will discuss this kind
of behaviour in more detail (for a light dark photon mediator)
in Section~\ref{sec_wd_dpl}.

\begin{figure}[t]
	\begin{center}
		WD capture for heavy dark photon mediator\par\medskip
\includegraphics[width=0.8\textwidth]{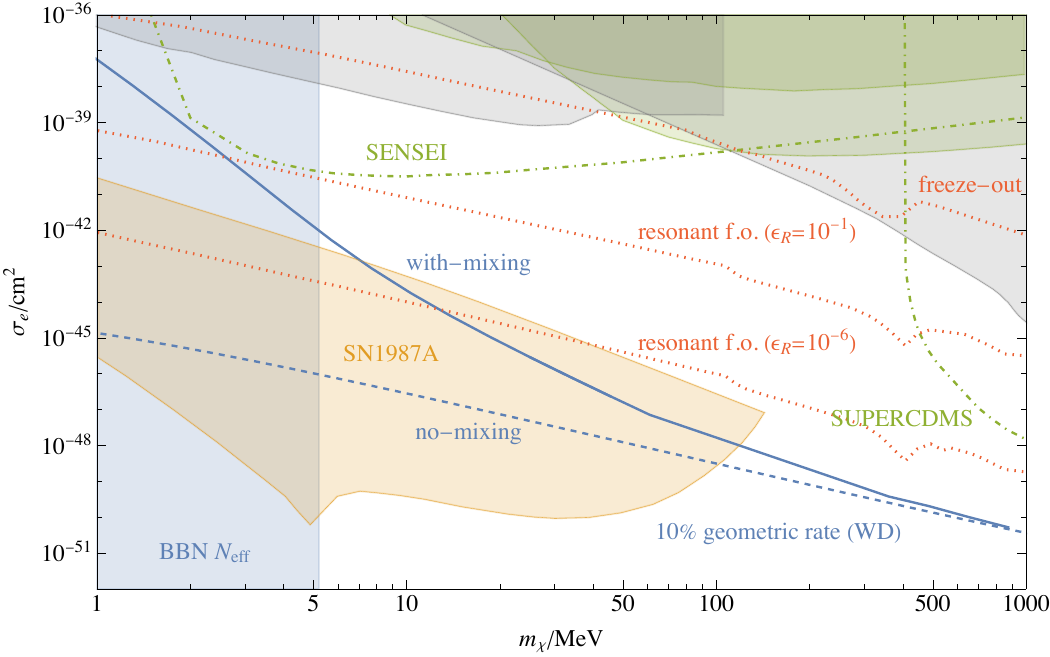}
		\caption{Plot of constraints
		on a scalar DM particle coupled via a heavy dark photon
		mediator, taking the mediator
		to have mass $m_X \sim 2 m_\chi$,
		with coupling $\alpha_\chi \equiv \frac{g_\chi^2}{4\pi}
		= 0.005$ to the DM $\chi$.
		The blue curve shows the coupling required
		to obtain $10\%$ of the geometric
		capture rate for our nominal WD
		model from Section~\ref{sec_wd};
		this corresponds to the approximate capture
		rate required for observable heating
		from DM annihilations, under the assumptions in~\cite{bell_wd}.
		The dashed blue curve shows the same calculation
		ignoring mixing effects.
		The other curves correspond to existing constraints
		on the DM model --- see Section~\ref{sec_wd_hdp}
		for details.
		}
		\label{fig_hdpwd}
	\end{center}
\end{figure}

\subsection{Heavy dark photon mediator}
\label{sec_wd_hdp}

Using the calculations outlined above, we can compare the potential 
reach of DM scattering signatures in WDs
to other constraints on DM parameter space.
This comparison will depend on the
DM model. Here, we will consider the illustrative
example of a `heavy' dark photon mediator;
specifically, $m_X \sim 2 m_\chi$.
This is heavy enough that 
the mediator mass is significantly larger
than scattering momentum transfers, but light enough
that cross-sections are not suppressed
by a parametrically higher scale.

Figure~\ref{fig_hdpwd} shows the constraints on this model,
in terms of the standard $\sigma_e$ parameter used in the 
direct detection literature,
\begin{equation}
	\sigma_e \equiv \frac{g_\chi^2 g_e^2}{\pi}
	\frac{\mu_{\chi e}^2}{((\alpha m_e)^2 + 
	m_X^2)^2}
\end{equation}
(this roughly corresponds to the DM-electron
cross section appropriate for direct detection
experiments relying on electron excitations).
The green shaded regions correspond to constraints
from existing direct detection experiments~\cite{1802.06998,1104.3088,1605.06262,1707.06749,1711.07692,1509.02448,1608.07648}, while the 
dashed green curves corresponds to sensitivity projections
for future direct detection experiments
(specifically, SENSEI~\cite{10.1103/PhysRevLett.121.0618038,10.1103/PhysRevLett.122.161801,2004.11378} and SuperCDMS~\cite{1610.00006}).

To place firm constraints on DM models from 
WD observations, we would need to observe
old, sufficiently cold white dwarfs in a location
where we can be confident the DM density is high.
So far, it is not possible to be certain that these
criteria are met. However, \cite{bell_wd}
argues that, from observations of WDs in the M4 globular
cluster, one can set plausible constraints
on the DM scattering rate for models in which in
the DM would subsequently annihilate in the WD,
assuming that the ambient DM density
in the globular cluster is $\OO(10^3 \GeV \cm^{-3})$.
To be in tension with observations, the capture
rate for such DM in heavy WDs would need 
to be $\gtrsim 10\%$ of the geometric capture rate~\cite{bell_wd}.
We indicate the coupling values
needed to attain this capture rate, for the nominal WD
model introduced above, via
the solid blue curve in Figure~\ref{fig_hdpwd}.

To calculate the DM capture rate in our nominal WD,
we take the DM halo velocity dispersion to be $\sigma = 8 \kms$,
and the velocity of the WD relative to the halo to be $v_\star = 20 \kms$,
as in \cite{bell_wd}. We approximate the capture
rate over the whole star by taking the per-volume capture
rate in the core, and multiplying it by the WD volume.
Since the core comprises the majority of the WD's volume,
and has a higher per-volume DM capture rate
than the less dense outer layers, this should be a good 
approximation.
It would obviously be possible to do a more proper calculation,
but the large uncertainties in parameters such as the halo DM density
in the globular cluster mean that other unknowns dominate the uncertainty in the calculation;
we leave a more careful treatment to future work.
As mentioned above, the main goal of our paper is to
illustrate the physics of collective effects.
To this end, we can compare our full calculation
to a calculation ignoring mixing effects,
shown as the dashed blue curve in Figure~\ref{fig_hdpwd}.
This illustrates how, for DM masses
$\lesssim \OO(100 \MeV)$, collective
effects suppress the scattering rate
via a heavy dark photon mediator by orders of magnitude.

The WD and direct detection signatures apply to a particle
that makes up $100\%$ of the DM abundance. While,
for a DM particle making up a fraction $f$ of the local DM
abundance, direct detection experiments are sensitive to
couplings $f^{-1/2}$ larger, WD heating signatures are
unobservable for sufficiently small $f$ (for \cite{bell_wd},
$f \lesssim 0.1$), due to the geometric capture rate ceiling.
Consequently, it is valuable to understand the parameter
space for which a candidate particle can make up $100\%$ of the
DM abundance.

The simplest way to produce the correct DM abundance
is via thermal freeze-out, i.e.\ for the DM
to be in chemical equilibrium with the SM
until $T \sim 0.1 m_\chi$, after which point
the exponentially Boltzmann-suppresssed
$\chi$ abundance means that $\chi$-$\chi$ annihilations
fall out of equilibrium. Since freeze-out relies
on $\chi$-$\chi$ annihilations being fast enough,
it implies a related $\chi$-$\chi$ annihilation
rate in the late universe, which could be detected through
cosmic rays or its effects on the CMB.
These constraints rule out standard freeze-out
scenarios for fermionic DM coupled
to a vector mediator, in which the annihilation
rate is dominated by a DM-velocity-independent component~\cite{10.1103/PhysRevD.96.095022}.
For scalar DM, on the other hand, 
the annihilation rate scales $\sim v_\chi^2$
since $s$-wave annihilation is forbidden by angular
momentum conservation~\cite{10.1103/PhysRevD.96.095022}.
Since the DM velocity is high during freeze-out,
$v_\chi \sim 0.1$, and much lower in the late
universe, scalar DM annihilating through a dark 
photon mediator can attain the correct
freeze-out abundance at couplings for which 
late-universe annihilation signatures are not constraining~\cite{10.1103/PhysRevD.96.095022}.
Consequently, for Figure~\ref{fig_hdpwd}, we have assumed
a scalar DM particle (though this does not make
a large difference to the WD calculations).

The upper red-dotted line in Figure~\ref{fig_hdpwd}
corresponds to the couplings required for the
correct freeze-out abundance, when $m_X$ is $\OO(1)$ different
from $2 m_\chi$ (e.g.\ $m_X = 3 m_\chi$).
As the figure shows, these couplings are large enough to be almost
excluded by existing experiments. 
To obtain the correct DM freeze-out abundance at smaller
couplings, we need to increase the annihilation
rate around freeze-out.
The simplest way to do this is to make the process resonant,
which occurs when $m_X \simeq 2 m_\chi$
(so that annihilation via an $s$-channel mediator is almost on-shell). 
The lower red-dotted lines in Figure~\ref{fig_hdpwd}
correspond to different resonant scenarios,
with $\epsilon_R \equiv \frac{m_X^2}{4 m_\chi^2} - 1$
quantifying how close to resonance a zero-velocity annihilation
is.
These curves illustrate that, by tuning the parameters,
we can obtain the correct DM abundance for a wide
range of couplings below current bounds.\footnote{For
scalar DM, $\epsilon_R \lesssim 10^{-6}$ results
in the dark photon dominantly decaying into SM
states, since the decay into DM particles is phase
space suppressed~\cite{10.1103/PhysRevD.96.095022},
changing the phenomenology. This can be avoided
in other models,
which can allow even smaller $\sigma_e$~\cite{10.1103/PhysRevD.96.095022}.}

Another constraint on DM freeze-out scenarios,
for sufficiently light DM, is the effect
of extra relativistic particles at $\sim \MeV$ temperatures
on BBN~\cite{2109.03246}. This rules
out thermal freeze-out DM for $m_\chi \lesssim 5.2 \MeV$~\cite{2109.03246}
(for scalar DM), as illustrated by the blue
shaded region in Figure~\ref{fig_hdpwd}.

At higher DM masses, the only astrophysical systems with sufficiently 
high temperatures to produce $\chi$ particles in large numbers are supernovae.
In particular, $\chi$ and $X$ production in SN1987A would
have modified the observed neutrino signal,
placing constraints on their couplings~\cite{1803.00993}.
Since $X$ production can also be significant (for $m_X$ not too large),
the SN1987A constraints will depend on
$g_\chi$ and $\kappa$ separately,
rather than just on their product.
In Figure~\ref{fig_hdpwd}, we show
the constraints calculated
in~\cite{1803.00993} for fermionic
DM, coupled to a dark photon with
mass $m_X = 3 m_\chi$, with $\alpha_\chi \equiv \frac{g_\chi^2}{4\pi}
= 0.005$ (this coupling is chosen
so that DM self-interactions are weak enough to be 
unconstrained~\cite{klz}). Though scalar DM with a
slightly different dark photon mass will behave
somewhat differently inside the SN, the overall
constraints should be similar~\cite{1803.00993}.

As well as being produced in supernovae,
$\chi$ and $X$ particles could also be produced
in accelerator experiments. Figure~\ref{fig_hdpwd}
show the constraints
from LSND~\cite{0906.5614}, E137~\cite{1406.2698}
and BaBar~\cite{1702.03327} as gray
shaded regions.
At DM masses $\gtrsim 40 \MeV$, where self-interaction
constraints are weaker,
the accelerator constraints can be relaxed
by taking larger $\alpha_\chi$
(since production is mostly
via on-shell $X$, which scales as $\kappa^2$ rather
than $\kappa^2 \alpha_\chi$).

Putting all of these constraints and projections
together, Figure~\ref{fig_hdpwd} illustrates
how DM scattering in WDs could potentiallly probe
much weaker couplings than even next-generation
direct detection experiments.
It also illustrates how, for DM masses 
$\lesssim \OO(100 \MeV)$, collective effects
suppress the capture rate by orders of magnitude
relative to a naive calculation.
While the difference between the 
mixing and no-mixing curves happens to 
be mostly within the SN1987A-disfavoured region
for this model, it is still important to 
have done the calculation correctly,
both to confirm that the rate is not even further suppressed,
and since the SN1987A constraints are 
subject to some uncertainties~\cite{1601.03422}.
For other dark matter models,
the relationship between WD signatures
and existing constraints will differ;
the main point of Figure~\ref{fig_hdpwd}
is to provide a worked example of how these
might fit together.

A possible caveat to our capture rate calculations is self-capture;
if the DM-mediator coupling $g_\chi$ is much larger than
the mediator-SM couplings
$\kappa e$, then even if only a small amount of 
DM is captured via SM interactions, it may come to dominate
the capture rate, and allow this to increase past the SM-only value.
For asymmetric DM models, in which DM does not annihilate,
but simply builds up in the WD, self-capture could
be important --- we leave investigations of
DM models and signatures in this regime
to future work. For symmetric DM, as we have been considering 
for WD heating signatures, the fairly short timescale
on which DM annihilates within the star means 
that, according to basic estimates, self-capture
should
not be significant for the mass range
in Figure~\ref{fig_hdpwd} (above the BBN bound).
We leave a more detailed treatment to future work.


\subsection{Light dark photon mediator}
\label{sec_wd_dpl}

The comparison between WD scattering signatures
and other constraints is somewhat different 
for a very light mediator, as low-momentum-transfer scatterings are enhanced.
In this Section, we will focus on the case
of an ultra-light dark photon mediator;
for other types of light mediators, there are typically
strong constraints on the mediator-SM couplings,
arising from high-energy 
experiments~\cite{1707.01503},
stellar production~\cite{ed}, or fifth force tests~\cite{10.1103/PhysRevD.93.075029}.

\begin{figure}[t]
	\begin{center}
		DM capture rate in WD (light dark photon mediator)\par\medskip
\includegraphics[width=0.7\textwidth]{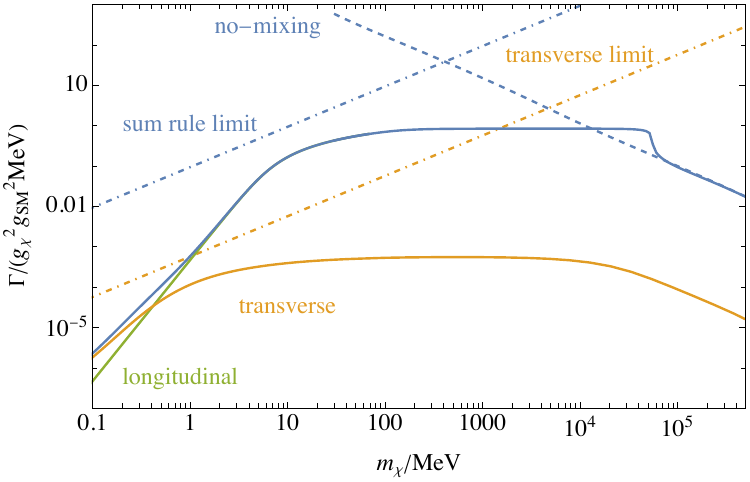}
		\caption{Rates for down-scattering to a gravitationally-bound
		orbit in the core of a heavy WD, for a DM particle with $v_\infty = 10^{-3}$, 
		coupling through a light dark photon mediator.
		The dashed blue curve corresponds to the rate calculated
		ignoring collective effects, while the solid
		blue curve incorporates collective effects,
		which become significant for
		$m_\chi \lesssim {\rm few} \times 10^4 \MeV$ 
		(corresponding to the bump in the solid blue curve).
		The dot-dashed blue curve corresponds to the EM
		sum rule limit for the scattering rate (Section~\ref{sec_emsum}),
		while the dot-dashed orange curve corresponds to the sum
		rule limit for scattering through transverse 
		dark photon exchange.
		The orange curve shows the contribution of
		scattering through transverse dark photon exchange 
		--- this is relatively suppressed at higher
		DM masses, but dominates at lower DM masses,
		due to the lack of screening at
		small momenta for transverse excitations.}
		\label{fig_wddpl}
	\end{center}
\end{figure}

Collisions in a WD occur at significantly higher
relative velocities
($\sim 0.1 c$) than typical halo DM velocities
($\lesssim 10^{-3} c$), so we expect the WD scattering
rate to be relatively enhanced for heavy mediators,
but not for light mediators.
Consequently, we expect that WD scattering signatures
for light mediators should be less competitive
with other probes, such as direct detection experiments,
than in the heavy mediator case. 

To illustrate the physics of scattering via a light dark photon
mediator (or equivalently,
for millicharged DM --- see e.g.\ Appendix D of~\cite{klz}),
Figure~\ref{fig_wddpl} shows the capture rate
for DM with a particular $v_\infty$ ($ = 10^{-3}$),
comparing the full calculation to a naive particle-by-particle one
(we show the results for spin-1/2 DM, but those for spin-0 DM are very similar).
For the naive calculation, the capture rate increases as we decrease
$m_\chi$, since smaller momentum transfers are required
to lose enough energy for capture.
Specifically, for capture to a bound orbit,
we need $q_0 > \frac{1}{2} m_\chi v_\infty^2$ 
(taking velocities to be non-relativistic).
For Yukawa scattering off stationary ions,
we have $q_0 = \frac{\mu_{\chi i}^2 v_\chi^2}{m_i} (1 - \cos\theta)$,
where $\theta$ is the scattering angle in the CoM frame,
so we need
$(1 - \cos\theta) > \frac{m_i m_\chi}{2 \mu^2} \frac{v_\infty^2}{v_\chi^2}
\equiv \epsilon$ for capture.
So, the total cross section for sufficiently hard scatterings
is
\begin{equation}
	\sigma = \frac{1}{8\pi} \frac{g_\chi^2 \kappa^2 e^2}{\mu^2 v_\chi^4}
	\left(\frac{1}{\epsilon} - \frac{1}{2 - \epsilon}\right)
	\overset{\epsilon \ll 1}{\simeq} 
	\frac{g_\chi^2 \kappa^2 e^2}{4 \pi} \frac{1}{m_i m_\chi}
	\frac{1}{v_\chi^2 v_\infty^2}
\end{equation}
showing that the small-momentum-transfer enhancement
more than compensates for the smaller $\mu_{\chi i}$
as $m_\chi$ is decreased, leading to a $\sigma \propto 1/m_\chi$
dependence, as per the dashed blue line in
Figure~\ref{fig_wddpl}.
Recalling that the total capture rate, from Eq.~\ref{eq_ceq1},
behaves as
\begin{equation}
	\frac{dC}{dV} \propto \int_0^\infty dv_\infty v_\infty f(v_\infty) \Gamma_{v_\infty}
\end{equation}
we see that attempting to calculate the total
capture rate, integrating over all $v_\infty$ using
the no-mixing rates, would result in a $\int \frac{dv_\infty}{v_\infty}$
logarithmic divergence driven by small $v_\infty$~\cite{1312.1336}.
This is why we displayed the capture
rate for a \emph{fixed} $v_\infty$ in Figure~\ref{fig_wddpl};
to obtain a finite total capture rate, we always need
to take into account mixing effects in some way,
and there is no sensible comparison to a fully naive calculation.

From Figure~\ref{fig_wddpl}, we can see that once
the momentum transfer required for capture
is smaller than the screening scale in the WD
(i.e.\ $m_\chi v_\infty^2 / v_{\rm esc} \lesssim \MeV$, rather
roughly),
the small-momentum enhancement is cut off by screening effects.\footnote{Even though Figure~\ref{fig_wddpl} shows DM masses up to 
$m_\chi \gg \GeV$, capture is dominated
by low-momentum-transfer scatterings, so nuclear
form factors are not important (they only become
important for $m_\chi \gtrsim 100 \TeV$).}
This results in a roughly constant scattering
rate as $m_\chi$ is decreased (since the rate is dominated by momentum transfers around
the screening scale),
until $m_\chi$ becomes
small enough that all of the accessible momentum
transfer range is below the screening scale,
after which the scattering rate decreases with $m_\chi$.
For small enough $m_\chi$, the weaker screening of transverse
modes is enough to make up for
the extra $v_\chi^2$ suppression,
and transverse scattering starts to dominate the capture
rate, as illustrated in Figure~\ref{fig_wddpl}.\footnote{
	As discussed in Section~\ref{sec_wd_scalc},
	a proper treatment of the ion lattice is somewhat complicated,
	and this is even more true for its transverse mode response.
	Transverse phonons (shear waves) can be supported by
	the lattice, even without dynamical electrons, so we
	cannot determine the transverse phonon
	properties via calculations like those in
	Section~\ref{sec_wd_phonon}. However, at
	the relevant momentum transfer scales,
	the total transverse response function will be dominated
	by the electron response, and different models
	for the ion response will make little difference.
	For example, we compared two different toy models
	for the ion contribution to the transverse response;
	treating ions as a free gas (Appendix~\ref{app_dnr}),
	and taking a single-pole response function for transverse
	phonons, assuming the shear wave velocity is 
	$\sim \sqrt{2}$ of the compression wave velocity~\cite{girvin}, with quality factor $Q = 100$.
	For the $m_\chi$ range shown in Figure~\ref{fig_wddpl}, these models
	result in almost the same behaviour.}
The figure also displays the EM sum rules limits
from Section~\ref{sec_emsum}, illustrating how 
the calculated rate gets within a factor few of this limit,
for $m_\chi$ such that the typical momentum transfer approximately
matches the appropriate screening scale in the WD.

\begin{figure}[t]
	\begin{center}
		WD capture for light dark photon mediator\par\medskip
\includegraphics[width=0.7\textwidth]{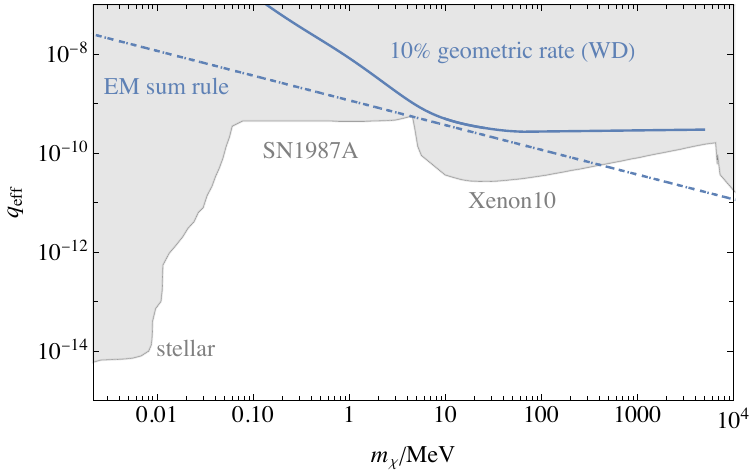}
		\caption{Bounds on the effective millicharge $q_{\rm eff}$
		for a DM particle coupling through an ultra-light
		dark photon mediator (or via the SM photon).
		The gray shaded region corresponds to existing constraints
		from stellar cooling~\cite{10.1088/1475-7516/2014/02/029}, SN1987A~\cite{1803.00993}, and the Xenon10 experiment~\cite{10.1103/PhysRevLett.109.021301,10.1103/PhysRevD.96.043017}. The blue solid curve corresponds to the
		approximate $q_{\rm eff}$ required to obtain $10\%$ of the geometric
		capture rate for the heavy WD model 
		in Section~\ref{sec_wd_scalc}. This
		corresponds to the approximate capture rate
		that could be constrained by DM capture and then annihilation,
		for WDs in the M4 globular cluster, according
		to~\cite{bell_wd}. The dashed line
		corresponds to the EM sum rule limit for this
		capture rate.}
		\label{fig_wd_dp_light}
	\end{center}
\end{figure}

Using the formulae from Section~\ref{sec_capture_rates}, 
we can compute the total capture rate, integrated
over the $v_\infty$ distribution.
Figure~\ref{fig_wd_dp_light} illustrates the results,
showing the coupling required
to obtain $10\%$ of the geometric capture rate
(as per the WD heating discussion in Section~\ref{sec_wd_hdp}),
using the same assumptions and approximations as described in the previous section\footnote{While the dependence
of the per-volume DM scattering rate on medium density is weaker
for a light dark photon mediator than for a heavy mediator,
it still generally increases with density
(as per the limits in Section~\ref{sec_emsum}).
This is in contrast to the case of dark 
photon \emph{production}, where the per-volume production rate
of very light dark photons is larger in less dense
media~\cite{1611.05852,1302.3884} ---
intuitively, the sterile mode decouples
from SM particles as $m_{A'} \rightarrow 0$, so its production
rate vanishes,
but DM-SM scattering is mediated by the active mode. 
Since the per-volume scattering rate
in the denser core is larger than in the less dense
outer layers of the WD, the approximation
of only using the scattering rate in the core
should be good.} (we display the results for fermionic DM
--- the rates for scalar DM are very similar).
The coupling is shown in terms of the `effective
millicharge' $q_{\rm eff} \equiv \kappa g_\chi / e$.
The gray shaded regions correspond to existing constraints;
at $m_\chi \lesssim 10 \MeV$, these arise
from production in SN1987A~\cite{1803.00993} or in stellar cores~\cite{10.1088/1475-7516/2014/02/029}, while at higher
masses, they come from direct detection experiments~\cite{10.1103/PhysRevLett.109.021301,10.1103/PhysRevD.96.043017}.
Figure~\ref{fig_wd_dp_light} illustrates that, over the whole $m_\chi$ range,
the couplings required for close-to-geometric capture
rates in WDs
lie within already-excluded parameter space.
The EM sum rule bound, indicated by the dashed line, shows
that this statement is almost independent of our model of the 
WD interior; for DM masses $\lesssim \GeV$, 
any causal model of the WD medium would result
in significantly-below-geometric capture rates, 
for unconstrained couplings (modulo 
the anti-screening issues discussed in Appendix~\ref{app_sumrules}).

One possible caveat to this statement is self-capture.
For millicharged DM (or somewhat equivalently, for a very light dark
photon mediator with $\kappa \sim \OO(1)$), self-capture is not enhanced.
If $\kappa \ll 1$, then $g_\chi$ can be much larger (for a given
$q_{\rm eff}$), and it might be possible for self-scatterings
to become important. However, it is worth noting that 
light mediators somewhat complicate the possible signatures of
DM scattering in WDs --- in particular, if DM annihilates
to a light mediator, this generally escapes from the WD,
so does not deposit its energy. Millicharged DM, which interacts
with the SM photon, would lead to energy depositions from annihilations,
but could only make up a small sub-component of the DM density
(at the relevant couplings),
since long-range interactions with SM matter would alter
the galactic dynamics of DM halos~\cite{1908.05275,10.1088/1475-7516/2009/07/014,10.1103/PhysRevD.83.063509,1602.04009,10.1016/j.physletb.2017.02.035,10.1088/1475-7516/2019/07/015,2007.00667}.
In fact, as demonstrated in recent work~\cite{2007.00667},
it is likely that dark photon models
are also constrained by galactic dynamics,
and are only viable as DM subcomponents at these couplings.

Overall, there are a number of points worth emphasizing.
For scattering via a light mediator, collective effects are important
even for large DM masses,
since the naive calculation for the capture rate
is dominated by soft scatterings. However, since
soft scatterings are enhanced, we obtain less benefit from the high
velocity of collisions in a WD, so obtaining
close-to-geometric capture rates requires
couplings above those allowed by other constraints. In addition,
it can be more complicated to obtain observable signatures,
such as WD heating, from light mediator models.

\subsection{Scalar mediator}
\label{sec_scalar_wd}

\begin{figure}[t]
	\begin{center}
		DM capture rate in WD (heavy scalar mediator)\par\medskip
\includegraphics[width=0.7\textwidth]{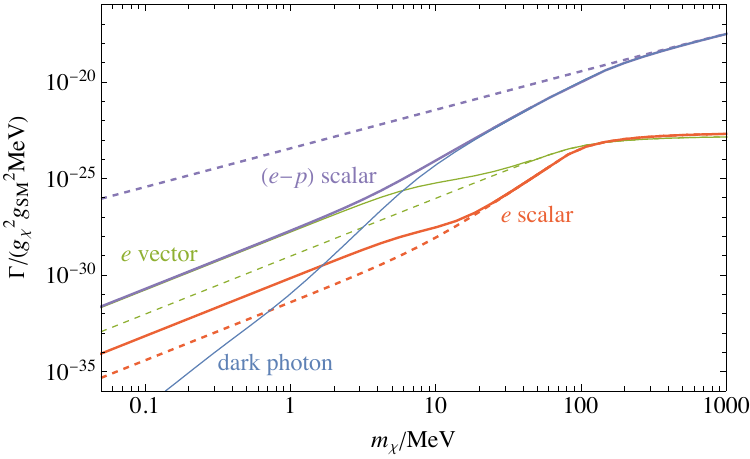}
		\caption{Rates for down-scattering to a gravitationally-bound
		orbit in the core of a heavy WD for a DM particle
		with $v_\infty = 10^{-3}$ coupling through
		a TeV-mass mediator of different types.
		The dashed curves show rates calculated ignoring
		collective effects, while the solid
		curves include collective effects.
		The red curves correspond 
		to a scalar mediator coupling to electrons.
		These can be compared to the rate for a electron-coupled
		vector mediator with the same mass
		(thin green curves).
		The purple curves show the rate for a scalar
		mediator coupling with opposite strengths
		to electrons and protons.
		The thin blue curve shows the rate
		for a heavy dark photon mediator; comparing
		to the $e-p$ scalar mediator,
		we see that the 
suppression at small DM masses is much less
		for the scalar mediator than for a dark photon mediator.}
		\label{fig_wdhs}
	\end{center}
\end{figure}

So far, we have been considering vector mediators,
with couplings of the form $X_\mu \bar f \gamma^\mu f$.
Our formalism can also be applied
to other types of mediators, such as scalars.
In Figure~\ref{fig_wdhs}, we show the scattering 
rates via different kinds of heavy scalar mediators
(with couplings of the form $\phi \bar f f$)
in our nominal WD core. 

For a scalar coupling to electrons,
$\LL \supset \phi \bar{e} e$, Figure~\ref{fig_wdhs}
shows that the scattering rate is the same as an electron-coupled
vector for large $m_\chi$, but significantly smaller for
small $m_\chi$. This is because, at large momentum transfers,
the relevant scale for both the vector-vector ($\Pi_L$) and
scalar-scalar ($\Pi^{\phi\phi}$) self-energies is $Q^2$,
whereas at small momentum transfer, the scale
for $\Pi^{\phi\phi}$ is $m_e^2$, whereas
the scale for $\Pi_L$ is $\sim E_F^2$ (for relativistic electrons).
Consequently, the scalar rate is suppressed by $\sim 1/\gamma_F^2$.
We see this same parametric difference between the 
no-mixing rates (shown as dashed lines), and the
full rates, which are dominated by resonant scattering
off longitudinal phonons, since the mixing
self-energy $\Pi^{\phi L} \sim m_e E_F$ is also suppressed compared to
$\Pi^{XL} \sim E_F^2$.

Figure~\ref{fig_wdhs} also shows the scattering rate for a scalar
with opposite couplings to electrons and protons,
$\LL \supset \phi (\bar{e}e - \bar{p}p)$. The scattering
rate is
almost exactly the same as the proton-coupled vector
in Figure~\ref{fig_wdhv}; this is because the ions
in the WD core are non-relativistic, whereas the electrons are relativistic,
so the scalar coupling to electrons is relatively suppressed,
but not the coupling to ions. This illustrates how,
even though an $e-p$ scalar mediator --- which
for non-relativistic SM matter, couples in
the same way as a dark photon ---
can somewhat suppress
scattering rates in e.g.\ direct detection experiments~\cite{2006.13909,ani},
this suppression is easily lifted by relativistic effects.


\section{Neutron stars}
\label{sec_ns}

Many of the possible observational signatures of DM scatterings
in WDs also apply to DM scattering in neutron stars (NSs),
including black hole formation~\cite{10.1103/PhysRevD.40.3221,1012.2039,1104.0382,1103.5472,1201.2400,1301.6811,1310.3509,1309.1721,1812.08773,2111.02414} and stellar heating~\cite{0708.2362,0709.1485,1004.0586,1005.5699,1704.01577,1707.09442,1807.02840,1901.05474,1904.09803,1906.10145,1911.06334,1911.13293}.
A benefit of the increased density of NSs is that the 
much higher escape velocity
(up to $\sim 0.9 c$ in a NS core) means that
the kinetic energy of infalling DM is comparable
to its rest mass energy,
so purely kinetic heating from DM scattering events could
potentially result in observable heating (if an old, cold neutron star
is observed in a region of sufficiently large DM density)
\cite{10.1103/PhysRevLett.119.131801,1911.06334}.

The major issue for predicting the results of DM scatterings
in NSs is our poor understanding of the physics of NS interiors.
The `crust' of a NS --- effectively, the region in which the
physics is somewhat understood --- extends to a depth of
$\sim 1 \km$ (compared to a total radius $\sim 10 \km$ for the star).
Its outer layers consist of an ionic lattice embedded within
a degenerate electron gas;
as the depth and density increase, this transitions
to a complicated `nuclear pasta' structure of nucleons~\cite{1911.06334}.\footnote{\cite{1911.06334} considers some collective effects
for DM scattering in NS crust, e.g.\ scattering off phonons,
but does not seem to perform a systematic calculation
including screening effects.}
At smaller radii (the `core'), where the densities are higher,
a range of possible behaviours have been proposed,
including various types of meson condensates, or
unconfined quark matter~\cite{10.1146/annurev.nucl.50.1.481,10.1016/j.ppnp.2004.07.001}.
In the latter case, it is even possible that
the QCD matter is neutral by itself,
forbidding the presence of an electron gas
\cite{10.1016/j.ppnp.2004.07.001,hep-ph/0012039}.
Consequently, the behaviour of the lepton
and QCD sectors in the NS core, which makes up the vast
majority of the star's mass and volume, are
very uncertain.

Given these uncertainties, one can either attempt to
place bounds on 
DM scattering rates, e.g.\
by considering scattering only within the NS crust~\cite{1911.06334},
or one can consider toy models of the NS core. Here, we adopt
the latter approach; we will consider a very simple toy model,
in which the electrons, muons and protons
are all represented by free, degenerate Fermi gases,
while QCD dynamics are ignored.
We emphasize that, given this unrealistic toy model,
the following calculations
should not be used to place
constraints or make precise predictions ---
their main point is
to illustrate how mixing effects are likely to have a significant
impact on scattering and capture rates.
As well as this toy model, we can calculate the EM sum rule
bounds for a dark photon mediator, which will be valid for
any physical model of the NS interior (modulo 
the anti-screening issues discussed in Appendix~\ref{app_sumrules}).

For our nominal NS parameters, we will take the heaviest
NS model from~\cite{bell_ns}, which has
\begin{equation*}
	M_* \simeq 2.2 M_{\odot}
	\quad
	R_* \simeq 12 {\rm \, km}
	\quad
	v_{\rm esc,core} \simeq 0.91
\end{equation*}
\begin{equation}
	\rho_{\rm core} \simeq 1.4 \times 10^{15} {\rm \, g \, cm^{-3}}
	\quad 
	\mu_{e, \rm core} \simeq \mu_{\mu,\rm core} \simeq 300 \MeV
\end{equation}
 The muon and electron
chemical potential determine the proton number density,
which corresponds to a proton chemical potential 
of $\sim 70 \MeV$ (i.e. $\mu_p \simeq m_p + 70 \MeV$).
We will approximate the photon self-energy
as the combination of the free, degenerate
Fermi gas contributions from these three species
(see Appendix~\ref{app_degen_gas}).
Neutrons, while uncharged, do interact with photons
via their magnetic moments~\cite{nieves}, but rough estimates indicate
that this contribution will be subdominant.
The more worrying omission is that protons (if they are even
a sensible degree of freedom in the NS core) strongly
interact with themselves, and with the rest of the strongly-interacting
QCD matter in the core.
We leave a more careful treatment of such effects to future work.

Neutron stars are born at high temperatures in supernovae,
but cool down over time; if heating is insignificant,
old and isolated NSs are expected to reach temperatures
$\sim \OO(100 \kelvin)$~\cite{10.1103/PhysRevLett.119.131801,1911.06334}. For kinetic heating via DM
scattering to be visible with projected instruments,
it would need to maintain a NS at a temperature $\gtrsim 10^3 \kelvin$~\cite{1704.01577}.
In either case, the temperatures of old NSs will generally
be negligible for the purposes of our scattering calculations.
These low temperatures, and the very high escape velocity,
mean that DM evaporation will only be significant
at very low DM masses~\cite{bell_ns_l,2104.12757}.

\begin{figure}[t]
	\begin{center}
		NS capture for heavy dark photon mediator\par\medskip
\includegraphics[width=0.8\textwidth]{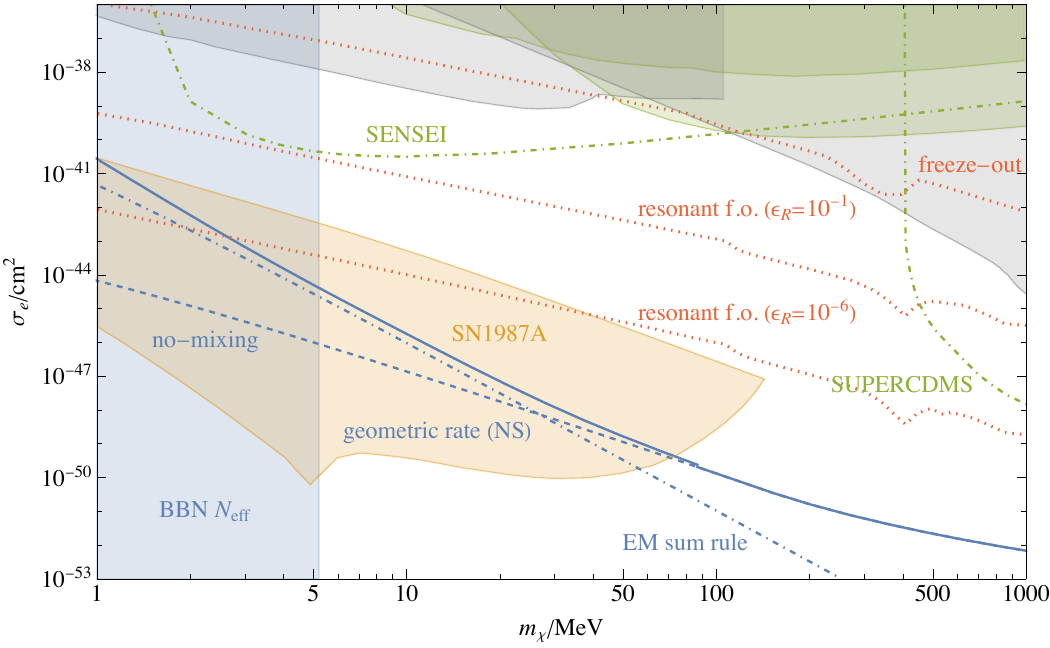}
		\caption{Plot of constraints
		on a scalar DM particle coupled via a heavy dark photon
		mediator, taking the mediator
		to have mass $m_X \sim 2 m_\chi$,
		with coupling $\alpha_\chi \equiv \frac{g_\chi^2}{4\pi}
		= 0.005$ to the DM $\chi$. 
		The solid blue curve corresponds to the approximate
		coupling required to obtain of order the geometric
		capture
		rate for DM in the toy NS model described in
		Section~\ref{sec_ns}.
		The dot-dashed blue curve
		corresponds to the EM sum rule limit
		for the capture rate (Section~\ref{sec_emsum}).
		The dashed blue curve corresponds to
		the capture rate computed ignoring
		mixing effects.
		The other curves and shaded areas are as per Figure~\ref{fig_hdpwd}.
}
		\label{fig_ns_hv}
	\end{center}
\end{figure}

Figure~\ref{fig_ns_hv} takes the same heavy dark photon mediator
model that we considered in Section~\ref{sec_wd_hdp},
and shows the approximate couplings for which one would obtain
$\sim$ geometric DM capture rates in our nominal NS,
for our toy model.
For reasonable DM densities, one generally needs close-to-geometric
DM capture rates for signatures such as kinetic heating to be
potentially observable~\cite{1704.01577}
(the DM capture rate
required for other signatures such as BH
formation is more model-dependent).
We calculate the capture rate assuming
a DM velocity distribution similar
to the Milky Way halo, with velocity dispersion
$v_0 \simeq 160 \kms$ and velocity offset
$v_\star \simeq 240 \kms$~\cite{10.1103/phys-revd.82.023530}
(in the notation of Appendix~\ref{app_vel_dist}).
Figure~\ref{fig_ns_hv} shows
that, for $m_\chi \lesssim 100 \MeV$, collective effects
start to make a significant difference to the capture
rate, even for a heavy mediator; for $m_\chi \lesssim 20 \MeV$,
the EM sum rule bounds show that almost any NS interior physics
leads to scattering rates below the naive particle-by-particle value,
for a heavy dark photon mediator.
Since the SN1987A constraints are subject
to considerable systematic uncertainties~\cite{1601.03422},
these rates are important to compute correctly,
even within the nominally disfavoured parameter space.

\begin{figure}[t]
	\begin{center}
		NS capture for light dark photon mediator\par\medskip
\includegraphics[width=0.7\textwidth]{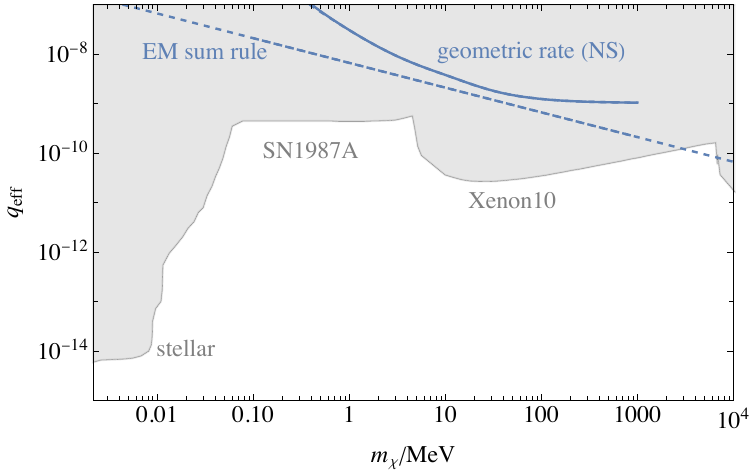}
		\caption{Bounds on the effective millicharge $q_{\rm eff}$
		for a DM particle coupling through an ultra-light
		dark photon mediator (or via the SM photon).
		The gray shaded region corresponds to existing constraints
		from stellar cooling~\cite{10.1088/1475-7516/2014/02/029}, SN1987A~\cite{1803.00993}, and the Xenon10 experiment~\cite{10.1103/PhysRevLett.109.021301,10.1103/PhysRevD.96.043017}. The blue solid curve corresponds to the
		approximate $q_{\rm eff}$ required to obtain of order the geometric
		capture rate for the heavy NS model in Section~\ref{sec_ns}. This
		corresponds to the approximate capture rate
		that could be constrained by kinetic
		heating signatures~\cite{1704.01577}. The dashed line
		corresponds to the EM sum rule limit for this
		capture rate.}
		\label{fig_ns_dpl}
	\end{center}
\end{figure}

Figure~\ref{fig_ns_dpl} shows the parameter space
for an ultra-light dark photon mediator (or millicharged
DM), as per Section~\ref{sec_wd_dpl}. Since
the NS escape velocity is even higher than for a WD,
we expect constraints involving lower DM velocities
(such as direction detection experiments)
to be comparatively even better relative to NS scattering,
for an ultralight mediator.
Correspondingly, our toy model gives a capture rate
that is always well below geometric, for couplings
in unconstrained parameter space, while the 
EM sum rule limit is sub-geometric for $m_\chi \lesssim 10 \GeV$.

\section{Dilute stellar plasmas}
\label{sec_dilute}

As well as the compact remnants discussed
in previous sections, DM scattering in
other types of stars could also lead to observable
signatures. In this Section, we will give
a brief overview of such signatures, and
then sketch how collective effects can 
modify DM behaviour, for a number of
illustrative cases involving our Sun.

The dilute, non-relativistic, weakly-coupled plasma that makes up
stars such as the Sun is rather simpler than the dense media
in WDs and NSs. 
However, the relatively high temperature inside the Sun ($T \sim \keV$
in the core) means that upscattering needs to be taken into account.
For light enough DM (not too much heavier than $m_e$),
the fact that the electron velocity in the Solar core is
much larger than
the escape velocity ($v_{\rm th} \sim \sqrt{3 T / m_e} \simeq 2 \times 10^4 \kms \gg 
v_{\rm esc,core} \simeq 10^3 \kms$) means that
scattering of halo DM can result in a higher-energy
population of reflected DM, if the optical depth of the
Sun is small enough~\cite{1708.03642}. The flux of this reflected component at Earth
will be smaller than the halo DM flux, but its higher energy
means that it can be a useful signal in direct detection experiments~\cite{1708.03642,2108.10332}.

For heavier DM, the Solar core temperature means that if
$m_\chi \lesssim {\rm few} \times \GeV$, evaporation
of previously-captured DM (i.e.\ up-scattering
to an unbound orbit) is not Boltzmann-suppressed enough to be
ignorable~\cite{1506.04316,2104.12757}. Consequently, it is not possible for a large population
of captured DM to be built up, though the evaporating flux,
which includes some particles at very high velocities, may
be useful target for direct detection experiments~\cite{1506.04316}.

At yet higher DM masses, scattering of halo DM can lead
to a captured population, which thermalizes
down to a smaller volume within the Sun.
If annihilation events occur, but the products
are absorbed and thermalize within the Sun, 
not enough energy is deposited to have an observable effect
on the Sun (even if DM is captured at the maximum possible geometric
rate). However, if the annihilation products escape,
such as neutrinos or long-lived hidden sector particles,
then these could give signatures in detectors on Earth~\cite{1108.3384,1503.04858,1602.07000,1603.02228,1612.05949,0910.1602,1102.2958,1602.01465,1703.04629}.
For DM heavy enough for this scenario to be of interest,
collective effects are only likely to be important for
the capture rate if scattering is via a light mediator.

If the dark matter is asymmetric, or otherwise
unable to efficiently annihilate, then other types 
of observational signatures are possible. The simplest
of these is anomalous heat transport. If DM has a significantly
longer mean free path than photons in the star, then despite
its much lower number density, it could contribute to heat transport,
potentially changing the structure of the star
by a detectable amount~\cite{2009.00663,1212.2985,1505.01362,1701.03928}.
For Population III stars, which may have formed at
the cores of dense DM halos, other possible signatures 
such as BH formation have been proposed~\cite{2111.02414}.

These potential observational signatures, in the Sun
and in other stars, motivate proper calculations of DM
scattering rates.
In Appendix~\ref{app_dnr}, we derive formulae
for vector self-energies in dilute non-relativistic plasmas (which translate
simply to scalar self-energies,
since all particles are non-relativistic); as far as we are aware,
these formulae are novel.
Below, we use these results to derive some features
of DM scattering inside the Sun, illustrating how 
collective effects can be important, and how
our formalism simplifies scattering calculations.
We do not attempt a comprehensive investigation of DM
scattering and its signatures in the Sun, though
our methods should be useful for future work on these topics.

\subsection{Solar capture with light mediator}
\label{sec_solar_dpl}

\begin{figure}[t]
	\begin{center}
		DM capture rate in Sun for dark photon mediator\par\medskip
\includegraphics[width=0.65\textwidth]{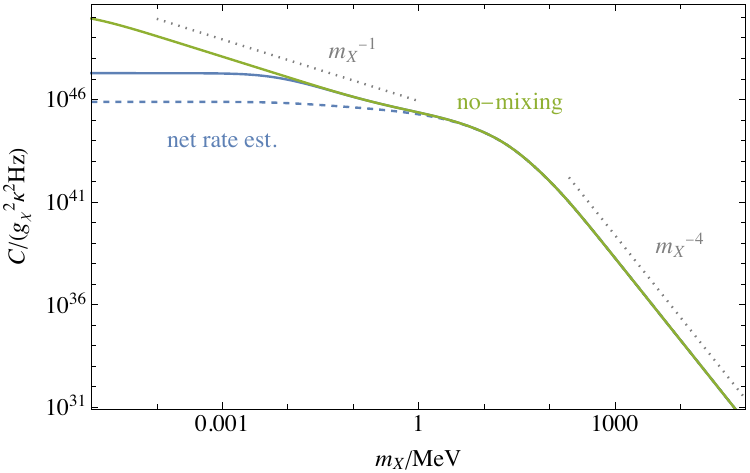}
		\caption{Estimate of total DM capture
		rate in the Sun for DM particles
		of mass $m_\chi = 50 \GeV$
		coupling through a dark photon mediator of 
		mass $m_X$. 
		The solid blue curve corresponds
		to the total down-scattering rate from unbound orbits
		into gravitational bound orbits, while
		the green curve corresponds to the capture rate
		ignoring mixing effects.
		The dashed blue curve corresponds to
		the approximate  net capture
		rate described in Section~\ref{sec_solar_dpl}.
		The dotted gray lines indicate different power law dependences
		on $m_X$.
		}
		\label{fig_solar_capture}
	\end{center}
\end{figure}

Since, as mentioned above, collective effects
(during the capture of halo DM particles) are most likely to be
important for light mediators, we will
investigate the effects of mediator mass on the capture rate
of DM in the Sun.
For DM coupling through a dark photon mediator,
Figure~\ref{fig_solar_capture} shows the 
capture rate for halo DM in the Sun, as a function of mediator mass
(taking $m_\chi = 50 \GeV$). This rate is estimated
via calculating the per-volume capture rate
in the core of the Sun, and multiplying this by the 
Solar volume. We take the halo DM velocity distribution
to be an offset Maxwell distribution as
described in Appendix~\ref{app_vel_dist},
with $v_0 \simeq 160 \kms$ and $v_\star \simeq 240 \kms$,
and use the BS2005 model~\cite{astro-ph/0412440}
for Solar core properties, taking into account
scattering off H, $^4$He, C, N, and O ions
(heavier ions can be important for hard scatterings
and heavier DM, but will not be significant for us).\footnote{More
up-to-date Solar models, such as AGSS09~\cite{10.1146/annurev.astro.46.060407.145222}, are available, but are not importantly different for
our purposes.}
More detailed computations
are easily done, but our calculations are mostly intended
to illustrate the physics involved. As we will see,
even when collective effects are not important,
the kinematics of scattering against a thermal
distribution of SM target particles can lead
to non-trivial behaviour, which our formalism makes
it simple to compute.

\begin{figure}[t]
	\begin{center}
		Imaginary part of longitudinal photon propagator in Solar core\par\medskip
\includegraphics[width=0.49\textwidth]{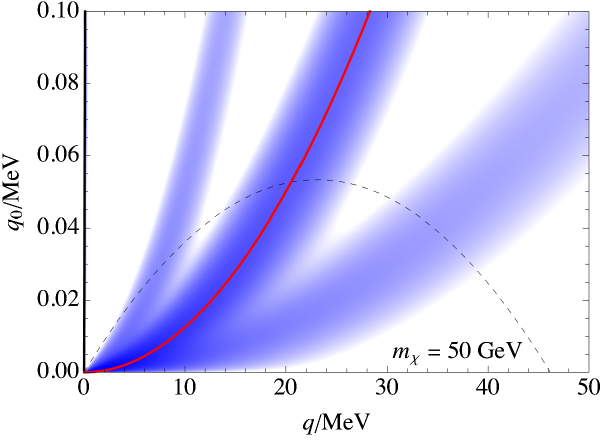}
\includegraphics[width=0.49\textwidth]{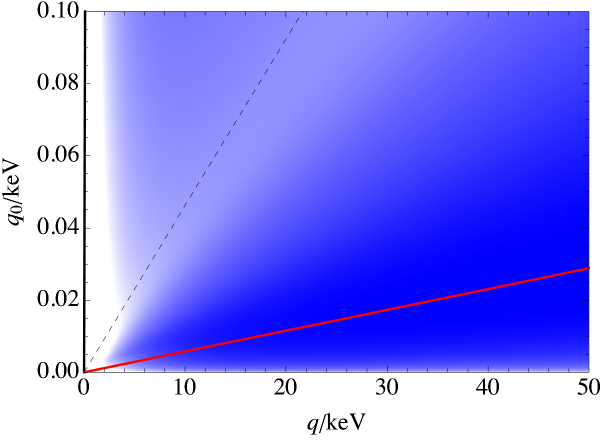}
		\caption{Density plot of the imaginary part of the longitudinal
		photon propagator $\Imag D_L(q_0,q)$
		in the Solar core. Darker shades
		correspond to larger values of the imaginary part,
		on a logarithmic scale.
		The left panel shows the imaginary part at larger momentum transfers;
		the three bands correspond to scattering
		off H, $^4$He, and heavier ions (from left to right).
		The dashed black line corresponds to $q_{0,\rm max}$
		for a DM particle of mass $m_\chi = 50 \GeV$,
		travelling at the Solar core escape velocity.
		The red line corresponds to the zero-velocity
		ion dispersion relation $q_0 = \frac{q^2}{2 m_i}$ for
		$^4$He.
		The right panel shows the imaginary part at
		smaller momentum transfers.
		Here, the dashed black line is as per the left panel,
		while the red line corresponds to
		$q_0 = q \sigma_i$, where $\sigma_i \equiv \sqrt{T/m_i}$
		is the thermal velocity dispersion
		for $^4$He. As discussed in Section~\ref{sec_solar_dpl},
		the phase space around the red curve in the left-hand panel
		dominates the scattering rate for
		large mediator masses, while the space around the red curve
		in the right-hand panel dominates for small mediator masses.}
		\label{fig_sdisp}
	\end{center}
\end{figure}

For high enough mediator masses, scattering is effectively
via contact interactions, and we have $C \sim m_X^{-4}$.\footnote{The DM mass $m_\chi = 50 \GeV$ is small
enough that nuclear form factor effects are not important
for Solar capture, even for a heavy mediator.}
We start to deviate from this behaviour once
$m_X$ is small enough that momentum transfers
$q \gtrsim m_X$ contribute significantly to the scattering
rate, i.e.\ $m_X \lesssim m_\chi v_{\rm esc,core} \simeq 3 \times 10^{-3} m_\chi$; this power-law break is visible in Figure~\ref{fig_solar_capture}.
If the Solar core were cold, then as calculated in
Section~\ref{sec_wd_dpl}, the capture rate
would scale as $\sim \log(1/m_X)$ for smaller $m_X$.
However, the Solar core temperature is high enough that
this does not hold for sufficiently small $m_X$; instead, as we can see from
Figure~\ref{fig_solar_capture}, the scaling
changes to $\sim m_X^{-1}$. In the full calculation, taking
into account mixing effects, this scaling holds until
$m_X$ becomes small compared to screening scales
(in the Solar core, the Debye scale $k_D \simeq 11 \keV$).

We can derive these scalings analytically.
Ignoring mixing effects, and taking the non-relativistic
limit of Eq.~\eqref{eq_cv1}, we have
\begin{equation}
	\frac{dC}{dV} \simeq \frac{2 g_\chi^2}{\pi}
	\int dq  
	\frac{q}{(q^2 + m_X^2)}
	\int_0^{q_{0,\rm max}} dq_0 \hat F_\infty(q_0)
	(1 + f(q_0)) \Imag \Pi_L
	\label{eq_dcdv}
\end{equation}
For $q_0 \ll q$ (as is true for non-relativistic scattering),
the contribution to the imaginary part of 
the photon self-energy from each plasma species is
(from Appendix~\ref{app_dnr})
\begin{equation}
	\Imag \Pi_L^{(i)} \simeq \frac{2 \sqrt{\pi} e^2 Z_i^2 n_i}{q \sigma_i} 
 \frac{1}{1 + f(q_0)}	e^{-(\xi_i - \delta_i)^2}
	\label{eq_pili}
\end{equation}
where the index $i$ ranges over ion species plus electrons,
and we have defined
$\xi_i \equiv \frac{q_0}{\sqrt{2} q \sigma_i}$,
$\delta_i \equiv \frac{q}{2 \sqrt{2} m_i \sigma_i}$,
where $m_i$ is the mass of the $i$-th species,
and $\sigma_i \equiv \sqrt{T/m_i}$ is its thermal velocity dispersion.
The $1/(1 + f(q_0))$ term in Eq.~\eqref{eq_pili}
cancels with the $1 + f(q_0)$ term in Eq.~\eqref{eq_dcdv},
so the $q_0$ dependence of the integrand in Eq.~\eqref{eq_dcdv}
is purely through the $\xi_i$ term in $e^{-(\xi_i - \delta_i)^2}$,
and the $\hat F_\infty(q_0)$ term.

For small $\sigma_i$, the $e^{-(\xi_i - \delta_i)^2}$ term gives
an exponential suppression unless the momentum transfer almost matches the on-shell ion dispersion relation, $q_0 \simeq q^2 / (2 m_i)$,
as we would expect.\footnote{Unlike in the WD case
analysed in Section~\ref{sec_wd_phonon},
there are not narrow acoustic resonances that we can resonantly
scatter into. Since the electron and ion velocities 
in the Sun are thermal, with the same temperature,
the imaginary part of $\Pi_L$ is comparable to the 
real part, along the $Q^2 - \Real \Pi_L = 0$ `dispersion relation'
\cite{fitzpatrick2015plasma}, and we do not have a narrow pole,
as illustrated in Figure~\ref{fig_sdisp}.
In a cold WD, the electrons have large velocities
due to the Fermi sea, and we do obtain a well-defined
pole for the ion acoustic mode, as per Figure~\ref{fig_phonon}.}
For $q_0 \ll m_\chi v_0^2$, where $v_0$ is the typical halo
DM velocity, we have $\hat F_\infty(q_0)  \underset{\sim}{\propto}
q_0$ (see Figure~\ref{fig_finf}), so the $\int dq_0$ integral in Eq.~\ref{eq_dcdv}
scales as $\sim q^2$.
Consequently, if we ignore mixing terms in $\Imag \Pi_L$,
the $\int dq$ integral in Eq.~\ref{eq_dcdv}
scales as $\int dq \frac{q^3}{(q^2 + m_X^2)^2}
\sim \log m_X^{-1}$ for small enough $m_X$, and we have re-derived the logarithmic
dependence we expected.

This derivation applies if we can always treat $\sigma_i$
as small. However, for $q$ small enough that $\delta_i \lesssim 1$,
we simply need $\xi_i \lesssim \OO(1)$ in order for $e^{-(\xi - \delta_i)^2}$
not to be an exponential suppression, rather
than $\xi_i \simeq \delta_i$. 
That is, we simply need the transfer phase velocity $v_{\rm ph} \equiv q_0/q$
to be $\lesssim \sigma_i$, rather than almost equal to
$q/(2 m_i)$, the phase velocity on the ion dispersion relation.
This is illustrated in Figure~\ref{fig_sdisp};
for $q \gtrsim 2 m_i \sigma_i$, the
dispersion relation bands are separated from the horizontal axis,
while for smaller $q$, they are not (as per the right-hand panel
of Figure~\ref{fig_sdisp}).

For these smaller $q$, 
the $\int dq_0$ integral is dominated
by $q_0 \sim q \sigma_i$ (assuming
$v_\chi > \sigma_i$,
which is the case for the Sun, as illustrated
in Figure~\ref{fig_sdisp}). If
this $q_0$ is small enough to be in the $\hat F_\infty(q_0) \sim q_0$
regime, then the $\int dq$ integral scales as $\int dq \frac{q^2}{(q^2 + m_X^2)^2} \sim m_X^{-1}$ overall. For the $m_\chi = 50 \GeV$ DM mass
considered in Figure~\ref{fig_solar_capture},
we have $m_i \sigma_i^2 = T  \simeq \keV
\ll m_\chi v_0^2 \simeq 14 \keV$, where $v_0$ is the halo
DM velocity dispersion,
so if $q \lesssim 2 m_i \sigma_i$, then
$q_0 \sim q \sigma_i$ is small enough for
$\hat F_\infty(q_0)$ to be in the linear regime (Figure~\ref{fig_finf}).
Consequently, for $m_X$ small enough that $\delta_i \lesssim 1$ dominates
the capture rate integral, the total capture rate scales
as $m_X^{-1}$.
In the Solar core, $m_i \sigma_i \sim {\rm few} \times \MeV$
(for the different ion species), so
the capture rate should scale as $m_X^{-1}$ for
$m_X \ll \MeV$, as we observe in Figure~\ref{fig_solar_capture}.\footnote{In recent
work~\cite{2110.02234},
it was claimed that the capture rate
(from a no-mixing calculation) scales as $m_X^{-2}$
in this regime. \cite{2110.02234} appears to derive
this conclusion from somewhat complicated numerical
calculations, rather than analytic arguments, and it 
is not clear where they differ from our analyses.
We have confirmed that our calculations match the results of directly
simulating scattering events for many randomly-generated particles.}
This calculation illustrates how, even in settings
where collective effects are not important, our formalism
can make including thermal kinematic effects simple,
since the self-energies incorporate all of the necessary
information about the velocity distributions of the SM particles.

One caveat regarding these capture rates is that 
they apply to down-scattering of halo DM from unbound to bound
orbits. As discussed in Section~\ref{sec_capture_rates},
if the temperature of the star is important for capture, then we also
expect \emph{up-scattering} of captured DM into unbound orbits
to be significant, once enough captured DM has been built up.
Correspondingly, the solid curves
shown in Figure~\ref{fig_solar_capture} give
the net DM capture rate before DM has built up in the Sun.
To compute the net capture rate once dynamic equilibrium
has been reached would require a full treatment of DM scattering
and evolution post-capture.
We do not attempt that here --- instead,
Figure~\ref{fig_solar_capture} shows 
a plausible estimate for the equilibrium net capture
rate (the dashed blue curve), from assuming that
the DM phase space density in bound orbits is the same
as the halo DM phase space density in just-escaping orbits.
That is, we take
\begin{equation}
	\frac{dC_{\rm net}}{dV} \simeq \frac{2 g_\chi^2}{\pi}
	\int dq  
	\frac{q}{(q^2 + m_X^2)}
	\int_0^{q_{0,\rm max}} dq_0 \left(\hat F_\infty(q_0)
	(1 + f(q_0))
	- q_0 \hat f_\infty(0) f(q_0) \right)
    \Imag \Pi_L
	\label{eq_dcdv_net}
\end{equation}
In situations where the capture rate is dominated
by low-$q_0$ scatterings, it is plausible that 
the phase space density for bound orbits
just below $E_{\rm esc}$ is similar to the phase
space density for halo DM trajectories just
above $E_{\rm esc}$, in which case Eq.~\eqref{eq_dcdv_net}
should be a good approximation.
We leave a more careful treatment of DM build-up in the Sun
to future work.

\subsubsection{Solar basin scattering}

A non-DM example in which Solar scattering via a light
mediator can be important is the scenario
of a `Solar Basin' of millicharged particles~\cite{2111.01796}.
The idea is that light ($\lesssim \keV$ mass) particles
with a small coupling to the photon (or to a light dark photon mediator)
would be produced in the Sun, and in particular, some would be produced
on bound orbits. These may survive in the Solar
system over very long timescales, and so build up in density
until they are detectable in Earth-based experiments~\cite{10.1016/S0927-6505(02)00186-X,10.1016/j.astropartphys.2005.01.002,10.1093/ptep/ptx137,2006.12431,2008.08594}.
To calculate the present-day density of such particles 
at Earth, it is important to take into account possible scatterings
within the Sun, which may scatter
Solar basin particles into different orbits (or eject
them from the Solar system entirely).

Doing these scattering calculations properly requires taking into
account the thermal velocities of the SM particles
inside the Sun, and the $Q$-dependent screening
of the mediator, as our formalism automatically includes.
Compared to the estimate in~\cite{2111.01796},
which ignores thermal velocities and approximates
screening with a constant Debye scale,
preliminary calculation using our approach
give a scattering rate $\OO(100)$ times larger,
which could be significant for the evolution
of Solar basin particles. We leave a full calculation
to future work.

\subsection{Solar reflection}

As mentioned above, light enough DM can be significantly accelerated
by scattering with electrons in the Solar core,
resulting in a high-energy `reflected' flux which could
be detected on Earth.
The most recent calculations~\cite{2108.10332} for this Solar
reflection scenario consider models with DM coupling through a heavy
mediator, as well as models with DM coupling via a light dark photon.
For the latter, they do attempt to take screening into account.
In our terms, their formulae
correspond to $\Imag \Pi_L/(|Q^2 - \Pi_L^{\rm Vlasov}|^2)$,
compared to the full expression
$\Imag \Pi_L/(|Q^2 - \Pi_L|^2)$,
where $\Pi_L^{\rm Vlasov}$ is the fluid
approximation to
the self-energy, as reviewed in Appendix~\ref{app_dnr}.
However, since $\xi_e \lesssim 1$ and $\delta_e \lesssim 1$
in electron scattering events (the DM is accelerated to 
at most the electron speed), the Vlasov
approximation to $\Pi_L$ will be $\OO(1)$ correct, and 
using our full expressions makes only a small difference to
the overall rates.

For other mediators, \cite{2108.10332} neglected collective
effects, whereas we have emphasised that these apply
even to heavy mediators.
However, since the DM masses of interest are generally
$\gtrsim 100 \keV$, to avoid other constraints,
we need momentum transfers $\gtrsim 10 \keV$ to 
give the DM $\gtrsim \keV$ kinetic energies,
relevant for detection in experiments on Earth.
Since the Debye scale in the Sun is also $\sim 10 \keV$, we
would expect collective effects to have $\mathcal{O}(1)$ (rather than
parametrically large) effects on signal rates. We leave
more precise calculations to future work. Nevertheless,
it is likely that, for the lower end of the DM mass
range considered in~\cite{2108.10332},
collective effects could be quantitatively important even for heavy
mediator models.


\section{Conclusions}
\label{sec_conc}

In this paper, we have showed how collective
effects in dense stellar media can significantly
modify DM scattering rates. For DM scattering
via contact interactions, we illustrated
that scattering rates in WDs and NSs for DM masses
$\lesssim 100 \MeV$ can be either parametrically
above or parametrically below the results of naive
calculations, depending on the model.
For scattering via a light mediator, naive
capture rate calculations diverge due to long-range
interactions, so collective effects are always important;
our formalism allows them to be computed systematically.
While collective effects of these kinds have
been investigated in a range of contexts
relevant to laboratory DM detection experiments 
(though even there, our setup can offer new insights,
e.g.~\cite{ani}),
they have predominantly been ignored for scattering 
in astrophysical media.
Since DM scattering in stars could lead to a wide
range of potential observational signatures that could give strong constraints or discovery potential
for DM, understanding the physics of
such scatterings
is phenomenologically important.

The aim of this paper has been to 
present a systematic formalism
for computing scattering rates in
media, and to illustrate its application
with some simple examples. We have not attempted to 
perform comprehensive analyses of different stellar
models, different DM models, and their
phenomenological consequences. This is the most
obvious avenue for future work based
on our results.

For stellar media whose underlying physics is known,
performing such analyses is a fairly well-defined
exercise. Properly treating DM scattering in NSs,
on the other hand, is more difficult problem.
While our toy calculations,
and EM sum rule bounds for dark photon mediators, suggest
that collective effects will be significant
for DM masses $\lesssim 100 \MeV$ even for contact interactions,
they do not provide reliable predictions.
Given the current uncertainty surrounding the basic physics of neutron
star cores, making precise predictions
is likely to be difficult, but it may be possible
to derive useful bounds on scattering rates,
e.g.\ by considering scattering in the better-understood crust~\cite{1911.06334}.

Among different DM candidates, a class of models
where our techniques will apply, but may have qualitatively
different behaviour, is inelastic DM~\cite{10.1103/PhysRevD.64.043502}.
Compact stars are excellent potential probes of 
inelastic DM, since the large kinetic energy of infalling
DM enables upscattering to states separated
by larger energy splittings than in a terrestrial
detector~\cite{1704.01577}. Upscattering can also
be important in the Sun, from the high thermal velocities
of electrons in the Solar core~\cite{2006.13918}.
Inelastic scatterings involve different regions
of $(q_0,q)$ parameter space from the elastic scattering events
we have considered in this paper, but can be treated via obvious extensions
of our methods.

Another area in which our calculations should prove
useful is the computation of particle emission rates
from astrophysical media.
The impact of collective effects on such processes
has been explored in a wide range of papers,
and was presented systematically in~\cite{ed}.
However, there are a range of particle candidates
and emission environments for
which only partial or flawed calculations have
been presented (e.g.\ the example of resonant
scalar emission from supernovae
discussed in Appendix~\ref{sec_res_scalar}).
Important ingredients for all of these calculations
are particle self-energies,
as presented in Appendices~\ref{app_vse} and \ref{app_scalar}.

In addition to stars, another obvious astrophysical
setting for dense plasmas is the hot early universe.
At some level, collective
effects will apply to DM scattering with the SM
plasma in the early universe.
For contact interactions, a generic issue is that,
post-electron-freezeout,
the number density of charged particles in the universe
is very small (compared to the photon number density),
and the collective scales are correspondingly small.
Pre-electron-freezeout, relativistic abundances of charged
particles are present, and collective scales can be large;
however, effects from this era are often hard to observe
in the late universe, having been thermalised along the way.
For light mediators, on the other hand,
collective effects in DM-SM scattering can be important
even for small screening scales, as explored
in various papers~\cite{1311.2600,10.1103/PhysRevD.99.115009,10.1103/PhysRevLett.127.111301}.
These generally use a fixed `Debye mass' to incorporate
screening, rather than the full
thermal effects, which may sometimes make
an appreciable difference.

In some circumstances, it is useful to treat the hidden
sector itself as a medium with which particles
can scatter. For example, if we consider
DM interacting via a light mediator,
then DM self-scattering can be calculated using 
the self-energy of the mediator in the DM medium.
This provides a systematic way to perform such calculations,
and should allow discrepancies between existing
treatments~\cite{0810.5126,1610.04611} to be
resolved, as well as encompassing collective
effects such as those investigated in~\cite{2007.00667}.

All of these avenues remain open and are ripe
for further research. Regardless, the examples
presented in this paper illustrate the importance
of considering collective effects when studying
DM in astrophysical settings. We hope
that the formalism presented here, as well as the
equations presented in the Appendices, will be useful
to researchers seeking to robustly understand
the dynamics of DM in astrophysical
settings.


\acknowledgments{We thank Asimina Arvanitaki, Masha
Baryakhtar, Tongyan Lin, Tanner Trickle, Ken Van Tilburg, and Zhengkang Zhang
for helpful conversations, and Masha Baryakhtar and Maxim Pospelov 
for comments on a draft of this paper. 
Some of the computing for this project was performed on the Sherlock and Farmshare clusters. We would like to thank Stanford University and the Stanford Research Computing Center for providing computational resources and support that contributed to these research results.
RL and MG's research is supported in part by the National Science Foundation under Grant No.~PHYS-2014215, and the Gordon and Betty Moore Foundation Grant GBMF7946. WD is supported by the U.S.A. Department of Energy, Grant No.\ DE-SC0010107.}


\appendix


\section{In-medium scattering rates}
\label{app_srates}

As discussed in Section~\ref{sec_srate}, the
scattering rate for a particle
passing through a medium can be related to the imaginary
part of the particle's effective in-medium propagator.
For a weakly-coupled, thermal medium, we can calculate
this using the tools of thermal field theory~\cite{bellac}.
In this Appendix, we will give explicit formulae
for the scattering rate in some simple situations,
considering spin-0 and spin-1/2 particles,
coupled via spin-0 and spin-1 mediators (with scalar
and vector couplings).
Straightforward extensions of these formulae apply to other
kinds of DM-mediator couplings, e.g.\
pseudoscalar couplings, DM form factors, inelastic scattering, etc.

\subsection{Spin-$1/2$ DM}

\subsubsection{Vector mediator}

Suppose that we have a Dirac fermion $\chi$ (our DM particle),
coupled to a vector mediator $X$ of mass $m_X$,
$\LL \supset g_\chi X_\mu \bar\chi \gamma^\mu \chi$.
If a $\chi$ particle is passing through a uniform (though 
not necessarily isotropic) medium,
then at leading order in the (assumed weak) couplings
between $\chi$ and $X$ and between $X$ and the medium,
its scattering rate can be obtained from the imaginary
part of the self-energy diagram
\begin{center}
	\begin{tikzpicture}[line width=1]
          \begin{scope}[shift={(2.4,0)}]
            \draw[f] (2*0.33,0)  node[above] {$ \chi $}
            -- (2*1.2,0);
            \node[] at (1.7,0.8) {$X$};
            \node[] at (4.3,0.8) {$X$};
            \node[] at (1.4,-0.3) {$P$};
            \node[] at (3.0,-0.3) {$P-Q$};
            \node[] at (4.5,-0.3) {$P$};
			\draw[f] (2*1,0) -- (2*2,0);
			  \draw[f] (2*2,0) -- (2*2.66,0) node[above] {$\chi$};
            \draw[v] (2*2,0) arc (0:180:2*0.5);
            \draw[pattern=north east lines,preaction={fill=white}] (2*1.5,2*0.52) circle (2*0.13);
          \end{scope}
	\end{tikzpicture}
\end{center}
and is given by~\cite{bellac}
\begin{equation}
\Gamma =-\frac{2 g_\chi^2}{E} \int \frac{d^3 q}{(2\pi)^3}
	\frac{1}{2 E'}
	(1 + f(q_0))
\frac{1}{(Q^2 - m_X^2)^2}
	\Imag \Pi^X_{\mu\nu}(Q) (P \cdot Q \eta^{\mu\nu} + P^\mu (P - Q)^\nu
	+ P^\nu (P - Q)^\mu)
\end{equation}
where $\Pi^X_{\mu\nu}$ is the in-medium self-energy for
the mediator,
$P = (E,p)$, and $q_0$ is such that $P - Q = (E',p')$ 
puts $\chi$ on-shell.
If the SM current that the mediator couples
to is conserved (or more generally, if current non-conserving
processes are unimportant in the medium),
then we have the Ward identity $Q^\mu \Pi_{\mu\nu}(Q) = 0$,
which we can use to write
\begin{equation}
\Gamma =-\frac{2 g_\chi^2}{E} \int \frac{d^3 q}{(2\pi)^3}
	\frac{1}{2 E'}
	(1 + f(q_0))
\frac{1}{(Q^2 - m_X^2)^2}
	\Imag \Pi^X_{\mu\nu}(Q) \left(\frac{Q^2}{2} \eta^{\mu\nu} 
	+ 2 P^\mu P^\nu\right)
\end{equation}
Specializing to an isotropic medium, we can
decompose the self-energy
as
\begin{equation}
	\Pi_{\mu\nu} = \Pi_L \epsilon^L_\mu \epsilon^L_\nu
	+ \Pi_T \sum_{i = 1,2} \epsilon^i_\mu \epsilon^i_\nu
\end{equation}
where $\epsilon^L_\mu$ is the unit longitudinal polarization vector
perpendicular to $Q$, and $\epsilon^{1,2}_\mu$ are the unit
transverse polarization vectors (all quantities
are functions of $Q$).
Explicitly, if we take e.g.\ $Q_\mu = (q_0,0,0,q)$, then
we can take
\begin{equation}
	\epsilon^1_\mu = (0,1,0,0) 
	\quad , \quad
	\epsilon^2_\mu = (0,0,1,0) 
	\quad , \quad
	\epsilon^L_\mu = \frac{-1}{\sqrt{Q^2}}(q,0,0,q_0)
	= \frac{i}{\sqrt{|Q^2|}}(q,0,0,q_0)
\end{equation}
where the last equality holds for $Q^2 < 0$, as we are interested
in for scattering.
Using this decomposition, we have
$\Gamma = \Gamma_L + \Gamma_T$, where
\begin{equation}
	\Gamma_L =
	\frac{4 g_\chi^2}{E} \int \frac{d^3 q}{(2\pi)^3} \frac{1}{2E'}
	(1 + f(q_0)) \frac{1}{(Q^2 - m_X^2)^2} \left(E^2 - p^2 \cos^2\theta\right) \Imag \Pi_L^X(Q)
\end{equation}
and
\begin{equation}
	\Gamma_T =
	\frac{2 g_\chi^2}{E} \int \frac{d^3 q}{(2\pi)^3} \frac{1}{2E'}
	(1 + f(q_0)) \frac{1}{(Q^2 - m_X^2)^2}
	\left( -Q^2 + 2 p^2 \sin^2\theta \right)(-\Imag \Pi^X_T(Q))
	\label{eq_gt1}
\end{equation}
We can change the integration variables to $q_0,q$ to get
\begin{equation}
	\Gamma_L =
	\frac{g_\chi^2}{2 \pi^2}
	\frac{1}{E p} \int dq \, dq_0 \, q 
	(1 + f(q_0)) \frac{1}{(Q^2 - m_X^2)^2} \left(E^2 - p^2 \cos^2\theta\right) \Imag \Pi_L^X(Q)
	\label{eq_glqq0}
\end{equation}
and analogously for $\Gamma_T$.
For non-relativistic scattering,
the leading term in small $v_\chi$ is
\begin{equation}
	\Gamma_L \simeq
	2 g_\chi^2 \int \frac{d^3 q}{(2\pi)^3} 
	(1 + f(q_0)) \frac{1}{(q^2 + m_X^2)^2} \Imag \Pi_L^X(Q)
	\label{eq_gl_nr}
\end{equation}

\subsubsection{Scalar mediator}

For a scalar mediator $\phi$ of mass $m_\phi$,
coupling to $\chi$ via $\LL \supset g_\chi \phi \bar \chi \chi$, we have a similar formula for the scattering rate,
\begin{equation}
	\Gamma =
	-\frac{2 g_\chi^2}{E} \int \frac{d^3 q}{(2\pi)^3} \frac{1}{2E'}
	(1 + f(q_0)) \frac{1}{(Q^2 - m_\phi^2)^2} \left(2 m_\chi^2
	- Q^2/2\right) \Imag \Pi^\phi(Q)
\end{equation}
For a non-relativistic $\chi$ particle, this has leading-order form
\begin{equation}
	\Gamma \simeq
	- 2 g_\chi^2 \int \frac{d^3 q}{(2\pi)^3} 
	(1 + f(q_0)) \frac{1}{(Q^2 - m_\phi^2)^2} \Imag \Pi^\phi(Q)
\end{equation}
Since $\bar{f}f \simeq \bar{f} \gamma^0 f$, we expect
$\Pi^\phi \simeq \Pi^X_{00} = \frac{q^2}{Q^2} \Pi^X_L
\simeq - \Pi^X_L$ if $\phi$ and $X$ couple to SM fermions
in same way, and the SM fermions are non-relativistic. So,
in this limit, scattering via scalar and vectors mediators 
results in the same rate, as expected.
(Note that nucleons in nuclei move at
speeds $\OO(0.1 c)$, so we have deviations
at the $\OO(10^{-2})$ level even for non-relativistic matter
--- this can be important for lifting cancellations etc.~\cite{ani}).

\subsection{Spin-$0$ DM}

\subsubsection{Vector mediator}

For complex scalar DM $\chi$, interacting with a vector
mediator $X$ via $\LL \supset i g_\chi X_\mu (\chi^* 
\partial^\mu \chi - \chi \partial^\mu \chi^*)$
(writing the three-particle coupling, which is the only 
one that contributes to the leading-order scattering rate), 
the scattering rate is given by 
\begin{equation}
	\Gamma =
	-\frac{4 g_\chi^2}{E} \int \frac{d^3 q}{(2\pi)^3} \frac{1}{2E'}
	(1 + f(q_0)) \frac{1}{(Q^2 - m_X^2)^2} P^\mu P^\nu \Imag \Pi^X_{\mu\nu}(Q)
\end{equation}
where we have used the Ward identity $Q_\mu \Pi^{\mu\nu}(Q) = 0$ as above.
For an isotropic medium, splitting this into transverse
and longitudinal parts gives
\begin{equation}
	\Gamma =
	\frac{4 g_\chi^2}{E} \int \frac{d^3 q}{(2\pi)^3} \frac{1}{2E'}
	(1 + f(q_0)) \frac{1}{(Q^2 - m_X^2)^2} \left(
	p^2 \sin^2 \theta (-\Imag \Pi^X_T(Q))
	- \frac{Q^2}{q^2}(E - q_0/2)^2 \Imag \Pi^X_L(Q)\right)
\end{equation}
with leading non-relativistic form
\begin{equation}
	\Gamma \simeq
	2 g_\chi^2 \int \frac{d^3 q}{(2\pi)^3} 
	(1 + f(q_0)) \frac{1}{(Q^2 - m_X^2)^2}\Imag \Pi^X_L(Q)
\end{equation}
This matches the expression for Dirac fermion DM, as expected.

\subsubsection{Scalar mediator}

For complex scalar DM $\chi$, interacting with a scalar mediator
$\phi$
as $\LL \supset c_\chi \phi \chi^* \chi$, the scattering rate is given by
\begin{equation}
	\Gamma =
	-\frac{c_\chi^2}{E} \int \frac{d^3 q}{(2\pi)^3} \frac{1}{2E'}
	(1 + f(q_0)) \frac{1}{(Q^2 - m_\phi^2)^2}\Imag \Pi^\phi(Q)
\end{equation}
(note that $c_\chi$ has dimensions of energy)
with leading non-relativistic form
\begin{equation}
	\Gamma \simeq
	-\frac{c_\chi^2}{2 m_\chi^2} \int \frac{d^3 q}{(2\pi)^3}
	(1 + f(q_0)) \frac{1}{(Q^2 - m_\phi^2)^2}\Imag \Pi^\phi(Q)
\end{equation}


\section{Vector self-energies}
\label{app_vse}

As discussed in Section~\ref{sec_srate} and
Appendix~\ref{app_srates}, the scattering
rate for DM passing through a uniform medium can
be expressed in terms of the in-medium
self-energy of the mediating particle. 
In this Appendix, we calculate the 
self-energy for a vector mediator $X$, which couples to
SM Dirac fermions as $X_\mu \bar f \gamma^\mu f$,
at leading order (i.e. treating the SM fermions as free).
For weakly-coupled plasmas, this will give a good
approximation, with other contributions
being suppressed by higher powers of $\alpha_{\rm EM}$.
Even for strongly-coupled plasmas,
if there is an effective description
in terms of weakly interacting quasi-particles
(e.g.\ as per Fermi liquid theory~\cite{girvin}),
these calculations can still apply.

\subsection{One-loop free fermion}

The leading-order self-energy in a medium of free Dirac
fermions corresponds to the one-loop diagram
\begin{center}
	\begin{tikzpicture}[line width=1]
			  \draw[v] (-1.5,0) node[above] {$X$} -- (-0.5,0);
			  \draw[v] (0.5,0) -- (1.5,0) node[above] {$X$};
            \draw[preaction={fill=white}] (0,0) circle (0.5);
			  \draw[-stealth] (-1.4,-0.3) -- node[below] {$Q$} (-0.6,-0.3);
	\end{tikzpicture}
\end{center}
If we consider the electron contribution to the photon
self-energy, we have
\begin{equation}
	\Pi^{\mu\nu}(Q) = 4 e^2 \int \frac{d^3 k}{(2\pi)^3}
	\frac{1}{2 E_k}
	(f_e(E_k) + f_{\bar e}(E_k))
	\frac{Q \cdot K (K^\mu Q^\nu + Q^\mu K^\nu)
	- Q^2 K^\mu K^\nu - (Q \cdot K)^2 g^{\mu\nu}}{
		(Q \cdot K)^2 - Q^4/4}
	\label{eq_pimunu}
\end{equation}
where $f_e$ is the in-medium occupation number for
electrons, and $f_{\bar e}$ for positrons.
This expression agrees with those in~\cite{braaten}
and~\cite{bellac}.
In a thermal medium, with temperature $T$
and chemical potential $\mu_e$, we have 
$f_e(E) = (e^{(E - \mu_e)/T}+1)^{-1}$,
and $f_{\bar e}(E) = (e^{(E + \mu_e)/T}+1)^{-1}$.\footnote{our convention for the chemical potential
is different from that of~\cite{bell_ns}, which
takes $f_e(E) = (e^{(E - m_e - \mu_e)/T} + 1)^{-1}$ etc.}
For Dirac fermions with an anomalous magnetic moment (e.g. neutrons),
there are extra contributions, which are
given in~\cite{nieves}.

In~\cite{raffelt_stars,braaten}, it is mentioned
that the $-Q^4/4$ term in the denominator
of Eq.~\eqref{eq_pimunu}
`can be ignored at $\OO(e^2)$', and moreover,
that it should be neglected to avoid introducing a spurious
imaginary part. These statements are only
true for timelike $Q$ with $Q^2 = \OO(e E)$,
where $E$ is the typical electron energy scale.
For our scattering calculations, where we are interested in
$Q^2 < 0$, keeping the $-Q^4/4$ term does
not introduce any spurious imaginary part,
and it is important to retain it
to correctly treat scatterings with larger
momentum transfers.

While the expressions we give in this section are presented
for the electron contribution to the photon
self-energy, it is simple to find the contribution
to general vector mediator self-energies from general Dirac fermion
species. We simply substitute the fermion-mediator
coupling $g_{Xf}$ for $e$, the fermion mass $m_f$
for $m_e$, and the species' chemical potential $\mu_f$
for $\mu_e$.

\subsubsection{Imaginary part}
\label{sec_vec_imag}

We can evaluate the imaginary part of the self-energy
either directly from Eq.~\eqref{eq_pimunu},
or by evaluating the cut self-energy~\cite{bellac}.
From either method, we obtain (for the longitudinal mode)
\begin{align}
	\Imag \Pi_L(Q)
	&= - \frac{e^2}{4\pi} \frac{Q^2}{q^2} \frac{1}{q} \frac{\sgn(q_0)}{1 + f(q_0)}
	\int_{R_Q}
	dE \sgn(E) \sgn(E') \times 
	\nonumber \\
	&(1 - \tilde f(E)) \tilde f(E')
	\left(\frac{Q^2}{2}
	+ 2 E E'\right)
	\label{eq_PiL_degen}
\end{align}
where $E' = E - q_0$,
and $\tilde{f}(E) \equiv (e^{(E-\mu_e)/T}+1)^{-1}$ is the fermionic
occupation number for a medium with electron chemical potential $\mu_e$
and temperature $T$.
The integration is over the range of $E$ for which 
$K$ and $K-Q$ can be on mass-shell,
for appropriate directions of $\vec{k}$,
i.e.
$R_Q = \{ E : |2 E q_0 - Q^2| \le |2 k q|
\mbox{ and } |E| \ge m_e \}$, where $k = \sqrt{E^2 - m_e^2}$.
For the transverse modes,
\begin{align}
	\Imag \Pi_T(Q)
	&= - \frac{e^2}{4\pi}  \frac{1}{q} \frac{\sgn(q_0)}{1 + f(q_0)}
	\int_{R_Q}
	dE \sgn(E) \sgn(E') \times 
	\nonumber
	\\
	&(1 - \tilde f(E)) \tilde f(E')
	\left(-\frac{Q^2}{2} + k^2 - \frac{(2 E q_0 - Q^2)^2}{4 q^2}\right)
	\label{eq_PiT_degen}
\end{align}

\subsubsection{Real part}

To evaluate the longitudinal and transverse
self-energies in an isotropic
medium, we can reduce the integral in Eq.~\eqref{eq_pimunu}
to an integral over the fermion energy,
by doing the angular integrals analytically.
Taking $K = (E,k)$ to be the four-momentum of
the fermion in the loop, we 
can define $x \equiv Q \cdot K = E q_0 - k q \cos \theta$,
with extreme values $x_\pm \equiv E q_0 \pm k q$ (where
$E$ and $k$ are such that $K^2 = m_e^2$).
In terms of $x_\pm$, the real parts of the self-energy 
are
\begin{align}
	\Real \Pi_L(Q) &= \frac{Q^2}{q^2} \frac{e^2}{2\pi^2} \frac{1}{q} \int_{m_e}^\infty dE (f_e(E) + f_{\bar e}(E)) \times \nonumber \\
	&\left[- 2 k q
	- \left(E(E-q_0) + \frac{Q^2}{4}\right)
	\log\left|
	\frac{x_+ - \frac{Q^2}{2}}{x_- - \frac{Q^2}{2}}\right|
	\right.\nonumber \\
	&+ \left. \left(E(E+q_0) + \frac{Q^2}{4}\right) \log\left|
	\frac{x_+ + \frac{Q^2}{2}}{x_- + \frac{Q^2}{2}}\right|
	\right]
	\label{eqQL1}
\end{align}
and
\begin{align}
	\Real \Pi_T(Q) &= \frac{e^2}{2\pi^2} \frac{1}{q} \int_{m_e}^\infty dE (f_e(E) + f_{\bar e}(E)) \times \nonumber \\
	&\left[2 k q \left(1 + \frac{Q^2}{2 q^2}\right) \right.
	\nonumber \\
	&+ \left.
	\frac{1}{2}\left(-k^2 + \frac{Q^2}{2}
	+
	\frac{(E q_0 - Q^2/2)^2}{q^2} \right)
	\log\left|
	\frac{x_+ - \frac{Q^2}{2}}{x_- - \frac{Q^2}{2}}\right|
	\right. \nonumber\\
	&+ \left.
	\frac{1}{2}\left(k^2 - \frac{Q^2}{2}
	-
	\frac{(E q_0 + Q^2/2)^2}{q^2} \right)
	\log\left|
	\frac{x_+ + \frac{Q^2}{2}}{x_- + \frac{Q^2}{2}}\right|
	\right]
	\label{eqQT1}
\end{align}

\subsection{Small-$Q$ approximations}

If we expand Eq.~\eqref{eqQL1} and Eq.~\eqref{eqQT1}
in small $Q^2$, we obtain
\begin{equation}
	\Real \Pi_L(Q) \simeq\frac{Q^2}{q^2}  \frac{e^2}{\pi^2}
	\int_{m_e}^\infty dE \,  k \,  (f_e(E) + f_{\bar e}(E))
	\left[-1 + \frac{E q_0}{k q} \log \left|\frac{x_+}{x_-}
	\right| - \frac{Q^2 E^2}{x_+ x_-} \right]
	\label{eq:PiL_smallQ}
\end{equation}
in agreement with~\cite{braaten} (equation A17),
and with~\cite{bellac} (equation 9.5), and
\begin{equation}
	\Real \Pi_T(Q) \simeq
	\frac{e^2}{\pi^2}
	\int_{m_e}^\infty
	dE \, k \, (f_e(E) + f_{\bar e}(E))
	\left[
		\frac{q_0^2}{q^2}
		- \frac{Q^2}{q^2} \frac{E q_0}{2 k q} \log \left|\frac{x_+}{x_-}
	\right|
		\right]
		\label{eq:PiT_smallQ}
\end{equation}
in agreement with~\cite{bellac} (equation 9.4).
These approximations are valid if $|Q^2| \ll |x_\pm|$
for the $E$ that contribute significantly to the integral.
For example, if the distribution is dominated
by highly relativistic electrons, then this
condition becomes $q \ll E$.

\subsection{Degenerate fermion gas}
\label{app_degen_gas}

For simple forms of $\tilde f(E)$, we can evaluate the one-loop
self-energy expressions analytically. An example
is a fully degenerate
Fermi gas, for which $\tilde f(E) = \mathbb{1}_{0 \le E < \mu_e}$
(if $\mu_e$ is positive).
For physical systems, this is a good approximation 
if $T \ll \mu_e$, as we discuss in Section~\ref{sec_degen_T}.

We will mostly be interested in scattering events
resulting in energy loss, so we will 
assume $Q^2 < 0$ and $q_0 > 0$ for simplicity (other
kinematic regions can be analysed similarly).
To obtain $\Imag \Pi_L$, we can perform the integral
in Eq.~\eqref{eq_PiL_degen} over final state
energies in the range $[E_+,\mu_e + q_0]$,
where 
\begin{equation}
	E_+ \equiv \max \left(\mu_e, 
\frac{1}{2}\left(q_0 
	+ q \sqrt{1 - 4 \frac{m_e^2}{Q^2}}\right)\right)
\end{equation}
(if $E_+ \ge \mu_e + q_0$, the integral is zero).
This gives
\begin{equation}
	\Imag \Pi_L(Q) \simeq -\frac{e^2}{4\pi} {Q^2}{q^3}
	\sgn(q_0)
	\left(
	-\frac{2 E_+^3}{3} + \frac{2 \mu_e^3}{3} + E_+^2 q_0 + \mu_e^2 q_0 - \frac{q_0^3}{3} 
		 -\frac{E_+ Q^2}{2} + \frac{\mu_e Q^2}{2} + \frac{q_0 Q^2}{2}
	\right)
\end{equation}
For the transverse part, 
\begin{align}
	\Imag \Pi_T(Q) &\simeq -\frac{e^2}{4\pi} \frac{1}{q}
	\sgn(q_0)
	\Big(
	-m_e^2 (-E_+ + \mu_e + q_0) + \frac{-E_+^3 + (\mu_e + q_0)^3}{3}
	\nonumber
	\\
	&\quad - 
	\frac{q_0^2 (-E_+^3 + (\mu_e + q_0)^3))}{3 q^2} - \frac{(-E_+ + \mu_e + q_0) Q^2}{2} 
	\nonumber
	\\
	&\quad + \frac{ 
	q_0 (-E_+^2 + (\mu_e + q_0)^2) Q^2)}{2 q^2} - \frac{(-E_+ + \mu_e + q_0) Q^4}{4 q^2}
	\Bigg)
\end{align}
We can also integrate the expressions for the real parts
in Eq.~\eqref{eqQL1} and Eq.~\eqref{eqQT1} analytically,
but the expressions are significantly more complicated
--- they are available 
in the code repository associated with this paper (\href{https://github.com/wderocco/DarkScatter}{github.com/wderocco/DarkScatter}).

In the small $Q^2$ limit, Eq.~\eqref{eq:PiL_smallQ} gives:
\begin{equation}
    \Real \Pi_L(Q)\simeq\frac{e^2}{\pi^2}\frac{Q^2}{q^2}E_F p_F\pare{-1+\frac{z_F}{2}\log\left|\frac{1+z_F}{1-z_F}\right|}
\end{equation}
and the imaginary part is
\begin{equation}
    \Imag \Pi_L(Q)\simeq-\frac{e^2 E_F p_F}{2\pi}\frac{Q^2}{q^2}z_F\Theta(1-|z_F|)
\end{equation}
where $z_F\equiv\frac{q_0}{q v_F}$, with $v_F \equiv p_F/E_F$. The imaginary part is zero for $q_0<0$.
The transverse equivalents are 
\begin{equation}
    \Real \Pi_T(Q)\simeq-\frac{e^2E_Fp_F}{2\pi^2}\parea{-1+\frac{z_F}{2\gamma_F^2}\log\left|\frac{1+z_F}{1-z_F}\right|+\frac{Q^2}{q^2}\pare{-1+\frac{z_F}{2}\log\left|\frac{1+z_F}{1-z_F}\right|}}
	\label{eq_pitr}
\end{equation}
and
\begin{equation}
    \Imag \Pi_T(Q)\simeq
	\frac{e^2}{4\pi}E_F p_F z_F 
\pare{\frac{Q^2}{q^2}+\frac{1}{\gamma_F^2}}\Theta(1-|z_F|)
	\label{eq_piti}
\end{equation}
where $\gamma_F \equiv E_F/m_e$.

\subsubsection{Temperature effects}
\label{sec_degen_T}

The formulae above were computed using the
zero-temperature electron distribution functions.
It turns out that these are good approximations
when $T \ll \mu_e$, even for $q_0, q \ll T$.

For the real parts, this is fairly obvious ---
the formulae in Eqs.~\eqref{eqQL1} and \eqref{eqQT1}
do not depend strongly on the sharpness of
the step function in $f_e(E)$.
For the imaginary parts, 
the $(1 - \tilde f(E)) \tilde f(E - q_0)$ term in
Eqs.~\eqref{eq_PiL_degen} and \eqref{eq_PiT_degen}
has integral
\begin{equation}
	\int_{-\infty}^\infty dE (1 - \tilde f (E)) \tilde f(E - q_0)
	= \frac{q_0}{1 - e^{-q_0/T} }
	= q_0 (1 + f(q_0))
\end{equation}
Consequently, the effect of changing $T$
on 
the kernel
$\frac{(1 - \tilde f(E))\tilde f(E - q_0)}{1 + f(q_0)}$
is simply to spread out the function in $E$, 
while keeping its integral fixed.
Since the energy range $\sim T$ over which this kernel
gets spread is small compared to the energies
$\sim \mu_e$ at which it has support,
the zero-temperature result is a good 
approximation for $T \ll \mu_e$.

From Appendix~\ref{app_srates}, the total
scattering rates
depend on $\int \dots \int dq_0 (1 + f(q_0)) \Imag \Pi^X_L(Q)$.
Consequently, if the self-energy is dominated by
degenerate Fermi gas contributions, then 
only the $1 + f(q_0)$ part depends strongly on $T$,
and we have that scatterings
with $q_0 \ll T$ are enhanced by $\sim T/q_0$,
as observed in~\cite{bell_ns_l,bell_wd}.

\subsection{Dilute non-relativistic gas}
\label{app_dnr}

Another situation in which we can evaluate 
Eq.~\eqref{eq_pimunu} analytically is when our plasma
species are dilute and non-relativistic.
From Eq.~\eqref{eq_finfty}, the occupation number
is given by $f_e(E_v) = (2\pi)^3 \frac{n_e}{2 m_e^3} p(v)$,
where $E_v$ is the energy of an electron with velocity $v$, and
$p(v)$ is the probability distribution for electron velocities.
The longitudinal mode self-energy is given by
$\Pi_L = \frac{Q^2}{q^2} \Pi^{00}$, so
\begin{align}
	\Pi_L &\simeq  \omega_p^2 \frac{Q^2}{q^2}
	\int d^3 v \, p(v) \frac{2 q_0(q_0 - v\cdot q) - Q^2 - (q_0 - v\cdot q)^2}{(q_0 - v \cdot q)^2 - \frac{Q^4}{4 m_e^2}}
	\nonumber \\
	&=  \omega_p^2 \frac{Q^2}{q^2}
	\int d^3 v \, p(v) \frac{q^2 - (v \cdot q)^2}{(q_0 - v \cdot q)^2 - \frac{Q^4}{4 m_e^2}}
	\label{eq_pil11}
\end{align}
where $\omega_p^2 \equiv e^2 n_e / m_e$ is the plasma frequency.
Since we are assuming that $v$ is non-relativistic,
the $(v \cdot q)^2$ term in the denominator will be negligible
compared to the $q^2$ term.
The integrand in Eq.~\eqref{eq_pil11}
only depends on $v \cdot q \equiv q v_q$, where $v_q$ is the component
of the velocity parallel to $q$, so
\begin{equation}
	\Pi_L \simeq  \omega_p^2 Q^2
	\int dv_q \, p_q(v_q) \frac{1}{(q_0 - q v_q)^2 - \frac{Q^4}{4 m_e^2}}
	\equiv  \omega_p^2 Q^2 S
\end{equation}
where $p_q$ is the probability distribution for $v_q$.
We can write the integral
$S$
as
\begin{align}
	S &= \int dv \, p_q(v)  \frac{1}{(q_0 - q v)^2 - \frac{Q^4}{4 m_e^2}}\nonumber \\
	&= - \frac{m_e}{Q^2} \int dv \, p_q(v) 
	\left(
	\frac{1}{\frac{Q^2}{2 m_e} + q_0 - q v}
	- \frac{1}{-\frac{Q^2}{2 m_e} + q_0 - qv}
	\right)
\end{align}
For a Maxwell velocity distribution,
$p_q(v) = \frac{1}{\sqrt{2\pi\sigma^2}} e^{-v^2/(2 \sigma^2)}$, this gives 
\begin{equation}
	S = - \frac{m_e}{\sqrt{2} Q^2 \sigma} \int dv \frac{1}{\sqrt\pi}
	e^{-v^2/(2 \sigma^2)}
	\left(
	\frac{1}{\frac{Q^2}{2 m_e} + q_0 - q v}
	- \frac{1}{-\frac{Q^2}{2 m_e} + q_0 - qv}
	\right)
\end{equation}
Using the integral definition of the plasma dispersion function~\cite{fitzpatrick2015plasma},
\begin{equation}
	Z(\xi) = \frac{1}{\sqrt\pi} \int_{-\infty}^\infty dx \frac{e^{-x^2}}{x - \xi}
\end{equation}
and writing $\xi = \frac{q_0}{\sqrt{2} \sigma q}$, we have
\begin{align}
	S &= \frac{m_e}{q_0 Q^2} 
	\xi \left[
		Z\left(\xi\left(1+ \frac{Q^2}{2 q_0 m_e}\right)\right)
		- Z\left(\xi\left(1- \frac{Q^2}{2 q_0 m_e}\right)\right)
		\right]
	\nonumber \\
	&\equiv \frac{m_e}{q_0 Q^2} 
	\xi (Z(\xi - \delta) - Z(\xi + \delta))
\end{align}
where
$\delta = \frac{-Q^2}{2 \sqrt{2} q m_e \sigma}$.
So,
\begin{equation}
	\Pi_L \simeq 
	\omega_p^2 \frac{m_e}{q_0}
	\xi (Z(\xi - \delta) - Z(\xi + \delta))
	\label{eq_pil_dnr}
\end{equation}
To connect this to the usual approximations for 
$\Pi_L$, we can consider situations in which
$\delta$ is small (we make this more precise below). Then,
using the fact that 
$Z'(\xi) = - 2(1 + \xi Z(\xi))$, we have that
\begin{equation}
	Z(\xi - \delta) - Z(\xi + \delta)
	\simeq 
	- 2 \delta Z'(\xi)
	= 
	\frac{Q^2}{q_0 m_e} \xi Z'(\xi)
	= - \frac{2 Q^2}{q_0 m_e} \xi (1 + \xi Z(\xi)) 
\end{equation}
This is a good approximation
if $|\delta| \ll \max(1,|\xi|)$.
In this regime,
\begin{equation}
	\Pi_L \simeq -  \omega_p^2 \frac{Q^2}{q^2} 
	\frac{1}{\sigma^2} (1 + \xi Z(\xi))
	\label{eq_vlasov}
\end{equation}
This is the `Vlasov' approximation to
the self-energy~\cite{fitzpatrick2015plasma}, which only depends on $e$, $n_e$ and $m_e$ 
through $\omega_p^2$.
It corresponds to treating each plasma species
as a fluid, characterized by its charge-to-mass
ratio $e/m_e$ and its charge density $e n_e$,
rather than taking into account the kinematics of individual 
scattering events.
For small $|\xi|$, corresponding to phase velocities
small compared to electron velocities,
the condition on $\delta$ is equivalent to
$q \ll m_e \sigma$, so the Vlasov
approximation is valid for momentum transfers
much smaller than the typical electron momenta.
For large $|\xi|$, we need $\frac{q^2}{2 m_e} \ll q_0$.
The useful aspect of our full formula, Eq~\eqref{eq_pil_dnr},
is that it simultaneously includes the kinematics of `hard' scattering events, for which these conditions are not satisfied,
as well as the coherent response to low-wavenumber
perturbations.

\begin{figure}[t]
	\begin{center}
\includegraphics[width=0.49\textwidth]{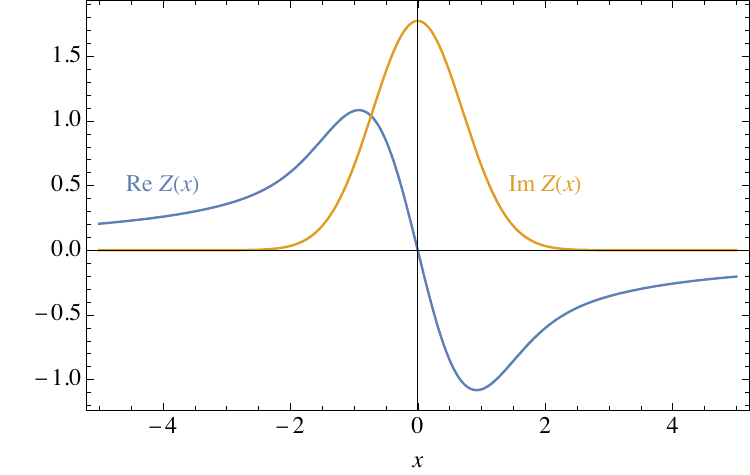}
\includegraphics[width=0.49\textwidth]{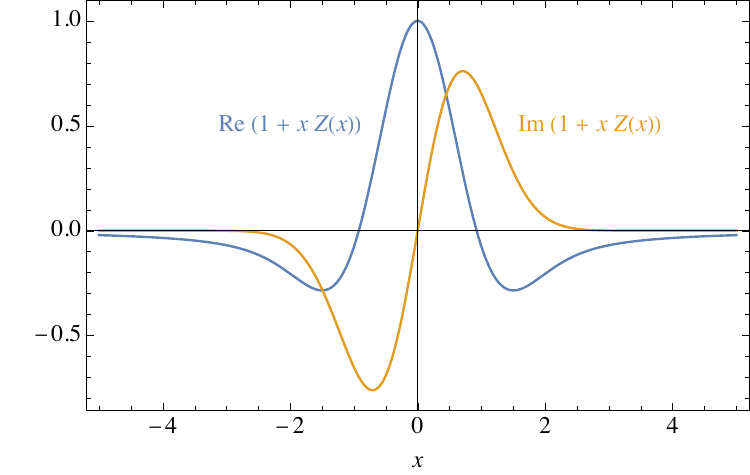}
		\caption{\emph{Left:} plots of real 
		(blue curve) and imaginary (orange curve) parts
		of the plasma dispersion
		function $Z(z) =i \sqrt\pi e^{-z^2} {\rm erfc}(-i z)$,
		evaluated for real arguments.
		\emph{Right:} real (blue)
		and imaginary (orange) parts of $1 + z Z(z)$,
		as arises in Vlasov approximations to
		self-energies (Eq.~\eqref{eq_vlasov}).}
    \label{fig_zfn}
	\end{center}
\end{figure}

To understand the behaviour of $\Pi_L$, we need the
behaviour of the plasma dispersion
function $Z$. In terms of the complex
error function, we can write
$Z(z) = i \sqrt{\pi} e^{-z^2} \erfc(-i z)$.
Evaluated at real arguments (as is the case above),
we have~\cite{10.1016/C2013-0-12176-9}
\begin{equation}
	Z(x) = i \sqrt{\pi} e^{-x^2} - 2 x Y(x)
\end{equation}
where
\begin{equation}
	Y(x) = \frac{e^{-x^2}}{x} \int_0^x dt \, e^{t^2}
\end{equation}
The real and imaginary parts of $Z(x)$ for real $x$ are plotted
in the left-hand panel of Figure~\ref{fig_zfn}.
For large $|x|$, we have
\begin{equation}
	\Real Z(x) \sim - \frac{1}{x}\left(1 + \frac{1}{2 x^2} + \dots\right)
\end{equation}
while for small $|x|$, 
\begin{equation}
	\Real Z(x) \sim - 2 x \left(1 - \frac{2 x^2}{3} + \dots\right)
\end{equation}
Figure~\ref{fig_zfn} also plots the real and imaginary parts
of $1 + x Z(x)$, which gives the Vlasov approximation
to the self-energy (Eq.~\eqref{eq_vlasov}).

It is easy to see that Eq.~\eqref{eq_vlasov} corresponds
to the expected plasma behaviour in the static and zero-momentum
limits. In the static ($q_0 \rightarrow 0$) limit,
we have $\xi \rightarrow 0$, 
so $\xi Z(\xi) \sim i \sqrt\pi \xi$, and
consequently $\Pi_L \simeq \omega_p^2 / \sigma^2$, giving
the usual Debye screening scale.
In the zero-momentum limit, $q\rightarrow 0$,
we have $\xi \rightarrow \infty$, so
$\xi Z(\xi) \sim - \left(1 + \frac{1}{2 \xi^2}\right)$,
so $\Pi_L \simeq \omega_p^2$, as expected.

We can obtain a simplified expression 
for the imaginary part of Eq.~\eqref{eq_pil_dnr}
when $q_0 \ll q$.
In that case, $\delta \simeq \frac{q}{2 \sqrt{2} m_e \sigma}$,
and we have $\xi \delta \simeq \frac{q_0}{4 T}$.
Consequently,
\begin{equation}
	\Imag ( Z(\xi - \delta) - Z(\xi + \delta))
	\simeq 2 \sqrt{\pi} 
	e^{-(\xi^2 + \delta^2)}
	\sinh \left(\frac{q_0}{2 T}\right)
	= \sqrt{\pi} e^{-(\xi - \delta)^2}
	\frac{1}{1 + f(q_0)}
\end{equation}
so
\begin{equation}
	\Imag \Pi_L \simeq 
	\frac{\pi}{\sqrt 2}
	\frac{e^2 n_e}{q \sigma} \frac{1}{1 + f(q_0)}
 e^{-(\xi - \delta)^2}
	\label{eq_im_dnr}
\end{equation}

\subsubsection{Yukawa scattering rate}

As an illustration, we can put together the scattering
rate expressions
from Appendix~\ref{app_srates} with the $\Imag \Pi_L$ expression
from Eq.~\eqref{eq_im_dnr}
to see how, for a sufficiently dilute non-relativistic plasma, we recover
the particle-by-particle Yukawa scattering rate.
The total Yukawa cross section for particle-by-particle scattering is
\begin{equation}
	\sigma_{\chi e} = \frac{g_\chi^2 g_e^2}{\pi} \frac{\mu_{\chi e}^2}
	{m_X^2 (m_X^2 + 4 k_{\rm CM}^2)}
\end{equation}
where $\mu_{\chi e}$ is the DM-electron reduced mass,
and $k_{\rm cm}$ is the momentum in the CoM frame.
If we consider a heavy mediator, and assume
that the relative velocity is dominated
by the DM velocity $v_\chi$ in the plasma rest frame, then
the scattering rate is
\begin{equation}
	\Gamma_L \simeq n_e \sigma_{\chi e} v_\chi
	\simeq \frac{g_\chi^2 g_e^2}{\pi} \frac{\mu_{\chi e}^2}{m_X^4}
	n_e v_{\chi}
\end{equation}
From Eq.~\eqref{eq_gl_nr}, the rate for non-relativistic DM
scattering, in terms of the mediator self-energy, is
\begin{equation}
	\Gamma_L \simeq
	\frac{g_\chi^2}{2 \pi^2}
	\frac{1}{v_\chi} \int dq \, dq_0 \, q 
	(1 + f(q_0))
	\frac{1}{(q^2 + m_X^2)^2} \Imag \Pi_L^X(Q)
\end{equation}
If the medium is dilute enough for the mixing
terms in $\Imag \Pi^X_L$ to be unimportant, then
using Eq.~\eqref{eq_im_dnr}, this gives
\begin{equation}
	\Gamma_L \simeq
	\frac{1}{2 \pi^2} \sqrt{\frac{\pi}{2}} 
	\frac{g_e^2 g_\chi^2 n_e}{v_\chi \sigma m_X^4}
	\int dq \int dq_0 \, e^{-(\xi - \delta)^2}
	= 
	\frac{1}{4 \pi} 
	\frac{g_e^2 g_\chi^2 n_e}{v_\chi m_X^4}
	q_{\rm max}^2
	= 
	\frac{g_\chi^2 g_e^2}{\pi} 
	\frac{\mu_{\chi e}^2 n_e v_\chi}{m_X^4}
\end{equation}
matching the standard Yukawa rate. A useful aspect of the 
self-energy calculation is that it is very simple to take into account
the electron velocity distribution and collective effects.

\subsubsection{Transverse modes}

For non-relativistic electrons, the transverse self-energy is given by
\begin{equation}
\Pi_T \simeq  \omega_p^2 \int d^3 v \, p(v)
\frac{-Q^2 v^2 \cos^2\phi \sin^2\theta
+ (q_0 - v q \cos\theta)^2}{(q_0 - v \cdot q)^2 - \frac{Q^4}{4 m_e^2}}
\end{equation}
where $\theta$ is the angle between $q$ and $v$.
Doing the angular integrals, we obtain
\begin{equation}
	\Pi_T \simeq \omega_p^2 \int dv_q p_1(v_q) 
	\frac{(q_0 - q v_q)^2 - Q^2 \sigma^2}{(q_0 - q v_q)^2 - \frac{Q^4}{4 m_e^2}}
\end{equation}
We can rewrite this in terms of $\Pi_L$,
\begin{align}
	\Pi_T &\simeq \omega_p^2 \int dv_q p_1(v_q) 
	\left(
	1 - \frac{Q^2 \sigma^2 - \frac{Q^4}{4 m_e^2}}{(q_0 - q v_q)^2 - \frac{Q^4}{4 m_e^2}}
	\right)
	= \omega_p^2 \left(1 - Q^2\left(\sigma^2 - \frac{Q^2}{4 m_e^2}\right)
	S\right)
	\nonumber \\
	&= \omega_p^2 \left(1 - \left(\sigma^2 - \frac{Q^2}{4m_e^2}\right)
	\frac{m_e}{q_0}\xi(Z(\xi - \delta) - Z(\xi + \delta))\right)
	\nonumber \\
	&= \omega_p^2  - \left(\sigma^2 - \frac{Q^2}{4m_e^2}\right)
	\Pi_L
\end{align}
A distinct qualitative feature of the transverse
modes is their lack of static screening.
Taking the limit of $\xi$ small, so $\xi Z(\xi) \sim -i \sqrt{\pi} \xi$,
we have
\begin{equation}
	\Pi_T \simeq - i \omega_p^2 \sqrt{\frac{\pi}{2}} \frac{q_0}{q \sigma}
\end{equation}
which goes to zero as $q_0 \rightarrow 0$, for fixed $q$.
Consequently, there is no screening of static fields,
only `dynamical screening' of finite-frequency perturbations
(this is actually a general property
of QED plasmas, which holds at all orders in perturbation theory~\cite{bellac}).


\section{Scalar self-energies}
\label{app_scalar}

In this section, we calculate the leading-order
(one-loop) self-energies
for a scalar mediator, coupling
as $\LL \supset \phi \bar f f$, to Dirac fermions $f$.
As well as the scalar-scalar self-energy $\Pi^{\pi\phi}$,
we also calculate the mixing self-energy
$\Pi^{\phi \mu}$ with the SM photon.
Similarly to the vector mediator case
considered in Appendix~\ref{app_vse},
the one-loop result will be a good approximation
for weakly-coupled plasmas. This
is less obvious in the scalar case, since
the mixing self-energy
is proportional to
the (in-vacuum) electron mass $m_e$;
in cases where $\alpha E^2 \gg m_e^2$,
where $E$ is a typical electron energy, one might
worry that higher-order contributions
dominate (as assumed in some papers~\cite{klz}).
However, we show that corrections
to the one-loop result are subleading
(with consequences as discussed in Appendix~\ref{sec_res_scalar}).

\subsection{One-loop free fermion}
\label{app:oneloop_free_f}
The scalar-scalar self-energy is given by
\begin{equation}
     \Pi^{\phi\phi}(Q)=4g_\phi^2\int\frac{\di^3 k}{(2\pi)^3}\frac{1}{2E_k}\pare{f_e(E_k)+f_{\bar{e}}(E_k)}\frac{(Q\cdot K)^2-m_e^2Q^2}{(Q\cdot K)^2-Q^4/4}
	 \label{eq_pi_pp}
\end{equation}
while the scalar-vector mixing self-energy is
\begin{equation}
	\Pi^{\phi\mu}(Q) = 4g_\phi e m_e \int \frac{d^3 k}{(2\pi)^3}
	\frac{1}{2 E_k}
	(f_e(E_k) - f_{\bar e}(E_k))
	\frac{(Q \cdot K) Q^\mu - Q^2 K^\mu}
		{(Q \cdot K)^2 - Q^4/4}
		\label{eq_pi_px}
\end{equation}
We note here that the expressions \eqref{eq_pi_pp} and \eqref{eq_pi_px} correct some typos in the corresponding expressions from \cite{1611.05852}. In particular, both have an extra overall factor of $4$, and \eqref{eq_pi_px} has a minus sign between the electron and positron phase-space distributions. One way to understand this relative sign is to remember that, in the non-relativistic limit, a Yukawa potential is universally attractive, whereas a vector mediator couples to particles and antiparticles with opposite charges. While in diagrams such as $\Pi^{AA}$ and $\Pi^{\phi\phi}$ this does not appear, as the relative sign gets squared, in the mixing diagram $\Pi^{\phi\mu}$ it survives.

\subsubsection{Imaginary parts}

For an isotropic medium, the imaginary parts are
\begin{equation}
\Imag \Pi^{\phi\phi}(Q) =- \frac{g_\phi^2}{4\pi}  \frac{\sgn(q_0)}{1 + f(q_0)} \frac{2 m_e^2 - Q^2/2}{q}
	\int dE \sgn(E) \sgn(E') (1 - \tilde f(E)) \tilde f(E')
	\label{phiphiim}
\end{equation}
and
\begin{align}
	\Imag\Pi_{\phi A}^{\mu}(Q)
	&=-\frac{\sgn(q_0)}{1+f(q_0)}\frac{g_\phi em_e}{4\pi q}\int\di E\,\pare{2K^\mu-Q^\mu}\sgn(E)\sgn(E') (1-\tilde{f}(E))\tilde{f}(E'),
\end{align}
where the integration is over $R_Q$, as defined
in Section~\ref{sec_vec_imag}.

The transverse part of $\Pi^\mu_{\phi,A}$ does not contribute in the corrections to the scalar propagator (there is no preferred direction). The longitudinal projection vector depends on the sign of $Q^2$. Using the convention

\begin{equation}
    e_{L,\mu}\equiv-\frac{1}{q\sqrt{Q^2}}(q^2,q_0 q_i)=-\frac{1}{q}(q^2,q_0 q_i)\times
    \left\{
\begin{array}{ll}
      \frac{1}{\sqrt{Q^2}}, & Q^2>0  \\
      \frac{1}{i\sqrt{|Q^2|}},& Q^2<0
\end{array} 
\right. 
\end{equation}
we find
\begin{align}
	\Imag \Pi_{\phi A}^L(Q) &=- \frac{g_\phi e}{4\pi}  \frac{\sgn(q_0)}{1 + f(q_0)} \frac{m_e Q^2}{q^2\sqrt{Q^2}}
	\int dE \pare{2E-q_0}\sgn(E) \sgn(E') (1 - \tilde f(E)) \tilde f(E')\\
	&\qquad\qquad\qquad\qquad\qquad\qquad\qquad\qquad\qquad\qquad\text{if } Q^2>0\nonumber\\
	&=-\frac{g_\phi em_e}{2\pi^2}\frac{Q^2}{q^2\sqrt{|Q^2|}}\int\di E\,\parea{f_{e}(E)-f_{\bar{e}}(E)}\left[E\log\left|\frac{\pare{x_--\frac{Q^2}{2}}\pare{x_++\frac{Q^2}{2}}}{\pare{x_-+\frac{Q^2}{2}}\pare{x_+-\frac{Q^2}{2}}}\right|+\right.\nonumber\\
    &\qquad\qquad\qquad\qquad\qquad\qquad\qquad+\left.\frac{q_0}{2}\log\left|\frac{\pare{x_++\frac{Q^2}{2}}\pare{x_+-\frac{Q^2}{2}}}{\pare{x_-+\frac{Q^2}{2}}\pare{x_--\frac{Q^2}{2}}}\right|\right]\\
    &\qquad\qquad\qquad\qquad\qquad\qquad\qquad\qquad\qquad\qquad \text{if } Q^2<0\nonumber
	\label{phiAim}
\end{align}


\subsubsection{Real parts}

The real part of the scalar-scalar self-energy in an
isotropic medium is
\begin{align}
    \text{Re} \Pi_{\phi\phi}=4g_\phi^2\int_{m_e}^{+\infty}\frac{\di E }{(2\pi)^2}\pare{\tilde{f}_e(E)+\tilde{f}_{\bar{e}}(E)}\parea{k+\frac{m_e^2 -\frac{Q^2}{4}}{2q}\log\left|\frac{\pare{x_+ + \frac{Q^2}{2}}\pare{x_- -\frac{Q^2}{2}}}{\pare{x_+ - \frac{Q^2}{2}}\pare{x_- +\frac{Q^2}{2}}}\right|}
\end{align}
while the mixing self-energy with a longitudinal vector mode is
\begin{align}
    \text{Re} \Pi^L_{\phi A}=\frac{g_\phi em_e}{2\pi^2}\frac{Q^2}{q^2\sqrt{Q^2}}\int\di E\,\pare{f_{e}(E)-f_{\bar{e}}(E)}&\left[E\log\left|\frac{\pare{x_--\frac{Q^2}{2}}\pare{x_++\frac{Q^2}{2}}}{\pare{x_-+\frac{Q^2}{2}}\pare{x_+-\frac{Q^2}{2}}}\right|+\right.\nonumber\\
    &+\left.\frac{q_0}{2}\log\left|\frac{\pare{x_++\frac{Q^2}{2}}\pare{x_+-\frac{Q^2}{2}}}{\pare{x_-+\frac{Q^2}{2}}\pare{x_--\frac{Q^2}{2}}}\right|\right]\\
    &\text{if } Q^2>0\nonumber\\
    =- \frac{g_\phi e}{4\pi}  \frac{\sgn(q_0)}{1 + f(q_0)} \frac{m_e Q^2}{q^2\sqrt{|Q^2|}}
	\int dE  \pare{2E-q_0}&\sgn(E) \sgn(E') (1 - \tilde f(E)) \tilde f(E')\\
	&\text{if } Q^2<0\nonumber
\end{align}


\subsection{Degenerate Fermi gas -- Small-$Q$ approximations}

Similarly to Appendix~\ref{app_degen_gas}, we can evaluate
the self-energies analytically for a degenerate Fermi gas.
Here, we write down the small-$Q$ approximations:
\begin{align}
\Real \Pi_{\phi\phi}&\simeq    \frac{g_\phi^2}{2\pi^2}\parea{E_F p_F+3m_e^2\log\frac{m_e}{E_F+p_F}+m_e^2\frac{q_0}{q}\log\left|\frac{E_F q_0+ p_F q}{E_F q_0-p_F q}\right|}\\
\Imag \Pi_{\phi\phi}&\simeq-\frac{g_\phi^2}{2\pi}m_e^2\frac{q_0}{q}\Theta\pare{\left|\frac{v_Fq}{q_0}\right|-1}
    \end{align}
For $Q^2 < 0$, the mixing self-energy is
\begin{align}
\Real\Pi_{\phi A}^L(Q)&\simeq-\frac{g_\phi^2}{2\pi}E_Fm_e\frac{q_0}{q}\frac{q}{\sqrt{|Q^2|}}\Theta\pare{\left|\frac{v_Fq}{q_0}\right|-1}\\
    \Imag\Pi_{\phi A}^L(Q)&\simeq-\frac{g_\phi em_e}{\pi^2}\frac{Q^2}{q^2}\frac{q}{\sqrt{|Q^2|}}p_F\pare{-1+\frac{z_F}{2}\log\left|\frac{1+z_F}{1-z_F}\right|}
\end{align}
while for $Q^2 > 0$, 
\begin{align}
    \Real\Pi_{\phi A}^L(Q)&\simeq\frac{g_\phi em_e}{\pi^2}\frac{Q^2}{q^2}\frac{q}{\sqrt{Q^2}}p_F\pare{-1+\frac{z_F}{2}\log\left|\frac{1+z_F}{1-z_F}\right|}\\
    \Imag\Pi_{\phi A}^L(Q)&\simeq -\frac{g_\phi^2}{2\pi}E_F m_e\frac{q_0}{q}\frac{q}{\sqrt{Q^2}}\Theta\pare{\left|\frac{v_Fq}{q_0}\right|-1}
\end{align}


\subsection{Electron mass corrections}

As mentioned at the start of this Appendix,
for weakly-coupled plasmas it is sufficient
to use the one-loop results for the self-energies.
In particular, the scalar-vector mixing self-energy
$\Pi^{\phi L}$ for an electron-coupled scalar is proportional to $m_e$, the in-vacuum
electron mass, even when $m_e$ is much smaller than collective
scales in the plasma. This is
because the scalar coupling
$\phi \bar \psi \psi = \phi (\bar \psi_R \psi_L
+ \bar \psi_L \psi_R)$ mixes chiralities, whereas
vector couplings preserve chirality,
$X_\mu\bar{\psi}\gamma^\mu \psi=X_\mu\bar{\psi}_L\gamma^\mu \psi_L+ X_\mu\bar{\psi}_R\gamma^\mu \psi_R$.
If electrons were massless, there would be no
way to match these chiralities, so the mixing self-energy
would be zero.

To demonstrate the stronger claim that the one-loop
result is quantitatively a good approximation,
rather than just scaling in the correct way with $m_e$, we
perform an illustrative calculation for
the case of a relativistic degenerate plasma.
The most important corrections to the one-loop
result come from replacing the electron propagators
in the loop with in-medium propagators.
The leading contribution to the electron self-energy
$\Sigma(K)$ is
\begin{center}
	\begin{tikzpicture}[line width=1]
          \begin{scope}[shift={(2.4,0)}]
            \draw[f] (2*0.33,0)  node[above] {$e$}
            -- (2*1.2,0);
            \node[] at (1.4,-0.3) {$K$};
            \node[] at (3.0,-0.3) {$P$};
            \node[] at (4.5,-0.3) {$K$};
			\draw[f] (2*1,0) -- (2*2,0);
			  \draw[f] (2*2,0) -- (2*2.66,0) node[above] {$e$};
            \draw[v] (2*2,0) arc (0:180:2*0.5);
          \end{scope}
	\end{tikzpicture}
\end{center}
where $K$ is the electron four-momentum.
The in-medium propagator $S(K)$ is given by
$-iS^{-1}(K)=\slashed{K}-m_e-\Sigma(K)$.
For a degenerate electron gas, the integral is dominated
by placing the internal electron with momentum $P$ on-shell
(rather than placing the photon on-shell, or considering positrons),
and the real part
of the self-energy $\Sigma$ at one-loop order is given by
\begin{equation}
    \Sigma(K)=2e^2\int\frac{\di^3 p}{(2\pi)^3}\frac{1}{2E_p}\frac{2m_e-\slashed{P}}{(K-P)^2}f_e(E_p),
\end{equation}
where $E_p=\sqrt{p^2+m_e^2}$ and $K\equiv(k_0,\bold{k})$. We define an expansion of $\Sigma$ in $\gamma$-matrices as $\Sigma(K)\equiv\gamma^0\Sigma_0+\gamma^ik_i\Sigma_k+m_e\Sigma_m
\equiv \slashed \Sigma + m_e \Sigma_m$. The three terms are given by
\begin{align}
    \Sigma_m&=\frac{\alpha_{\rm EM}}{\pi k}\int_{m_e}^{+\infty} \di E \log\left|\frac{m_e^2+k_0^2-k^2-2Ek_0+2pk}{m_e^2+k_0^2-k^2-2Ek_0-2pk}\right|f_e(E),
    \label{eq:Sigma_m}\\
    \Sigma_0&=-\frac{\alpha_{\rm EM}}{2\pi k}\int_{m_e}^{+\infty} E\di E \log\left|\frac{m_e^2+k_0^2-k^2-2Ek_0+2pk}{m_e^2+k_0^2-k^2-2Ek_0-2pk}\right|f_e(E),\label{eq:Sigma_0}\\
    \Sigma_k&=-\frac{1}{k^2}\parea{-\frac{m_e^2+K^2}{4}\Sigma_m-k_0\Sigma_0+\frac{\alpha_{\rm EM}}{\pi}\int\di E\, p f_e(E)},
    \label{eq:Sigma_q}
\end{align}
where $p\equiv \sqrt{E^2-m_e^2}$ and $k\equiv|\bold{k}|$. 
In the zero temperature limit, the electron phase-space distribution is the step function $f_e(E)=\Theta(E_F-E)$.

In the one-loop expression for $\Pi^{\phi L}$ with in-medium
electron propagators, the relevant Dirac trace is
(for electron momenta $K$ and $K-Q$ in the loop)
\begin{align}
	\Pi^{\phi \mu}(Q) &\propto
	\tr \left[ (\slashed K - \slashed \Sigma(K) + m_e (1 + \Sigma_m(K)))
	\gamma^\mu (\slashed K - \slashed Q - \slashed \Sigma(K-Q)
	+ m_e(1 + \Sigma_m(K-Q))\right]
	\nonumber \\
	&= 4 m_e \left((1 + \Sigma_m(K-Q)) (\slashed K - \slashed
	\Sigma (K)) + (1 + \Sigma_m(K)) (\slashed K - \slashed Q
	- \slashed \Sigma (K-Q))\right)
\end{align}
showing how this is proportional to the in-vacuum mass $m_e$.
We are interested in evaluating this for $K$ on 
the electron's in-medium dispersion
relation~\cite{10.1086/171405}. For $Q$ small, this means that the $\slashed \Sigma$
terms are $\sim e E_F$, so compared to the electron momenta
$\sim E_F$ that dominate the loop integral, they do not change
the result significantly (this is analogous to the
argument for why the one-loop vector self-energies
do not receive large corrections).
The remaining question is how large the $\Sigma_m$
terms are.

\begin{figure}[t]
	\begin{center}
\includegraphics[width=0.6\textwidth]{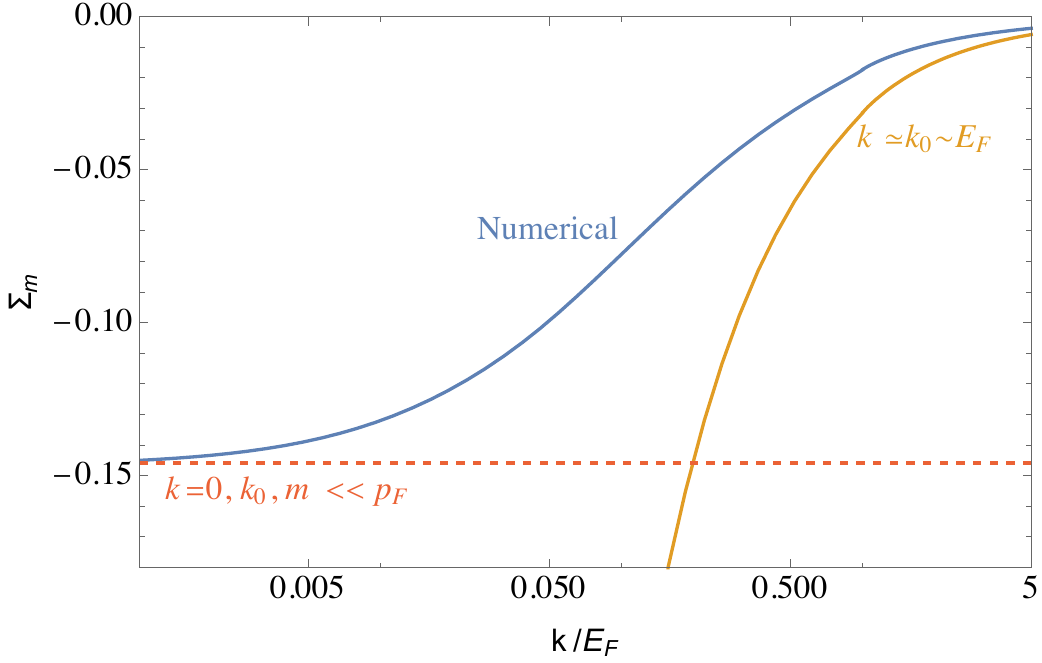}
		\caption{Plot of the quantity $\Sigma_m$, as defined in Eq.~\eqref{eq:Sigma_m}, for $\mu_e=350$ MeV. The dashed line corresponds to the zero momentum approximation of Eq.~\eqref{eq:Sigma_m_zero_k}, whereas the orange line corresponds to relativistic electrons and the approximation of Eq.~\eqref{eq:Sigma_m_q0=q}. The blue line uses the full integral of Eq.~\eqref{eq:Sigma_m} and the full numerical dispersion relation. This plot illustrates that corrections to the scalar-vector self energy $\Pi^{\phi L}$ from
		the `modified electron mass' are small.}
    \label{fig_Sigma_m}
	\end{center}
\end{figure}

In Figure~\ref{fig_Sigma_m}, we plot $\Sigma_m(k_0(k),k)$ (evaluated 
via Eq.~\eqref{eq:Sigma_m}), solving numerically for the dispersion
relation $k_0(k)$, for an ultra-relativistic degenerate electron gas.
It is clear that the correction to $\Pi^{\phi\mu}$ is small for the entire range of $k_0,k$.
We can understand some limiting cases analytically. For applications of interest, the electrons contributing the most have energies $\sim E_F\gg m_e$, for which $k_0\simeq k$. In the limit $k_0\simeq k\gg m_e$, we find
\begin{equation}
    \Sigma_m\simeq\frac{\alpha_{\rm EM}E_F}{\pi k}\parea{1+\log\frac{m_e^2}{4E_F^2}+\frac{E_F-k}{E_F}\log\frac{|E_F-k|}{k}}
    \label{eq:Sigma_m_q0=q}
\end{equation}
For smaller electron momenta $k$, the dispersion relation lies above the light-cone, so we can get some analytic handle in this case by considering the limit $k_0\gg k$. Taking also $ep_F\gg m_e$, the $k_0>0$ dispersion relation and $\Sigma_m$ are
\begin{equation}
    k_0\simeq \frac{ep_F}{2\sqrt{2}\pi},\quad\Sigma_m\simeq-\frac{2\alpha_{\rm EM} p_F}{\pi k_0}\pare{1+\frac{k_0}{2p_F}\log\frac{2p_F}{k_0}}\simeq- \frac{e\sqrt{2}}{\pi}\pare{1+\frac{e}{4\pi\sqrt{2}}\log\frac{4\pi\sqrt{2}}{e}},
    \label{eq:Sigma_m_zero_k}
\end{equation}
As we see from Figure~\ref{fig_Sigma_m}, these approximations
match the full integral well in their regions of validity.

\section{EM sum rule limits}
\label{app_sumrules}

Causality imposes constraints on the properties
of in-medium propagators, and as discussed in
Section~\ref{sec_emsum}, this allows
us to place constraints on
certain types of scattering rates. In particular,
for a dark photon mediator, we have
$\Imag (\Pi^{XX}_{\rm tot})_{\mu\nu} =  \kappa^2 Q^4 \Imag
\left(-i D_{\mu\nu}\right)$, where
$D_{\mu\nu}$ is the in-medium propagator
for the SM photon, in Lorenz gauge.
More specifically, $D_{\mu\nu}$ is the analytic continuation
of the in-medium imaginary-time propagator.
In the real-time formalism,
we can write the photon propagator $\left(D_{\mu\nu}\right)_{ab}$
as~\cite{bellac}
\begin{equation}
	\left(D_{\mu\nu}\right)_{ab} (Q)
	= \begin{pmatrix}
		\sqrt{n(q_0) + 1} & \sqrt{n(q_0)} \\
		\sqrt{n(q_0)} & \sqrt{n(q_0)+1} 
	\end{pmatrix}
	\begin{pmatrix}
		D_{\mu\nu} & 0 \\
		0 & \left(D_{\mu\nu}\right)^*
	\end{pmatrix}
	\begin{pmatrix}
		\sqrt{n(q_0) + 1} & \sqrt{n(q_0)} \\
		\sqrt{n(q_0)} & \sqrt{n(q_0)+1} 
	\end{pmatrix}
\end{equation}
where $n(q_0) \equiv (e^{|q_0|/T} - 1)^{-1}$,
so $D_{\mu\nu}$ corresponds to the `diagonal component'.


Consequently, to constrain the DM scattering rate
via a dark photon mediator, we are interested in the properties
of the SM photon propagator. In general, if we are allowed
to tune the medium properties and the initial DM velocity,
we can obtain arbitrarily large scattering rates, via
matching the on-shell momentum transfers possible
for the DM particle to the dispersion of weakly-damped
excitations in the medium (so that we obtain resonant
scattering at all momentum transfers).
However, if we are interested in the average scattering
rate across different incoming DM directions, such tuning is no longer
possible, and as we will see, it is possible to place
general limits on the scattering rate.
Equivalently, instead of thinking in terms of 
averaging over different DM directions, we can consider
a DM particle scattering from an isotropic medium (with 
mediator self-energy obtained by averaging over different
orientations of the original medium).

In an isotropic medium, we can decompose
$D_{\mu\nu}$ into transverse and longitudial parts,
\begin{equation}
	D_{\mu\nu}(Q)
	= \frac{i}{Q^2 - \Pi_T(Q)} P^T_{\mu\nu}
	+ \frac{i}{Q^2 - \Pi_L(Q)} P^L_{\mu\nu}
\end{equation}
where $P^{T,L}_{\mu\nu}$ are the transverse and longitudinal
projectors, and $\Pi_{T,L}$ are the transverse
and longitudinal photon self-energies, respectively
(we have elided the $i \epsilon$ prescription for evaluating
contour integrals, since these ambiguities do not affect
our calculations).
We can also define (following~\cite{bellac})
the `spectral density' quantities $\rho_L(Q) \equiv 2 \frac{Q^2}{q^2} \Imag D_L(Q)$,
and $\rho_T(Q) \equiv 2 \Imag D_T(Q)$,
where $D_L(Q) = \frac{-1}{Q^2 - \Pi_L(Q)}$,
$D_T(Q) = \frac{-1}{Q^2 - \Pi_T(Q)}$. 
Both $\rho_L$ and $\rho_T$ are always non-negative,
for both timelike and spacelike $Q$
(this is the point
of the $Q^2/q^2$ factor in the definition
of $\rho_L$).
From the Kramers-Kronig relations, we have the `sum rules'
\begin{equation}
	\int_0^\infty \frac{dq_0}{q_0} \rho_T(q_0,q) = \frac{\pi}{q^2}
	\label{eq_srt1}
\end{equation}
\begin{equation}
	\int_0^\infty \frac{dq_0}{q_0} \rho_L(q_0,q) = \pi
	\left(\frac{1}{q^2} - D_L(0,q)\right)
	= \frac{\pi k_S^2}{q^2 (q^2 + k_S^2)}
	\label{eq_srl1}
\end{equation}
which hold for any $q$, where $k_S^2 = \Pi_L(0,q)$ is
the static longitudinal screening scale (so $D_L(0,q) = \frac{1}{q^2 + k_S^2}$). Since $\rho_{L,T} \ge 0$, the integral
over any range of $q_0$ is also bounded by the corresponding RHS.
The $\rho_{L,T} \ge 0$ condition also implies that either $k_S^2 \ge 0$, or
$k_S^2 \le - q^2$ --- intermediate values would violate
the positivity of the LHS.
If $k_S^2 \ge 0$, corresponding to screening (rather than 
anti-screening) of static fields, then 
the RHS of Eq.~\eqref{eq_srl1} is $\le \pi/q^2$.
While $\Pi(0,q)$ should be non-negative at small enough $q$
for a stable system~\cite{10.1103/RevModPhys.53.81}, it is possible to have
$\Pi(0,q) < 0$ at $q$ comparable to lattice scales ---
for example, this is probably the case for some metals,
such as aluminium~\cite{10.1103/RevModPhys.53.81}. 
However, to obtain $k_S^2$ very slightly below
$-q^2$, which is required
for the RHS of Eq.~\eqref{eq_srl1} to be $\gg \pi/q^2$,
would require very strong antiscreening, which is most
likely only possible for a system very close to instability~\cite{10.1103/RevModPhys.53.81}.
In this work, we assume that $\pi/q^2$ is a good approximate
bound for the RHS of Eq.~\eqref{eq_srl1}
(as is the case for the toy models we consider), and leave
a more thorough investigation for future work.

Given Eqs.~\eqref{eq_srl1} and~\eqref{eq_srt1},
we can place bounds on the scattering rate via a dark photon mediator.
From Appendix~\ref{app_srates},
the longitudinal scattering rate is
\begin{equation}
	\Gamma_L = \frac{-1}{4\pi^2} \frac{g_\chi^2 \kappa^2}{E p}
	\int dq dq_0 \, q  
	(1 + f(q_0)) \frac{Q^2 q^2}{(Q^2 - m_X^2)^2} (E^2 - p^2 \cos^2 \theta)
	\rho_L(q_0,q)
	\label{eq_gl_sr1}
\end{equation}
For a given $q$, we can write the $\int dq_0$ integral as
\begin{equation}
	I_0 \equiv \int \frac{dq_0}{q_0} \rho_L(q_0,q) \left(
	\frac{-Q^2 q^2}{(Q^2 - m_X^2)^2} (1 + f(q_0)) q_0 (E^2 - p^2 \cos^2\theta)\right)
	\label{eq_q0int}
\end{equation}
From Eq.~\eqref{eq_srl1}, this can be bounded by $\pi/q^2$ times the maximum
of the bracketed term within the relevant $q_0$ interval.
To take a simple example, for non-relativistic scattering,
the bracketed term is approximately equal to
\begin{equation}
	 \frac{m_\chi^2 q^4}{(q^2 + m_X^2)^2}
	\frac{q_0}{1 - e^{-q_0/T}}
\end{equation}
This is maximized by taking $q_0$ to be as large as it can be,
for the given $q$.
If we take the mediator to be heavy,
and the temperature to be negligible, then the scattering rate is bounded
by
\begin{equation}
	\Gamma_L \lesssim \frac{1}{4\pi} \frac{g_\chi^2 \kappa^2}{v_\chi m_X^4}
	\int_0^{2 m_\chi v_\chi} dq \, q^3 \left(q v_\chi - \frac{q^2}{2 m_\chi}\right)
	= \frac{16}{15} 
	\frac{g_\chi^2 \kappa^2}{4\pi} m_\chi v_\chi
	\left(\frac{m_\chi v_\chi}{m_X}\right)^4
\end{equation}
as derived in~\cite{ani}. Similarly, for a light mediator, we have
\begin{equation}
	\Gamma_L \lesssim
	\frac{g_\chi^2 \kappa^2}{4\pi} m_\chi v_\chi
\end{equation}

For relativistic $\chi$, the bracketed term 
in Eq.~\eqref{eq_q0int} might not be maximized
at the maximum value $q_{0,\rm max}$ of $q_0$, for given $q$.
We attain $q_0 = q_{0,\rm max}$ at $\cos\theta = 1$, where
$\theta$ is the scattering angle.
Consequently, the $E^2 - p^2 \cos^2 \theta$ term in Eq.~\eqref{eq_gl_sr1}
is $= m_\chi^2$, which can be much smaller than its
maximum value of $E^2$, if $E \gg m_\chi$.
Consequently, it can be beneficial to take $\cos \theta$
somewhat smaller than 1. 
For example,
in the case of a heavy mediator, the derivative of the bracketed
term in Eq.~\eqref{eq_q0int} with respect to $q_0$,
evaluated at $q_0 = q_{0,\max}$, is 
\begin{equation}
	\frac{E^2 q^4}{m_X^4}\left( (1 - 6 v_\chi^2 + 5 v_\chi^4) + 
	\OO(q/E))
	\right)
\end{equation}
where we have expanded in large $E$. For small $q$, we can
see that if $v_\chi \ge 1/\sqrt{5} \simeq 0.45$, then
the derivative at $q_{0,\rm max}$ is negative,
so the quantity is maximized at some smaller $q_0$.
When this is true, we can still derive sum rule limits by optimizing
numerically.
Figure~\ref{fig_lims1} shows the results of such numerical calculations;
in the ultra-relativistic limit, the scattering rate
upper-bound scales $\sim g_\chi^2 \kappa^2 E^5 / m_X^4$
for a heavy mediator, and $\sim g_\chi^2 \kappa^2 E$ for a light mediator).

We can incorporate a non-zero medium temperature
by keeping the $1 + f(q_0)$ factor 
in Eq.~\eqref{eq_q0int}. However, as noted 
in Section~\ref{sec_capture_rates}, 
if the temperature is large enough to significantly
affect scattering rates,
then one is often interested in the \emph{net}
capture rate once upscattering has been taken
into account.

\subsection{Transverse limits}

\begin{figure}[t]
	\begin{center}
\includegraphics[width=0.6\textwidth]{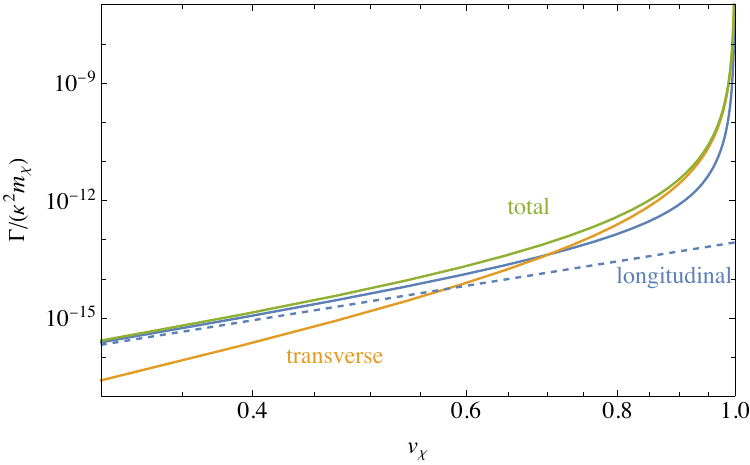}
		\caption{Plot of sum rule limits on the scattering
		rate for spin-$1/2$ DM via a heavy dark photon mediator
		(specifically, taking $m_X  = 10^3 m_\chi$),
		in a cold medium, as
		a function of the DM particle's velocity $v_\chi$.
		The green curve corresponds to the total
		limit, while the blue and orange curves
		correspond to the limits on the scattering
		rate via longitudinal and transverse modes
		respectively. The dotted
		blue line corresponds to the 
		$\Gamma \propto v_\chi^5$ limit
		for non-relativistic velocities.}
    \label{fig_lims1}
	\end{center}
\end{figure}

We can put similar bounds on the scattering rate
via transverse dark photon modes. The total scattering
rate is given by
\begin{equation}
	\Gamma_T = \frac{1}{8\pi^2} \frac{g_\chi^2 \kappa^2}{E p}
	\int dq dq_0 \, q  
	(1 + f(q_0)) \frac{Q^4}{(Q^2 - m_X^2)^2} (-Q^2 + 2 p^2 \sin^2 \theta)
	\rho_T(q_0,q)
\end{equation}
So, for given $q$, the $\int dq_0$ integral is
\begin{equation}
	I_0 \equiv \int \frac{dq_0}{q_0} \rho_T(q_0,q) \left(
	\frac{Q^4}{(Q^2 - m_X^2)^2} (1 + f(q_0)) q_0 (-Q^2 + 2 p^2 \sin^2\theta)\right)
\end{equation}
and we want to maximize the bracketed term over the relevant
$q_0$ range.
Since the $-Q^2 + 2p^2 \sin^2 \theta$ term is somewhat more complicated,
in the non-relativistic case, than the equivalent in the longitudinal case,
we simply optimize numerically. Doing so, as illustrated
in Figure~\ref{fig_lims1}, we find that in the non-relativistic
limit, the bound scales as 
\begin{equation}
	\Gamma_T \lesssim 1.16 \times \frac{g_\chi^2 \kappa^2}{4\pi}
	\frac{m_\chi^5 v_\chi^7}{m_X^4}
\end{equation}
for a heavy mediator (the parametric scaling can be obtained
from taking the loose bound $-Q^2 + 2 p^2 \sin^2 \theta \le q^2 + 2 p^2$;
the numerics provide the constant)
For a light mediator, we have
$\Gamma_T \lesssim 0.53 \times \frac{g_\chi^2 \kappa^2}{4\pi}$.
We can see that at small $v_\chi$, the transverse
bounds are suppressed by $\sim v_\chi^2$ relative
to the longitudinal bounds, as expected
from the forms of $\Gamma_L$ and $\Gamma_T$.
Numerically, the $\Gamma_T$ limit (for a heavy mediator) becomes larger
than the $\Gamma_L$ limit for $v \gtrsim 0.7$ ---
at ultra-relativistic velocities,
$\Gamma_T$ is $\sim 7$ times larger.

\section{DM velocity distribution}
\label{app_vel_dist}

For our purposes, the most important quantity related
to the DM halo velocity distribution is
\begin{equation}
	F_\infty(\delta) \equiv
	\int_0^\delta dE_K f_\infty(m_\chi + E_K)
\end{equation}
as introduced in Section~\ref{sec_capture_rates}.
This definition is phrased in a manner appropriate to isotropic
DM velocity distributions, for which $f_\infty$ 
is purely a function of the energy.
However, in most cases, a star will have 
some non-zero velocity
relative to the DM halo (which will not necessarily 
have an isotropic velocity distribution in any case~\cite{10.1103/PhysRevD.73.023524,10.1111/j.1365-2966.2008.13441.x,10.1111/j.1365-2966.2011.19008.x,10.1088/0004-637X/752/2/141,10.1088/1475-7516/2012/10/049,10.1093/mnras/stt1113}).
Nevertheless, if the orientation of the star is unimportant
for the capture rate
(e.g.\ if the star is spherical, and the
medium response is isotropic, both of which
are reasonable approximations in the parameter
space of interest to us), then we can imagine
averaging over different orientations of the halo
DM velocity distribution relative to the star,
to obtain an isotropic velocity distribution in
the rest frame of the star which would give the same capture rate.
Consequently, we can take the more general definition
\begin{equation}
	F_\infty(\delta)
	= \frac{m_\chi}{4\pi}
	\int_0^{v_\delta} d^3 v \frac{f_\infty(m_\chi v)}{v}
	 = \frac{2 \pi^2 n_\chi}{g_s m_\chi^2} \int_0^{v_\delta} d^3v
	 \frac{p_v(v)}{v}
	 \label{eq_fipv}
\end{equation}
where $v_\delta$ is such that $m_\chi v_\delta^2/2 = \delta$
(we assume that the DM halo velocities are non-relativistic),
and we view $f_\infty$ as a function of momentum.

\begin{figure}[t]
	\begin{center}
\includegraphics[width=0.6\textwidth]{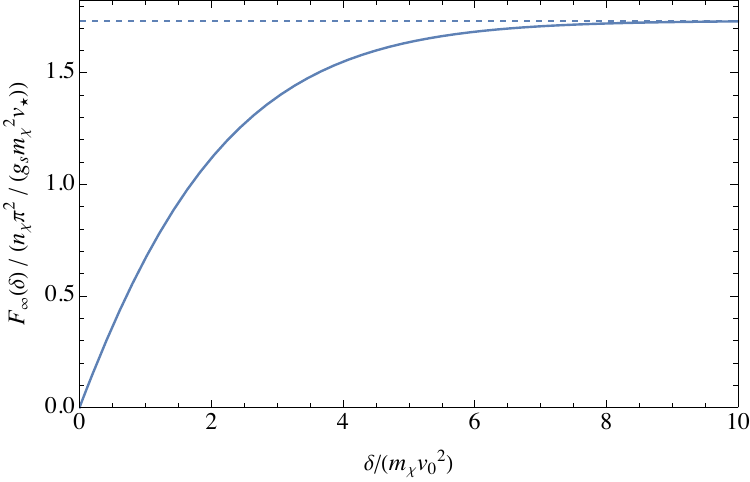}
		\caption{Plot of $F_\infty(\delta)$
		from Eq.~\eqref{eq_finfmw}, for
		DM halo parameters
		$v_0 = 160 \kms$ and $v_\star = 240 \kms$
		as used in the text. The
		dashed curve shows the asymptotic value of $F_\infty$,
		from Eq.~\eqref{eq_finfinf}.}
    \label{fig_finf}
	\end{center}
\end{figure}

In many circumstances, the halo DM velocity distribution in the rest
frame of the star
is reasonably well-modelled by an offset Maxwell distribution~\cite{10.1103/phys-revd.82.023530,10.1142/9789813149441_0007},
of the form
\begin{equation}
	p_v(v) = (2\pi v_0^2)^{-3/2} \exp\left(-\frac{(v - v_\star)^2}{2 v_0^2}\right)
\end{equation}
where $v_0$ is the velocity dispersion of the DM halo,
and $v_\star$ is the relative velocity of the star
(a truncated Maxwell distribution, which is cut
off for velocities higher than the halo escape velocity,
can be a better approximation, but for integrals
such as Eq.~\eqref{eq_fipv}, which are weighted towards smaller
$v$, this will make little difference). Using
this form of the velocity distribution,
we have
\begin{equation}
	F_\infty(\delta) = 
	\frac{n_\chi \pi^2}{g_s m_\chi^2 
	v_\star}
	\left( 2 \erf\left(\frac{v_\star}{\sqrt{2} v_0}\right)
	+ \erf\left(\frac{v_\delta - v_\star}{\sqrt{2} v_0}\right)
- \erf\left(\frac{v_\delta + v_\star}{\sqrt{2} v_0}\right)
	\right)
	\label{eq_finfmw}
\end{equation}
This function is plotted in Figure~\ref{fig_finf},
for the $v_0 = 160 \kms$, $v_\star = 240 \kms$ parameters
used in the text.
In the limit of very small $v_\star$, this becomes
\begin{equation}
	F_\infty(\delta) = \frac{(2\pi)^{3/2} n_\chi}{
		g_s m_\chi^2 v_0} 
	\left(1 - e^{-\delta/(m_\chi v_0^2)}
	\right)
\end{equation}
If capture is dominated by hard scatterings, then the quantity
of interest is 
\begin{equation}
F_\infty(\infty) = 
\frac{2\pi^2 n_\chi}{g_s m_\chi^2} \left\langle \frac{1}{v_\infty}\right\rangle
	\label{eq_finfinf}
\end{equation}
For an offset Maxwell distribution, we have
\begin{equation}
	\left\langle \frac{1}{v_\infty} \right\rangle
	= \frac{1}{v_\star}
 \erf\left(\frac{v_\star}{\sqrt{2} v_0}\right)
	\simeq  \sqrt{\frac{2}{\pi}} \frac{1}{v_0}
\end{equation}
where the last equality applies for $v_\star \ll v_0$.


\section{Resonant scalar emission}
\label{sec_res_scalar}

Though the main subject of this paper is dark matter
scattering, our self-energy calculations can also
be applied to particle emission rates, as per~\cite{ed}.
As an illustrative example, our calculations
show that the rate 
for resonant (electron-coupled) scalar emission from SN1987A in the literature,
calculated in~\cite{klz}, is parametrically too large.
This is because \cite{klz} takes the $\Pi^{\phi L}$ mixing self-energy
to be $\propto m_e^{\rm eff}$, where $m_e^{\rm eff} \simeq 12 \MeV$
is the `effective electron mass' in the supernova core~\cite{10.1086/171405}.
However, as we derived in Appendix~\ref{app_scalar},
the mixing self-energy is actually 
$\sim m_e(1 + \OO(e))$,
where $m_e$ is the in-vacuum electron mass. Physically, this is because
the mixing requires a chirality flip, and so
an insertion of the fermion mass (the in-medium `effective mass'
does not mix chiralities, but is simply a parameterization
of how far the dispersion relation lies from the light-cone).
Numerically, this will result in a resonant emission
rate $\sim 500$ times smaller than calculated
in \cite{klz}, for an equivalent scalar-electron coupling.
We leave a fuller calculation of the emission rate,
and its consequence for scalar coupling constraints,
to future work.


\bibliography{dm_medium}
\bibliographystyle{JHEP}

\end{document}